\documentclass[useAMS,usenatbib]{mn2e}
\usepackage{graphicx}
\title[A Spectroscopic Survey of WNL Stars in the LMC]{A Spectroscopic
Survey of WNL Stars in the LMC: \\ General Properties and Binary
Status} \author[O. Schnurr et al.]{O. Schnurr$^{1,2}$\thanks{Visiting
Astronomer at the Complejo Astron\'omico El Leoncito (CASLEO), the
Cerro Tololo Inter-American Observatory (CTIO), the South African
Astronomical Observatory (SAAO), and the Mount Stromlo and Siding
Spring Observatory (MSSSO)}\thanks{E-mail: o.schnurr@sheffield.ac.uk},
A. F. J. Moffat$^{1}$, N. St-Louis$^{1}$, N. I. Morrell$^{3}$,
\newauthor and M. A. Guerrero-Roncel$^{4}$ \vspace{3mm}\\
$^{1}$D\'ept. de Physique , Universit\'e de Montr\'eal, C. P. 6128,
succ. centre-ville, Montr\'eal (Qc) H3C 3J7, and Centre de
Recherche en\\ Astrophysique du Qu\'ebec, Canada\\ $^{2}$Dept. of
Physics and Astronomy, University of Sheffield, Hicks Building
Hounsfield Road, Sheffield S3 7RH, United Kingdom\\ $^{3}$Las Campanas
Observatory, Observatories of the Carnegie Institution of Washington,
Casilla 601, La Serena, Chile\\ $^{4}$Instituto de Astrof\'isica de
Andaluc\'ia, Consejo Superior de Investigaciones Cient\'ificas,
Apartado Correos 3004, E-18080 Granada,\\Spain}

\begin{document}

\date{Version 16 June 2008}

\pagerange{\pageref{firstpage}--\pageref{lastpage}} \pubyear{2008}

\maketitle

\label{firstpage}

\begin{abstract}
We report the results of an intense, spectroscopic survey of all 41
late-type, nitrogen-rich Wolf-Rayet (WR) stars in the Large Magellanic
Cloud (LMC) observable with ground-based telescopes. This survey
concludes the decade-long effort of the Montr\'eal Massive Star Group
to monitor every known WR star in the Magellanic Clouds except for the
6 crowded WNL stars in R136, which will be discussed elsewhere. The
focus of our survey was to monitor the so-called WNL stars for
radial-velocity (RV) variability in order to identify the short- to
intermediate-period ($P \la 200$ days) binaries among them. Our
results are in line with results of previous studies of other WR
subtypes, and show that the binary frequency among LMC WNL stars is
statistically consistent with that of WNL stars in the Milky Way. We
have identified four previously unknown binaries, bringing the total
number of known WNL binaries in the LMC to nine. Since it is very
likely that none but one of the binaries are classical, helium-burning
WNL stars, but rather superluminous, hence extremely massive,
hydrogen-burning objects, our study has dramatically increased the
number of known binaries harbouring such objects, and thus paved the
way to determine their masses through model-independent, Keplerian
orbits. It is expected that some of the stars in our binaries will be
among the most massive known. With the binary status of each WR star
now known, we also studied the photometric and X-ray properties of our
program stars using archival MACHO photometry as well as
\emph{Chandra} and \emph{ROSAT} data. We find that one of our
presumably single WNL stars is among the X-ray brightest WR sources
known. We also identify a binary candidate from its RV variability and
X-ray luminosity which harbours the most luminous WR star known in the
Local Group.
\end{abstract}

%
%


\section{Introduction}

The optical spectra of Wolf-Rayet (WR) stars feature broad emission
lines from highly-ionized elements. These emission lines arise in a
fast, hot, and dense stellar wind which is generally optically thick
in the inner part, thereby completely veiling the hydrostatic
photosphere of the WR star. Depending on the elements they show in
their optical spectra, WR stars are classified into three different
subtypes. If a WR star displays predominantly helium (He) and nitrogen
(N), which are formed during hydrogen burning via the CNO cycle, the
star is classified as WN; if the WR star displays, in addition to He,
predominantly carbon (C) or oxygen (O), which are the products of
$3\alpha$ He burning, it is classified as WC or WO.

Due to their chemical properties, it is now generally accepted that
classical WR stars are evolved objects, namely the almost bare,
hydrogen-depleted, helium-burning cores of stars whose initial mass on
the main sequence (MS) was, at solar metallicity, above $\sim 25
M_{\sun}$, i.e. that started their lives as O stars
(\citealt{Lamers91}; \citealt{MaedCont94}). Thus, the key question of
WR-star formation is how a WR-star progenitor loses its outer,
H-rich envelope to expose the CNO-enriched, deeper layers, and how
WR-star formation depends on ambient metallicity. For single stars,
three scenarios have been put forward, depending on the initial mass
of the star. Stars with initial masses $25 M_{\sun} \la M_{\rm i} \la
40 M_{\sun}$ are expected to become red supergiants (RSGs), as which
they experience continuous mass loss through winds. Stars with initial
masses $40 M_{\sun} \la M_{\rm i} \la 85 M_{\sun}$ are believed to
turn into luminous blue variables (LBVs), as which they experience
outbursts of violent, eruptive shell-ejections (\citealt{HD79},
\citealt{HD94}). Stars with even higher initial masses ($M_{\rm i} \ga
85 M_{\sun}$ at solar metallicity) are believed to reach the WN phase
while they are still core-hydrogen burning, and to \emph{directly}
evolve into classical, He-burning WR stars (\citealt{Conti76}) without
going through the LBV phase.

In close binaries, Roche-lobe overflow (RLOF) is suspected to enhance
WR-star formation by removing the H-rich envelope of the (more
evolved) primary (\citealt{KippWeig67}). \citet{Pacz67} was the first
to note that a thus stripped He-burning core would very likely
resemble a WR star. Since RSGs descend from the initially least
massive O stars (see above), it follows that RLOF might thus
considerably contribute to the total WR numbers, in particular in
low-metallicity environments, where radiatively-driven mass-loss rates
are expected to be very low (see \citealt{Vink00};
\citealt{Vink01}). Indeed, older, non-rotating stellar-evolution
models were unable to reproduce the observed WR populations at
different $Z$ without either enhancing by a factor of two the
then-higher mass-loss rates through stellar winds or increasing the
fraction of interacting binaries (\citealt{MaedMey94}). While updated
models now include stellar rotation, and much better reproduce the
observations without the need of increased binary interaction
(\citealt{MaedMey00a}), the influence of binarity on WR-star
formation remains unclear.

From model calculations, \citet{Vanbev98} found that close O+O
binaries with initial periods $P \la 1000$ d cannot escape RLOF if the
primary star reaches the RSG stage, resulting in WR+O binaries with
post-RLOF periods $\la 200$d (\citealt{Wellstein99}). Hence, in
environments where RLOF is expected to become the increasingly
important WR-star formation channel (i.e. at lower ambient $Z$), the
fraction of WR binaries with present-day periods of up to 200 days
should be higher. This prediction is accessible to observational
tests. For such a study, the Magellanic Clouds are the ideal
laboratory: $i$) The distances to both the SMC and the LMC are well
established and $\sim$ constant for all stars
(e.g. \citealt{KellerWood06}); $ii$) reddening towards the Clouds is
low and fairly constant, contrary to the Galaxy
(e.g. \citealt{Niko04}), $iii$) the WR populations in both Clouds are
nearly complete (e.g. \citealt{MassDuff01}, but see
\citealt{Massey03}), $iv$) the WR population is large enough to allow
for reasonable statistics. In total, the LMC harbours 132 WR stars
(\citealt{BAT99}, hereafter BAT99\footnote{BAT99 lists 134 WR stars in
the LMC, but see \citet{Moff91} and \citet{Niemela01} for the revised
non-WR status of BAT99-4 and BAT99-6, respectively.}), while 12 WR
stars are known in the SMC (\citealt{MassDuff01};
\citealt{Massey03}). To search for binaries, the total sample of 144
stars has been split into three distinct studies: \citet{Bartz01}
reported the results on the 25 Magellanic Cloud WC/WO stars, while
Foellmi et al. (2003a,b) studied the 71 then-known, early-type WNE
(=WN2-WN5) stars, with the exception of the H-rich WNE stars in and
around the R136 cluster in the 30 Dor region. The observations of a
72nd WNE star, which was newly discovered by \citet{Massey03} in the
SMC, were reported by \citet{Foell04}.

In the present paper, we describe our intense, spectroscopic survey of
the remaining 41, late-type WNL (=WN5-11) stars in the LMC. The two
WN6 stars in the SMC were already studied by \citet{Foell03a}. Our
survey includes those WNL stars in the periphery of R136 which could
be observed with ground-based telescopes without adaptive optics
(AO). For the 6 luminous, WN5-7ha stars in the very core of R136,
AO-assisted, near-infrared spectroscopy using VLT/SINFONI was used;
those results will be reported elsewhere (Schnurr et al., in
preparation). The aim of our study is manifold. First and foremost, we
will assess the binary status ($P \la 200$ days) of each of our 41
program stars. This will conclude the decade-long effort of the
Montr\'eal Massive Star Group to study spectroscopically the entire WR
population in the Magellanic Clouds, and pave the way to obtain a much
clearer view of the role binarity plays in the evolution of massive
stars at different metallicities. Secondly, binaries identified in
this study can be used, in the future, to determine their respective
masses by using model-independent, Keplerian orbits. Masses of WR
stars are of greatest importance in the context of the calibration of
both atmospheric and evolutionary models, in particular for the most
massive stars. We have ample reason to believe that at least some
H-rich WNL stars in our sample belong to the subgroup of very massive,
possibly even the most massive stars known in the Local Group
(\citealt{Rauw96b}; \citealt{Schweick99}; \citealt{Rauw04};
\citealt{Bonanos04}). Thirdly and as a side effect, we will be able to
put publicly available, archival X-ray data from \emph{Chandra} and
\emph{ROSAT} into context with the binary status of our stars, since
massive binaries with colliding stellar winds are known to be strong
X-ray emitters.

The paper is organized as follows: In Section \ref{wnlsection2}, we
will describe the observations of our program stars. In Section
\ref{wnlsection3}, we will briefly describe the data reduction. In
Section \ref{wnlsection4}, we will describe in detail how we analysed
our spectroscopic and X-ray data and the results we
obtained. These results will be discussed in Section
\ref{wnlsection5}. Section \ref{wnlsection6} summarizes and concludes
this paper.

\section{Data Acquisition}
\label{wnlsection2}

\subsection{Spectroscopic Observations}
\label{spectrodata}

Target stars were selected from {\it The Fourth Catalogue of
Population I Wolf-Rayet Stars in the Large Magellanic Cloud\/}
(BAT99). This catalogue lists 134 WR stars of which 47 fall into the
WNL (= WN6-11) class, including the two ``slash-star'' types O3If/WN6
(supposedly H-burning; \citealt{CroBo97}) and Ofpe/WN9 (supposedly
linked to the LBV phenomenon; see \citealt{Stahl83};
\citealt{Stahl87}; \citealt{Cro95}; \citealt{Nota96};
\citealt{Pasqua97}). 41 of these 47 stars (see Table \ref{targets})
were observed with conventional spectrographs, and the results of
those observations are reported here.

\begin{table}
\caption{List of our program stars as taken from the BAT99
catalogue. BAT99 numbers as well as cross-identification with the
Brey, Radcliffe and/or Melnick numbers are given, together with $v$
magnitudes and spectral types, according to BAT99. Stars without a
Brey denomination are those that have been newly included into the
BAT99 catalogue.}
\label{targets}
\begin{tabular}{@{}lccrr}
\hline
BAT99 & Brey & other  &   $v$ mag & spectral type \\
\hline
12    &  10a & ...    &  13.72    & O3If*/WN6-A  \\
13    &  ... & ...    &  12.89    & WN10  \\
16    &  13  & ...    &  12.73    & WN8h  \\
22    &  18  & R84    &  12.09    & WN9h  \\
30    &  24  & ...    &  13.40    & WN6h  \\
32    &  26  & ...    &  12.72    & WN6(h)  \\
33    &  ..  & R99    &  11.54    & Ofpe/WN9  \\
44    &  36  & ...    &  13.47    & WN8h  \\
45    &  ..  & BE294  &  12.80    & WN10h  \\
54    &  44a & ...    &  14.32    & WN9h  \\
55    &  ... & ...    &  11.99    & WN11h  \\
58    &  47  & ...    &  15.1:    & WN6h  \\
68    &  58  & ...    &  14.43    & WN5-6  \\
76    &  64  & ...    &  13.46    & WN9h  \\
77    &  65  & ...    &  13.34    & WN7  \\
79    &  57  & ...    &  13.58    & WN7+OB  \\
80    &  65c & ...    &  13.24    & O4If/WN6  \\
83    &  ... & R127   &  $10.5^{a}$     & Ofpe/WN9  \\
89    &  71  & ...    &  14.28    & WN7h  \\
91    &  73  & ...    &  13.98    & WN7  \\
92    &  72  & R130   &  11.51    & WN6+B1Ia  \\
93    &  74a & ...    &  13.83    & O3If/WN6  \\
95    &  80  & R135   &  13.16    & WN7h  \\
96    &  81  & Mk53   &  13.74    & WN8(h)  \\
97    &  ... & Mk51   &  13.77    & O3If*/WN7-A   \\
98    &  79  & Mk49   &  12.96    & WN6(h)   \\
99    &  78  & Mk39   &  12.96    & O3If*/WN6-A   \\
100   &  75  & R134   &  12.85    & WN6h  \\
102   &  87  & R140a2 &  12.99    & WN6+O  \\
103   &  87  & R140b  &  13.01    & WN6  \\ 
104   &  76  & Mk37Wb &  14.58    & O3If*/WN6-A  \\
105   &  77  & Mk42   &  12.80    & O3If*/WN6-A  \\
107   &  86  & R139   &  12.12    & WNL/Of  \\
113   &  ... & Mk30   &  13.57    & O3If*/WN6-A   \\
114   &  ... & Mk35   &  13.59    & O3If*/WN6-A  \\
116   &  84  & Mk34   &  13.65    & WN5h  \\
118   &  89  & ...    &  11.15    & WN6h  \\
119   &  90  & ...    &  12.16    & WN6(h)  \\
120   &  91  & ...    &  12.59    & WN9h   \\
130   &  ... & ...    &  12.82    & WN11h  \\ 
133   &  ... & ...    &  12.10    & WN11h  \\
\hline
\end{tabular}
\flushleft{$^{a}$Known LBV with large photometric variations,
cf. \citet{Stahl83}.}
\end{table}

Our observations were organized in three different campaigns or
``seasons'' between 2001 and 2003 to maximize the time coverage, and
were carried out during 13 runs at 6 different, 2m-class, southern
telescopes. The following observatories were used: Complejo
Astron\'omico El Leoncito (CASLEO), Argentina; Mount Stromlo
Observatory (MSO), Australia; Cerro Tololo Inter-American Observatory
(CTIO), Chile; South African Astronomical Observatory (SAAO), South
Africa; Siding Spring Observatory, Australia; Las Campanas Observatory
(LCO), Chile. In total, 99 nights were allocated, but due to bad
weather conditions or technical problems, not all nights were
useful. Long-slit spectrographs were used at all telescopes. The exact
wavelength coverage of the spectra depended on the respective
instrument used, but all sets of data included the wavelength range
from 4000 to 5000 \AA~, thereby comprising the strategic emission
lines He\textsc{ii} $\lambda$4686 and N\textsc{iv}$ \lambda$4058
(however, see below).

To maximize the flux reaching the detector and thus achieve good S/N,
a slit width of 1.5$''$ was used during the first two years, but under
the very good seeing conditions on Cerro Tololo during summer, this
yielded a relatively large RV scatter; we will come back to this
problem later in this paper. Therefore, the slit width was reduced to
1$''$ for the last season (2003/04, only at CTIO). The obtained linear
dispersion varied from 0.65 \AA/pixel (SSO) to 1.64 \AA/pixel
(CASLEO), with the 3-pixel spectral (velocity) resolution at 4686 \AA~
ranging from $R \sim 2400$ ($\sim 125$kms$^{-1}$) to $R \sim 950$
($\sim 315$kms$^{-1}$). Exposure times were chosen to provide a
signal-to-noise ratio (S/N) of $\sim$50-70 per collapsed pixel in the
blue continuum around 4500 \AA, and were adapted to the respective
telescope efficiencies and weather conditions. For our faintest stars,
long exposures were broken into two or three to facilitate cosmic-ray
rejection.

Even under best conditions, not more than two thirds to three quarters
of our sample (``sequence A'') could be observed in one night. The
remaining stars (``sequence B'') were then observed the following
night before sequence A was restarted again, and so forth. To further
break integer-day sampling ($\sim$2-day sampling in particular), we
also employed a scheme to re-observe the stars of a given sequence the
following night before resuming observations of the other sequence,
thereby obtaining a $\sim$1-day sampling (i.e. observations were
carried out as ABBAABAB etc.). At CTIO, where we had relatively long
runs of clear contiguous nights so that we could sufficiently plan
ahead, this was particularly successfully employed. Moreover, some
stars were observed more frequently (i.e. once or twice every
night) by default because they were of particular interest
(short-period and/or supposedly very massive binaries).

At the beginning of each night, bias frames and high-S/N, internal
(Quartz lamp) flatfield frames were taken. For reliable RVs, a
comparison-arc exposure was taken before and after each science
exposure, the exception to this rule being the stars in the periphery
of R136, because telescope slews were only very minor ($\la1'$) and
differential flexure of the Cassegrain spectrographs not an
issue. Neither dark nor twilight-flatfield frames were taken.

The observation journal is summarized in Table \ref{journal}, while the
instrumental properties are summarized in Table \ref{instruments}.

\begin{table*}
\caption{Journal of observations.}
\label{journal}
\begin{center}
\begin{tabular}{@{}llrrrr}
\hline
Run\# & Observatory &   Start (UT) & End (UT)  & Clear nights & Season\\
\hline
1  & CASLEO &  02 Nov 2001 & 09 Nov 2001 & 6  & 1 \\
2  & MSO    &  09 Nov 2001 & 06 Dec 2001 & 14 & 1 \\
3  & CTIO   &  26 Dec 2001 & 31 Dec 2001 & 6  & 1 \\
4  & CTIO   &  23 Jan 2002 & 27 Jan 2002 & 4  & 1 \\
5  & SAAO   &  05 Nov 2002 & 12 Nov 2002 & 7  & 2 \\
6  & CTIO   &  19 Nov 2002 & 27 Nov 2002 & 8  & 2 \\
7  & CTIO   &  16 Dec 2002 & 24 Dec 2002 & 8  & 2 \\
8  & MSO    &  03 Jan 2003 & 18 Jan 2003 & 10 & 2 \\
9  & SSO    &  27 Jan 2003 & 31 Jan 2003 & 3  & 2 \\
10 & LCO    &  27 Jan 2003 & 02 Feb 2003 & 6  & 2 \\
11 & CTIO   &  19 Dec 2003 & 29 Dec 2003 & 10 & 3 \\
12 & CTIO   &  02 Jan 2004 & 06 Jan 2004 & 4  & 3 \\
13 & CTIO   &  07 Jan 2004 & 09 Jan 2004 & 2  & 3 \\
\hline
\end{tabular}
\end{center}
\end{table*}

\begin{table*}
\footnotesize
\caption{List of observatories and instruments used.}
\label{instruments}
\begin{tabular}{l|cccccc}
\hline
Observatory               & CASLEO & MSO   & CTIO & SAAO  & SSO & LCO \\
Telescope                 & 2.15m  & 1.88m & 1.5m & 1.88m & 2.3m &
2.5m \\
Spectrograph              & REOSC & CassSpec & RCSpec & GratingSpec &
DBS$^{a}$ & WFCCD\\
Grating [l/mm]            & 600    & 600   & 600  &  600  & 600 & 600 \\
Dispersion [\AA/pix]      & 1.64   & 1.35  & 1.48 & 1.10  & 0.65 &
1.38 \\
$R$ @ 4686 \AA (3 pixels) & 945    & 1160  & 1055 & 1420  & 2400 & 1130 \\ 
Spectral range [\AA] & 3970-5645 & 3885-5515 & 3980-5730 &
3485-5415 & 3905-4875 & 3775-5600 \\
Comparison lamp           & HeNeAr & FeAr  & HeNeAr & CuAr & CuHeAr &
HeH\\
Overscan                  & no     & no    & yes    & yes  & yes & yes \\
\hline
\end{tabular}\\
\flushleft{{\bf Notes:\/} $^{a}$Blue arm only.}
\end{table*}
\normalsize

\subsection{Archival X-ray Data}

We made use of public archives to retrieve X-ray data for our program
stars. The \emph{Chandra} Archive\footnote{The \emph{Chandra} Archive
is available using the \emph{Chandra} Search and Retrieval Interface
(Chaser) at the \emph{Chandra} X-ray Observatory site; see at
\texttt{http://cxc.harvard.edu}.} available from October 2004 and the
entire \emph{ROSAT} Archive\footnote{ The \emph{ROSAT} Archive is
supported by the High Energy Astrophysics Science Archive Research
Center (HEASARC) of Goddard Space Flight Center, NASA. It can be
accessed at the site \texttt{http://heasarc.gsfc.nasa.gov/W3Browse}.}
have been searched for all \emph{Chandra} ACIS and \emph{ROSAT} PSPC
and HRI observations including any of the 41 WNL stars studied in this
paper. We de-archived \emph{Chandra} ACIS observations for 25 WNL
stars, and \emph{ROSAT} PSPC and HRI observations for 13 and 3 WNL
stars, respectively. A summarizing table of these observations is
provided in Section \ref{4.11} (Table \ref{xrayall}). Further details
of the selection and reduction of these observations are provided by
\citet{Guerrero06}.

\subsection{Archival photometry}

Several microlensing surveys carry out repeated photometry towards the
LMC and make their data publicly available. In order to obtain
repeated photometry of our program stars, we thus browsed through the
public archives of the Massive Compact Halo Object (MACHO; see Alcock
et al. 1996) experiment, and the Optical Gravitational Lensing
Experiment (OGLE; see \citealt{Udalski00}). To search the
databases\footnote{Web interfaces for lightcurve retrieval from the
respective databases are accessible at
\texttt{http://www.macho.macmaster.ca/} for MACHO, and at
\texttt{http://bulge.princeton.edu/$^{\sim}$ogle/} for OGLE}, J2000.0
coordinates as given in BAT99 were entered, and a search radius of
10$''$ was applied in both cases. While MACHO yielded photometry of
some of our targets (OGLE did not observe our programme stars), the
data showed too large scatter to prove useful, most likely due to the
brightness of our target stars. Thus, we did not carry out any further
analysis of the photometry.

%
%

\section{Data Reduction}
\label{wnlsection3}

Some of our observation runs, notably those at the Cerro Tololo
Inter-American Observatory (CTIO), were interrupted
by other, time-critical observations, and thus split into several
runs. Since the spectrograph had to be set up again each time, we
treated each such run individually to allow for effects of different
grating positions, angles, alignments, etc. Thus, a total of 13
individual runs were processed.

Data reduction was carried out in the usual way using the NOAO-IRAF
software package\footnote{IRAF is distributed by the National Optical
Astronomy Observatories, which are operated by the Association of
Universities for Research in Astronomy, Inc., under cooperative
agreement with the National Science Foundation.}. Science data were
corrected with average bias and flatfield frames, and object spectra
were optimally extracted using APALL with the 2D trace-fit option
enabled. Sky background was fitted and subtracted from the stellar
spectrum for all stars with the exception of those in the periphery of
R136. This cumbersome task (cf. \citealt{Selman05}) was deemed
unnecessary because despite its strength, the discrete emission
spectrum of the 30 Dor nebula does not affect the strategic WNL
emission line, He\textsc{ii} $\lambda4686$. It does however affect the
(intrinsic) Balmer lines of these WNL stars, which renders the
determination of the hydrogen content in the star highly
uncertain. This has consequences which will be examined later.

Arc frames taken before and after each science exposure were averaged
and then extracted. Unfortunately, due to the lack of good comparison
lines at the blue edge of our spectra, the dispersion solution is less
reliable in the region bluewards of 4400~\AA. This had severe
consequences for the precision of the radial-velocities measured from
the N\textsc{iv}$ \lambda$4058 emission line, which yielded very large
scatter. Extracted spectra were then rebinned to uniform stepwidth,
corrected for heliocentric velocities, and rectified.

The spectra were then cleaned of cosmic rays and bad pixels as far as
possible; where applicable, spectra from split exposures were
combined. If a cosmic ray or bad pixel fell onto a strategic emission
line, great care was taken not to significantly alter the original
line profile. If this was not possible, the spectrum was
discarded. The final stellar spectra were then rebinned to a uniform
step-width of 1.65 \AA/pixel, thereby yielding a conservative 3-pixel
resolving power of $R\sim1000$. The achieved S/N ratio per rebinned
pixel, calculated for the spectral region from 5050 to 5350~\AA, is 85
on average.

%
%

\section{Data Analysis and Results}
\label{wnlsection4}

\subsection{Radial-Velocity Measurements}
\label{rvmeasures}

Since in principle, we are mainly interested in \emph{changes} in
radial velocities (RVs) to identify the binaries, cross-correlation,
which measures shifts relative to a template, was used as our main
tool, since it yielded the smallest RV scatter.

In most of our WNL stars, the spectrum is dominated by the
He\textsc{ii} $\lambda$4686 emission, which is the strongest line by a
wide margin. Thus, we concentrated our RV measurements on this
emission line, restricting the cross correlation to the spectral
region from 4600 to 4800 \AA, which almost exclusively contains
He\textsc{ii} $\lambda$4686. To guard against an unlucky choice of
template, we used an iterative approach for cross-correlation: For a
first cross-correlation pass, a real, high-S/N spectrum from the data
set of a given star was chosen as template. This template was
cross-correlated with all other spectra. RV shifts were measured by
fitting a Gaussian plus a (linear, rarely parabolic) continuum
function to the cross-correlation profile. Resulting RVs were then
used to shift all spectra into the template's frame of
reference. Then, a S/N-weighted average spectrum was computed using
all shifted spectra, and used as a new template for a second cross
correlation pass. Again, all spectra were shifted and added, and with
this ``supertemplate'', a final cross-correlation pass was made.

The advantage of this approach is that the resulting supertemplate has
a higher S/N than a single spectrum, and that (stochastic or cyclical)
line-profile variations are mostly averaged out. The same is true for
the potentially present spectral signature of a companion, which is
maximally smeared out, because it moves in anti-phase to the WNL
star\footnote{This is particularly the case if the WN star is the more
massive component of the binary system, as is possible in luminous WNL
systems. The O star then shows the larger RV amplitude; its absorption
lines are hence smeared out over a larger spectral region.}. While
there were noticeable differences in RVs between the first and the
second cross-correlation iteration, in most cases the third
cross-correlation pass did not improve the RVs; still, it was carried
out in all cases for the sake of uniformity.

\begin{figure}
\begin{minipage}{85mm}
\includegraphics[width=47mm,angle=-90,trim= 10 10 0 280,clip]{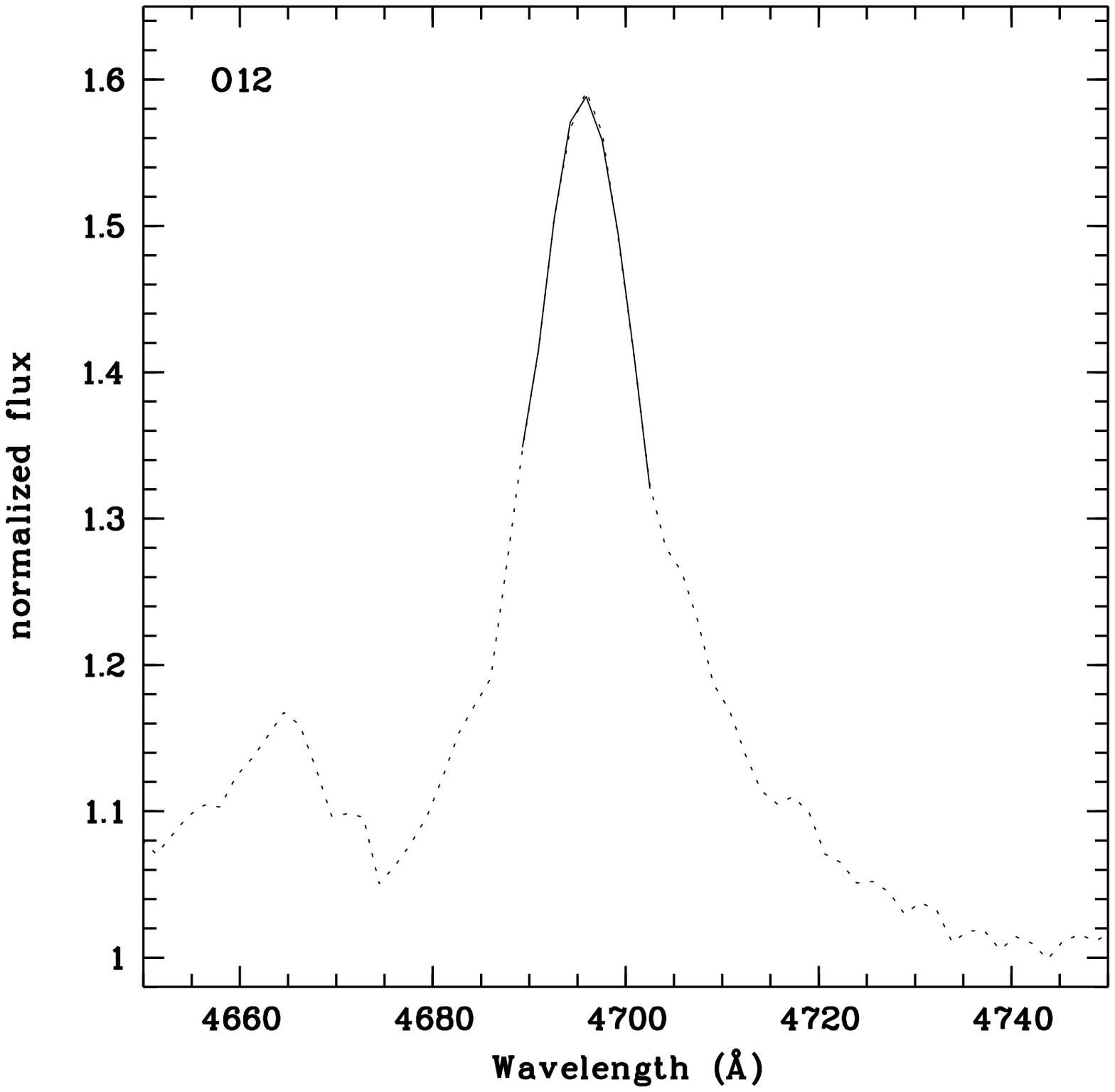}\hfill
\includegraphics[width=47mm,angle=-90,trim= 10 60 0 280,clip]{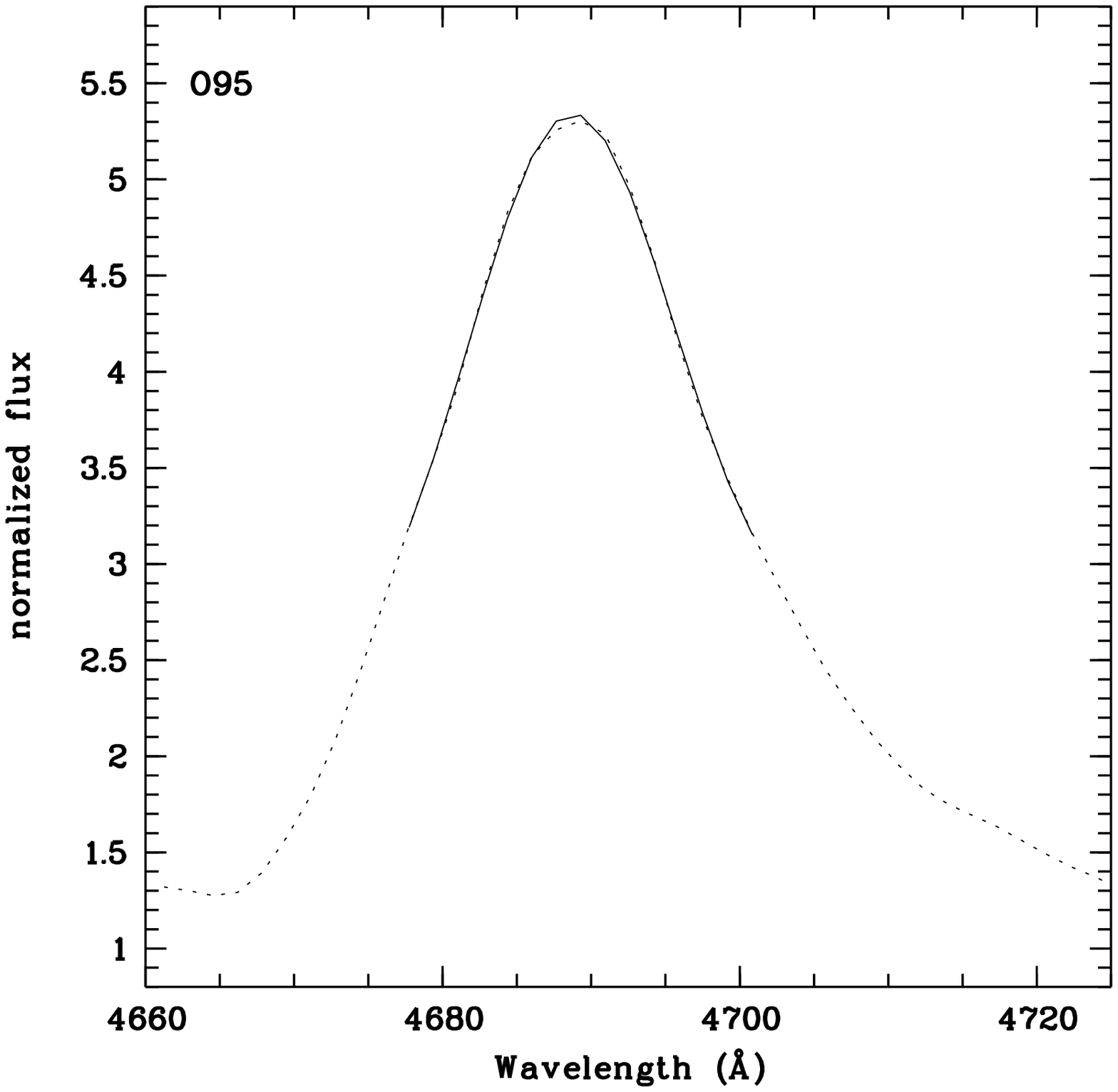}\hfill\\
\includegraphics[width=47mm,angle=-90,trim= 10 10 0 280,clip]{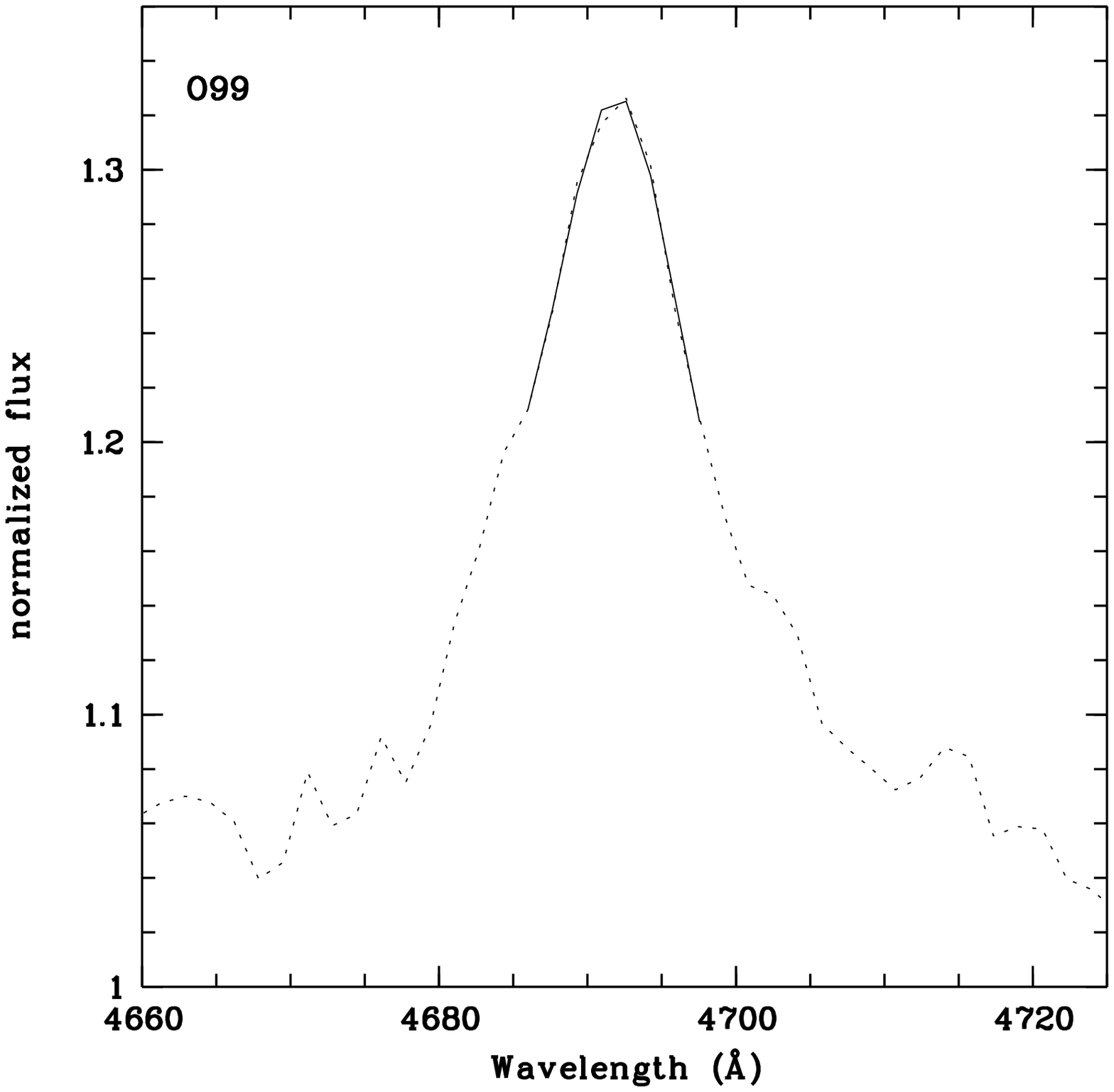}\hfill
\includegraphics[width=47mm,angle=-90,trim= 10 60 0 280,clip]{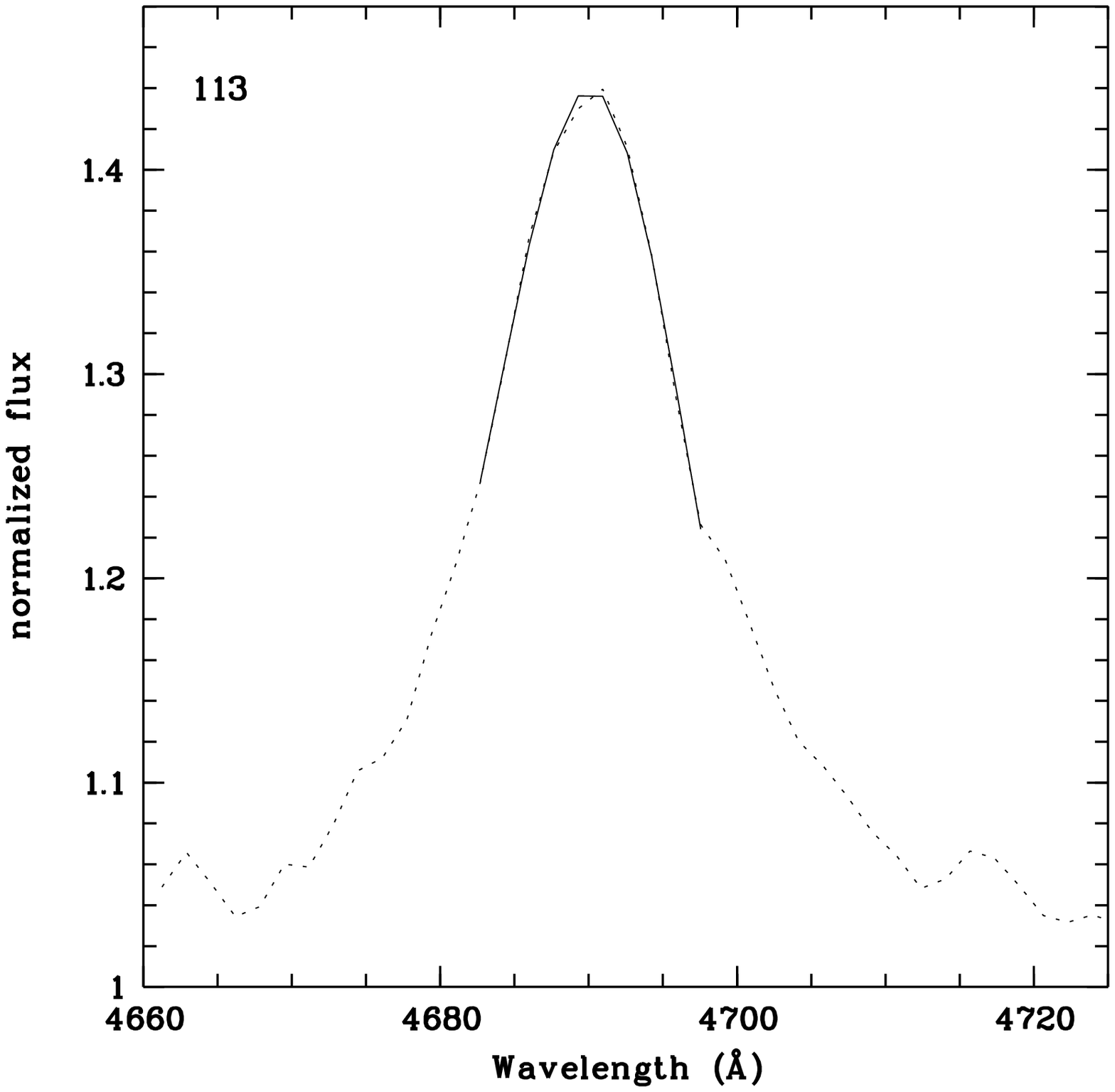}\hfill\\
\end{minipage}
\caption{Examples of the He\textsc{ii} $\lambda$4686 emission line
(dotted) and Gaussian fits (solid) which were restricted to the upper
part of the line. Shown are individual spectra with average S/N ratio
of the four binaries which have been newly discovered in this study,
BAT99-12, 95, 99 and 113.}
\label{gaussfits}
\end{figure}

Absolute (systemic) RVs were measured by fitting the He\textsc{ii}
$\lambda$4686 emission, whose peak was reasonably well reproduced by a
single Gauss function (see Figure \ref{gaussfits}). Other lines were
too weak and too noisy to be used, which is particularly unfortunate
for the narrow, clean and rather symmetric N\textsc{iv}$ \lambda$4058
emission at the very blue edge of our spectra. This line also suffered
from the lack of useful comparison-arc lines, which rendered
impossible a reliable wavelength calibration (see Section
\ref{wnlsection3}), yielding a RV scatter more than twice as large as
what was obtained from the He\textsc{ii} $\lambda$4686 line.

While differences between RVs obtained from fits and from cross
correlation remained remarkably small, direct fitting yielded a
slightly larger RV scatter than cross correlation, most likely because
of noise (in particular in weak-lined stars) and the line being
slightly asymmetric. Potential intrinsic line-profile variability
(e.g. due to wind-wind collision) might also add to the
difference. Thus, whenever it was possible, we relied on RVs obtained
by cross correlation rather than on those obtained by line-fitting.

\subsection{Standard Stars and Systematic Shifts Between Observatories}
\label{standards}

Most stars show a small to very small RV scatter around or below 20
kms$^{-1}$ (obtained by cross correlation). However, before
identifying the binary stars in our sample, another potential problem
had first to be taken care of. Foellmi et al. (2003a,b) reported
systematic shifts between CASLEO and SAAO. Both observatories together
with others were also used for our study; therefore, we had to make
sure that any RV variation is of stellar origin and not a consequence
of different (and possibly variable) instrumental zero points. In
order to compute systematic shifts among the respective runs, we
proceeded as follows: Of our 41 program stars, we selected those $i$)
which are well isolated (no composite spectrum due to crowding), $ii$)
for which RVs could be measured using cross correlation, $iii$) which
displayed very small RV scatter, and $iv$) for which a preliminary
period search in the period range from 1 d $\le P \le 200$ d did not
yield cyclical RV variability (the period-search method is described
more comprehensively below). Stars meeting these criteria are most
likely true single stars or binaries with sufficiently long periods,
large eccentricities, and/or low inclination angles that they can
serve effectively, within our detection limits, as constant-RV
standard stars. By construction, the only shifts these stars display
are then solely due to systematic shifts among different
observatories.

Twenty-three reference stars were thus selected (see Table
\ref{refstars}). For each star, mean RVs were computed {\it for each
individual run\/}. As described above, during the 2003/2004 campaign
at CTIO (i.e. runs 11 through 13), a 1$''$ slit width was used, hence
the RV scatter is very small in these data sets, reaching a minimum
during run 11 (see below). SSO displays a similarly small scatter, but
there are many fewer points available. Thus, for better statistics,
run 11 was chosen to provide the overall zero point. For each
reference star, the mean RV of run 11, $\overline{RV}_{11}$, was
subtracted from all RVs in order to normalize RVs to run 11. This was
necessary because the cross-correlation method only yields {\it
relative\/} RVs with respect to an {\it arbitrarily\/} chosen
template, i.e. one that does not necessarily come from run 11. (Note
that the shift-and-add method described above also suffers from this
effect, because it starts out with one arbitrarily chosen spectrum.)
Thus, for every individual star, $\overline{RV}_{11}=0$, whereas for
all other runs $j \neq 11$ in general $\overline{RV}_{j} \neq 0$. By
construction, the respective values of these $\overline{RV}_{j}$ are
purely due, and directly correspond, to the systematic shifts among
run 11 and all other runs.

\begin{table}
\caption{List of the 23 fiducial RV standard stars together with
their unweighted RV scatter.}
\label{refstars}
\centering
\begin{tabular}{rcrcrc}
\hline
BAT99 & $\sigma_{\rm RV}$ & BAT99 & $\sigma_{\rm RV}$ & BAT99 & $\sigma_{\rm RV}$ \\
      &  kms$^{-1}$ & & kms$^{-1}$ & & kms$^{-1}$\\
\hline
13  & 12.5  & 55  & 14.6  & 93  & 20.5\\
16  & 13.3  & 58  & 14.7  & 96  & 18.7\\
22  & 13.3  & 76  & 15.5  & 97  & 19.6\\
30  & 17.3  & 79  & 14.3  & 100 & 15.4\\
33  & 16.2  & 80  & 20.5  & 120 & 21.8\\
44  & 15.8  & 83  & 14.9  & 130 & 13.2\\
45  & 11.4  & 89  & 14.3  & 133 & 15.4\\
54  & 18.2  & 91  & 13.7  & ... & ... \\
\hline
\end{tabular}
\end{table}

However, not every star will display exactly the same shifts, but the
shifts will be randomly distributed due to noise. By combining the 23
reference stars into one ``super-reference'', we obtained the
respective {\it average\/} mean velocities $\overline{RV}_{j}$ per run
-- and thus directly the systematic shifts among run 11 and all other
runs -- as well as the standard deviations $\sigma_{j}$, which were
adopted as empirical {\it a posteriori\/} measurement errors for the
RVs of the respective run $j$. Values range from 23.8 kms$^{-1}$ at
LCO (run 10) to 10.3 kms$^{-1}$ at CTIO (run 11). Points from a given
run $j$ deviating from $\overline{RV}_{j}$ by more than
$\pm3\sigma_{j}$ were rejected. (This $\sigma$-clipping on a per-run
basis could not be done per individual star because there are not
enough data points per run for meaningful statistics.) Both
$\overline{RV}_{j}$ and $\sigma_{j}$ were then re-calculated; final
values are given in Table \ref{shifts}. Systematic shifts could now
easily be removed in {\it all\/} program stars by simply subtracting
the respective values from all RVs of a given star and run.

RVs obtained by cross-correlation and corrected for systematic shifts
are shown, for each season separately, in Figure \ref{allepochs}. The
dashed line indicates the zero mean velocity to which all RVs were
normalized.


\begin{figure*}
\begin{minipage}{165mm}
\includegraphics[width=39mm,angle=-90,trim= 0 60 40 38,clip]{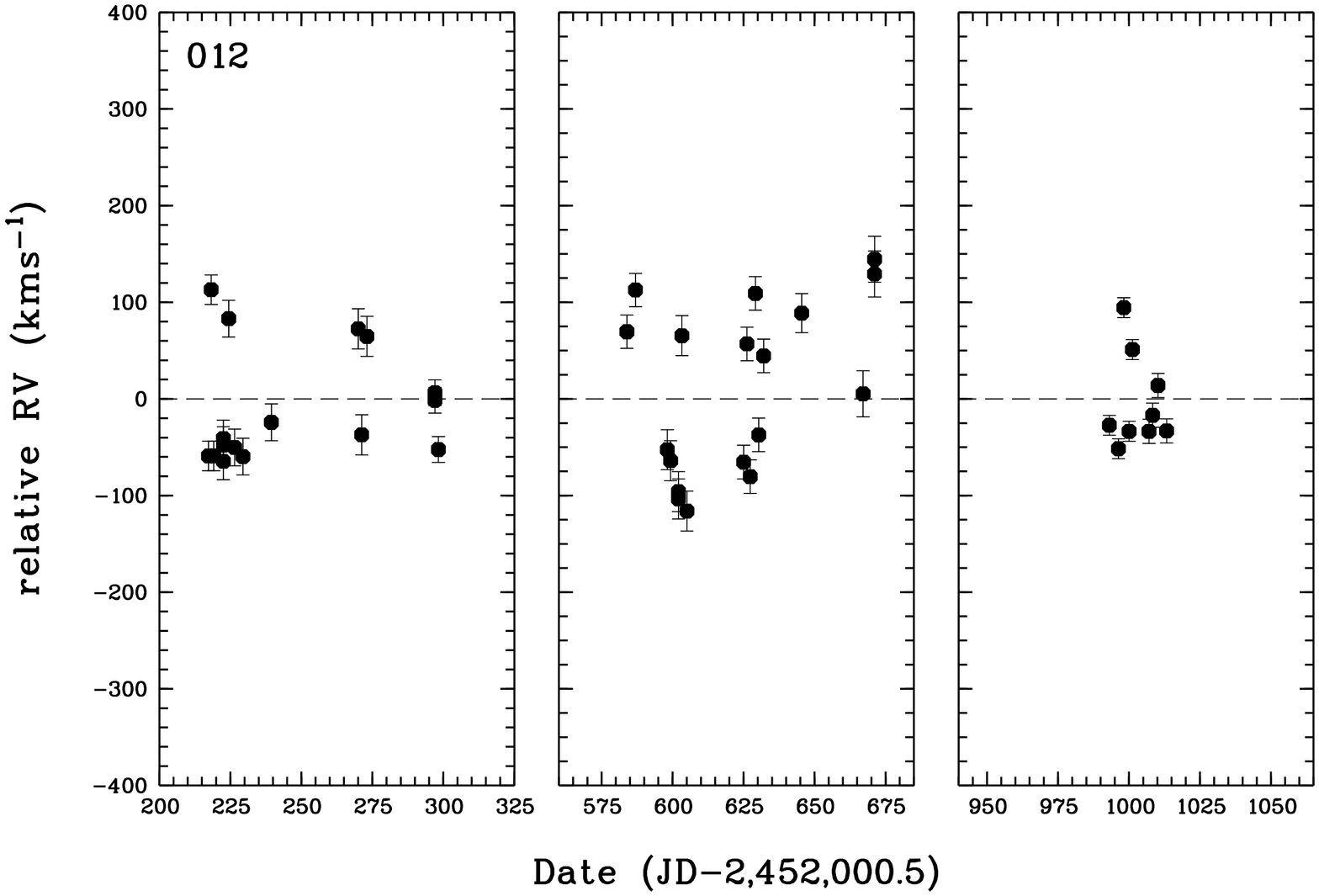}\hfill
\includegraphics[width=39mm,angle=-90,trim= 0 90 40 38,clip]{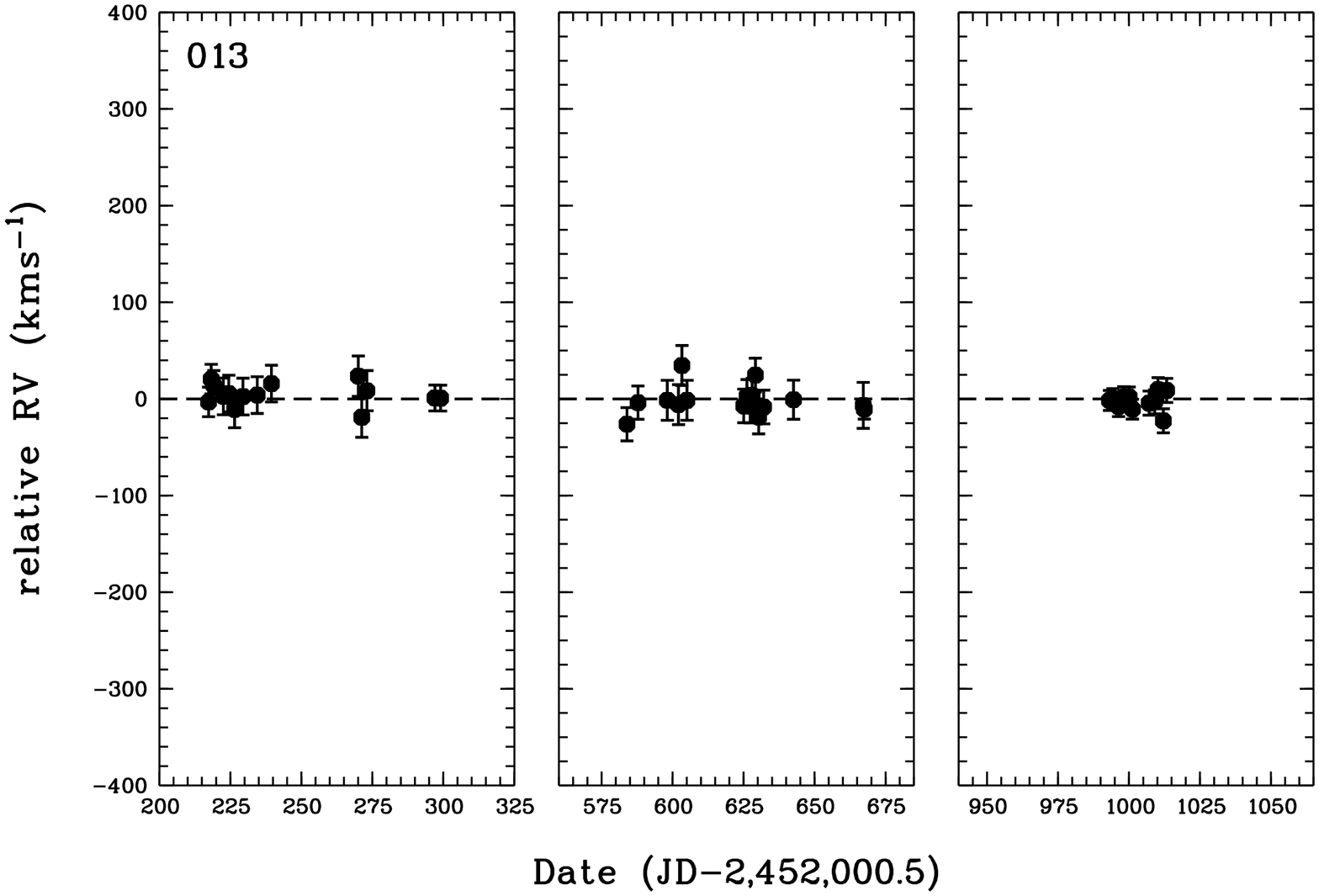}\hfill
\includegraphics[width=39mm,angle=-90,trim= 0 90 40 38,clip]{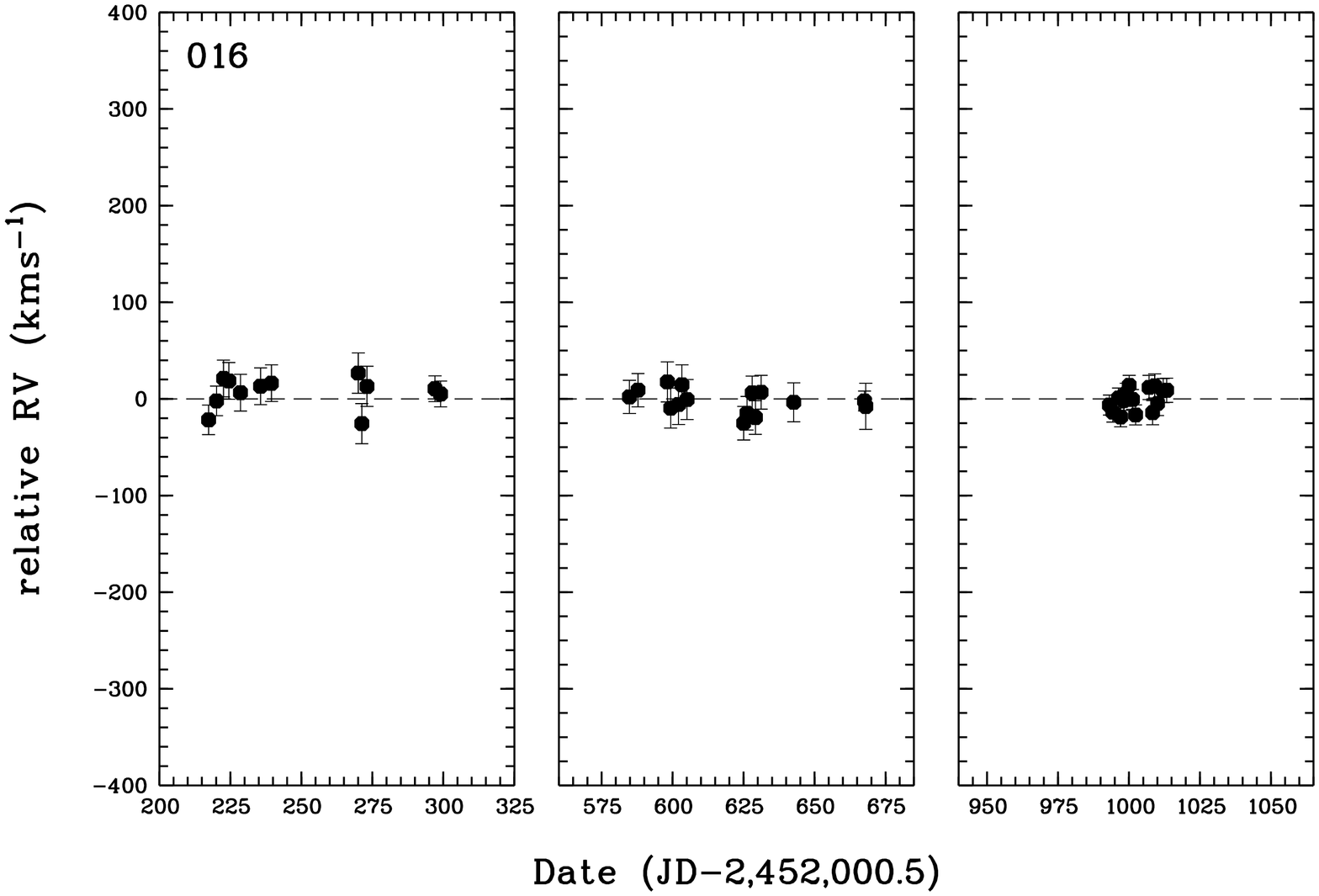}\hfill\\
\includegraphics[width=39mm,angle=-90,trim= 0 60 40 38,clip]{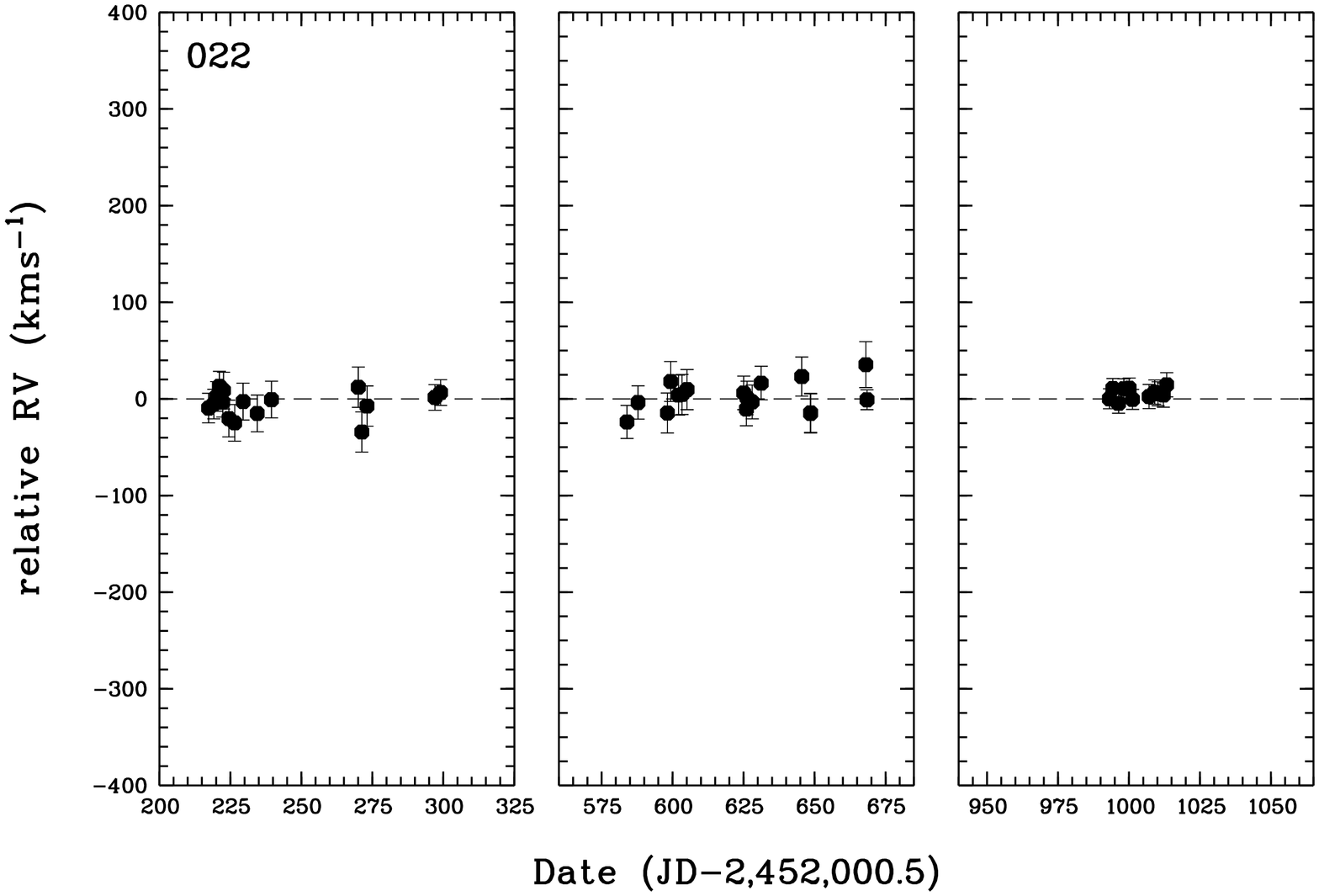}\hfill
\includegraphics[width=39mm,angle=-90,trim= 0 90 40 38,clip]{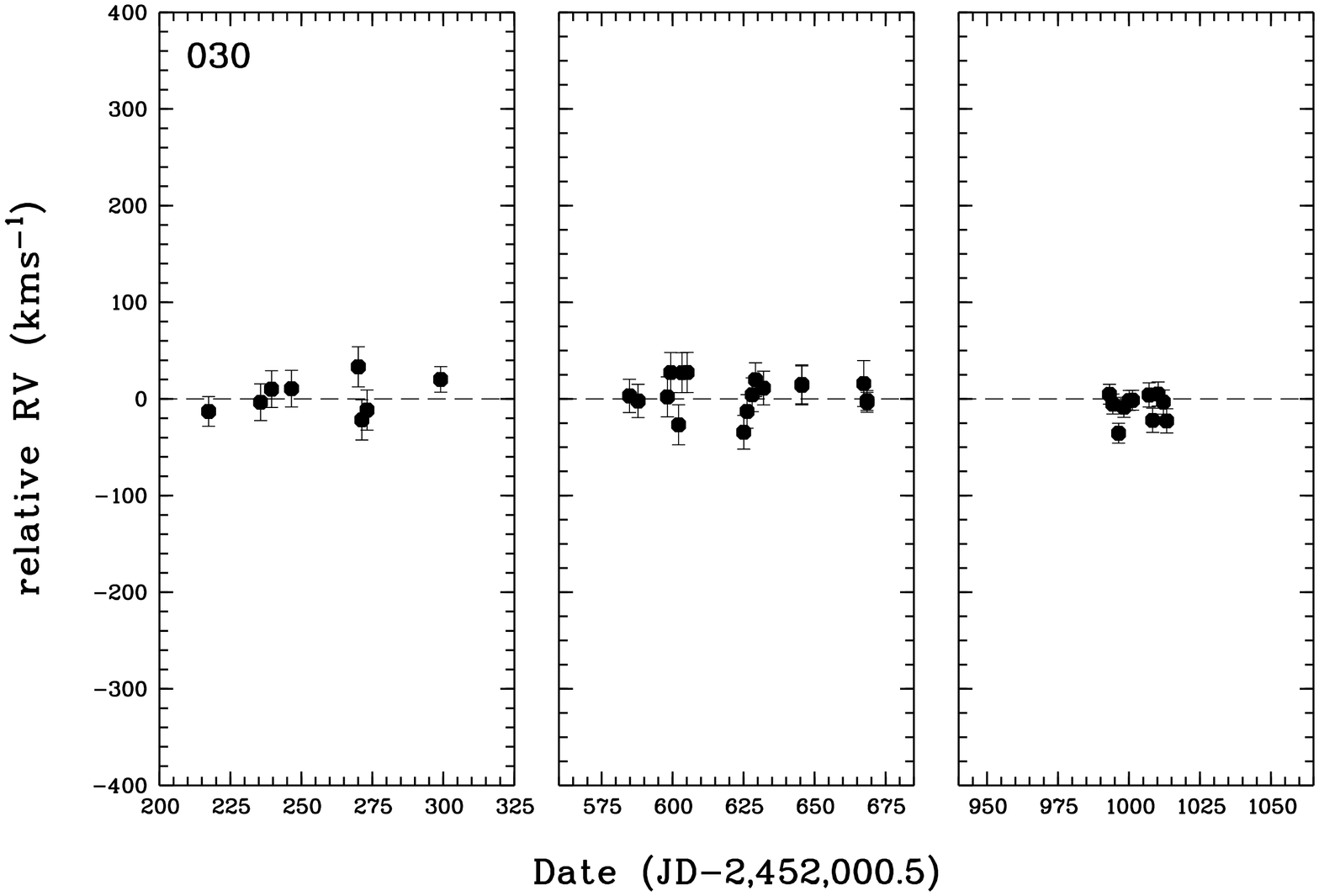}\hfill
\includegraphics[width=39mm,angle=-90,trim= 0 90 40 38,clip]{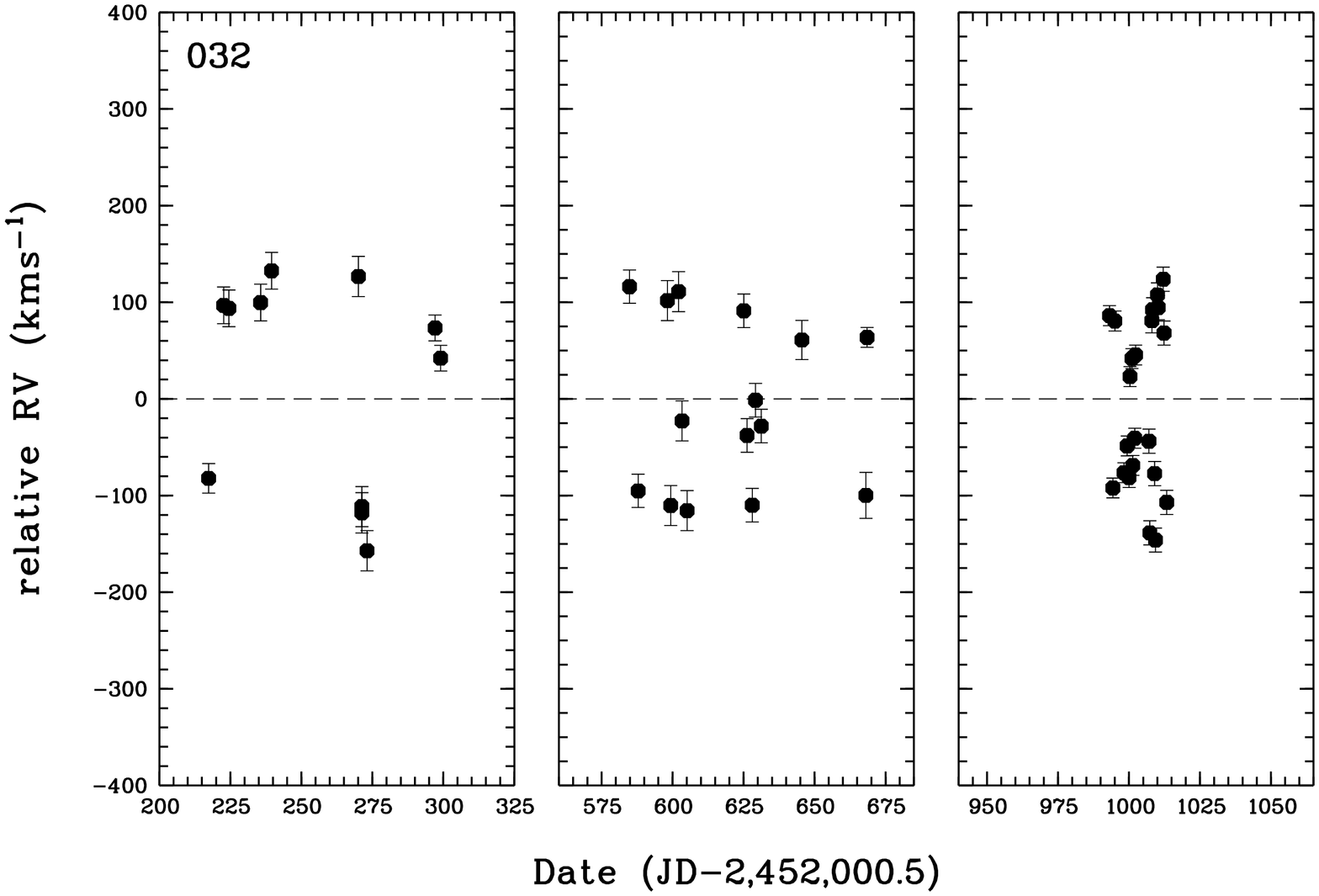}\hfill\\
\includegraphics[width=39mm,angle=-90,trim= 0 60 40 38,clip]{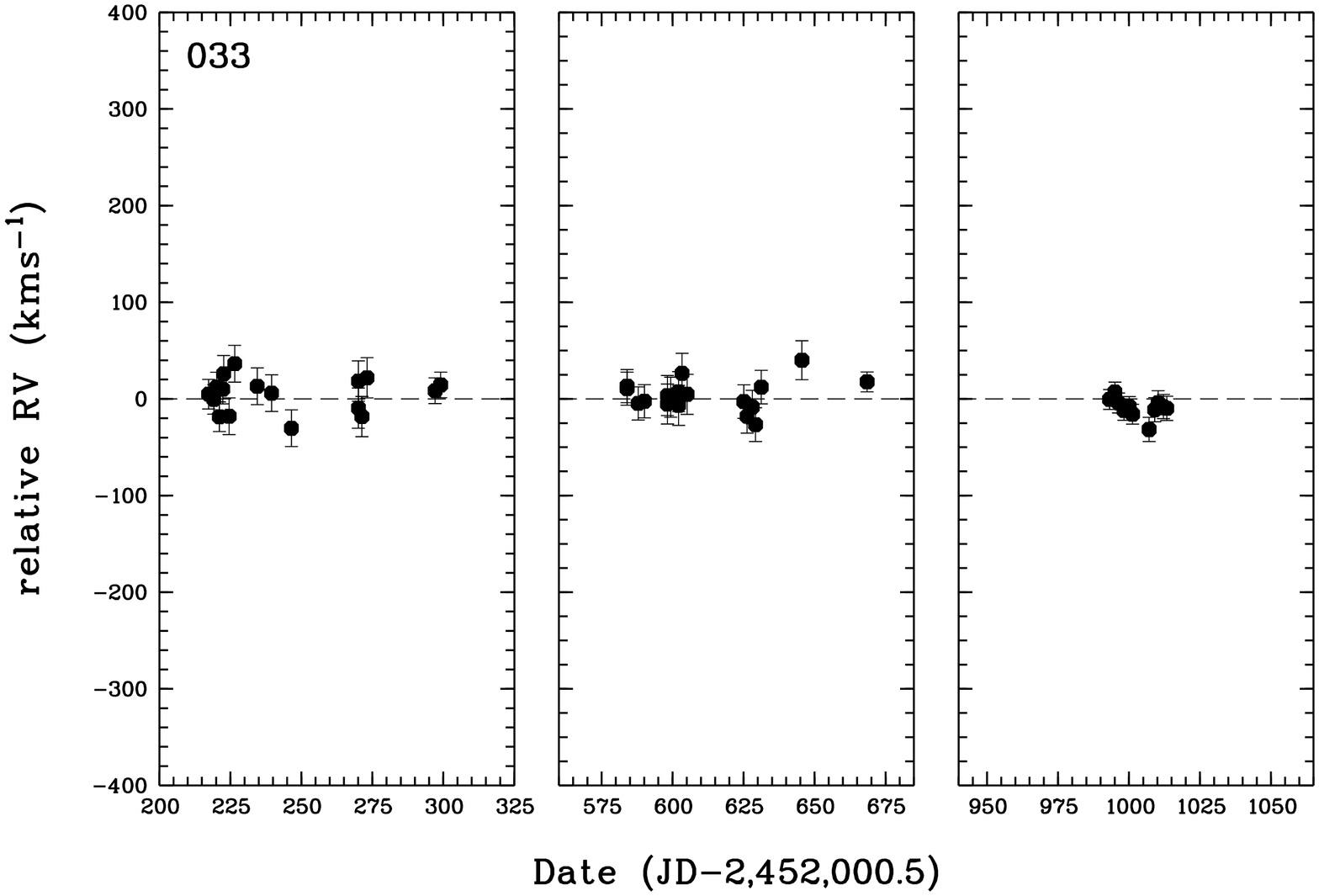}\hfill
\includegraphics[width=39mm,angle=-90,trim= 0 90 40 38,clip]{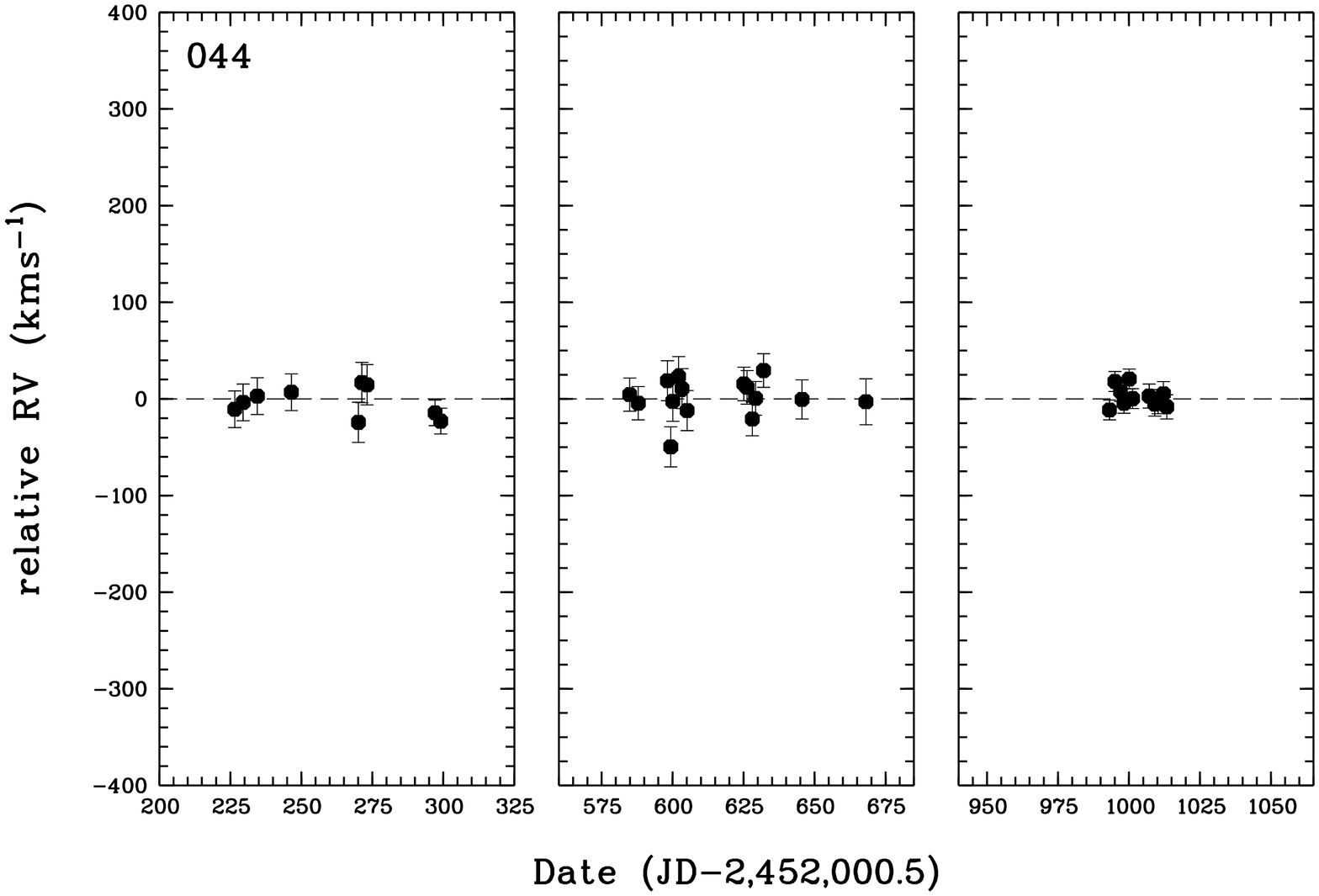}\hfill
\includegraphics[width=39mm,angle=-90,trim= 0 90 40 38,clip]{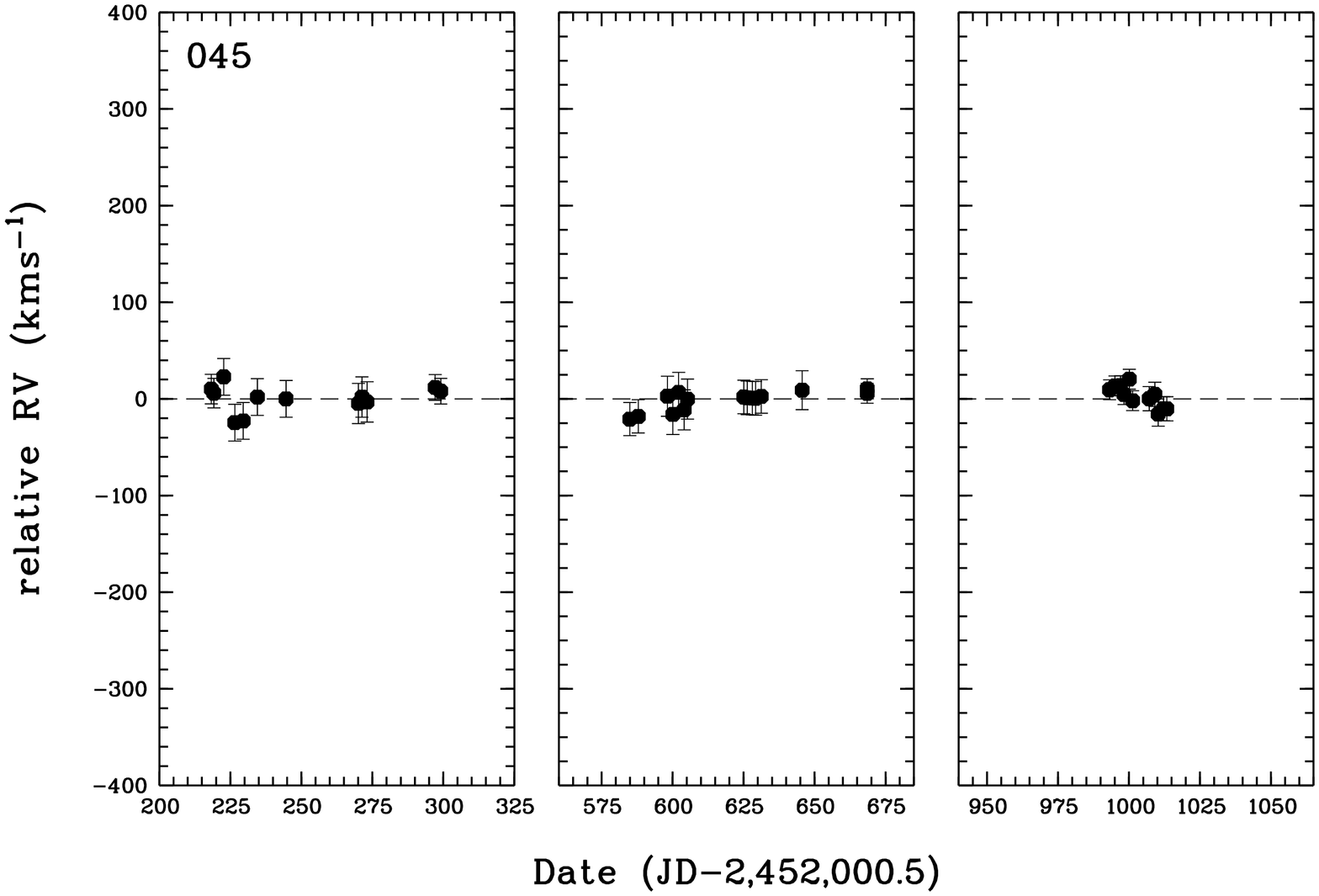}\hfill\\
\includegraphics[width=39mm,angle=-90,trim= 0 60 40 38,clip]{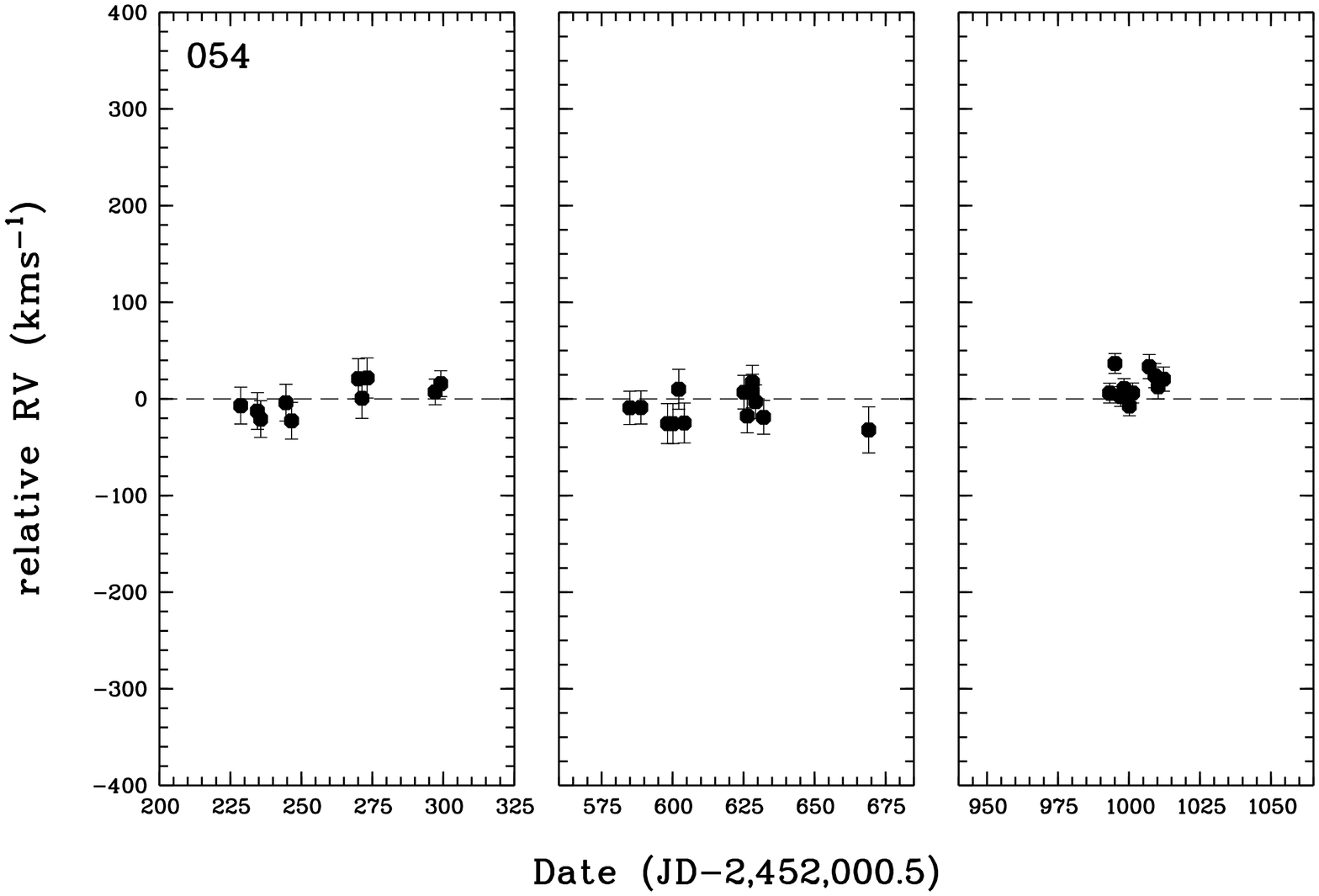}\hfill
\includegraphics[width=39mm,angle=-90,trim= 0 90 40 38,clip]{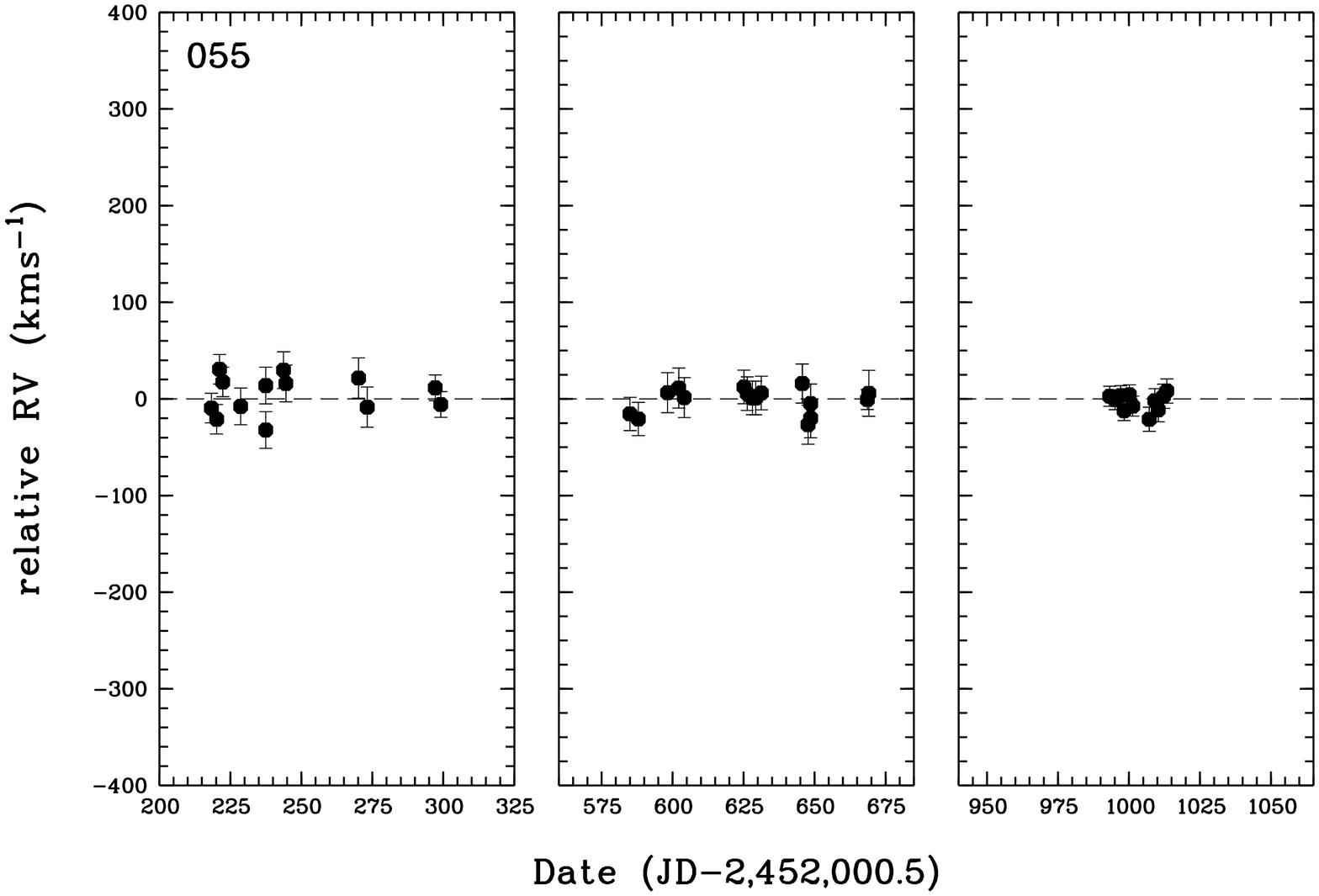}\hfill
\includegraphics[width=39mm,angle=-90,trim= 0 90 40 38,clip]{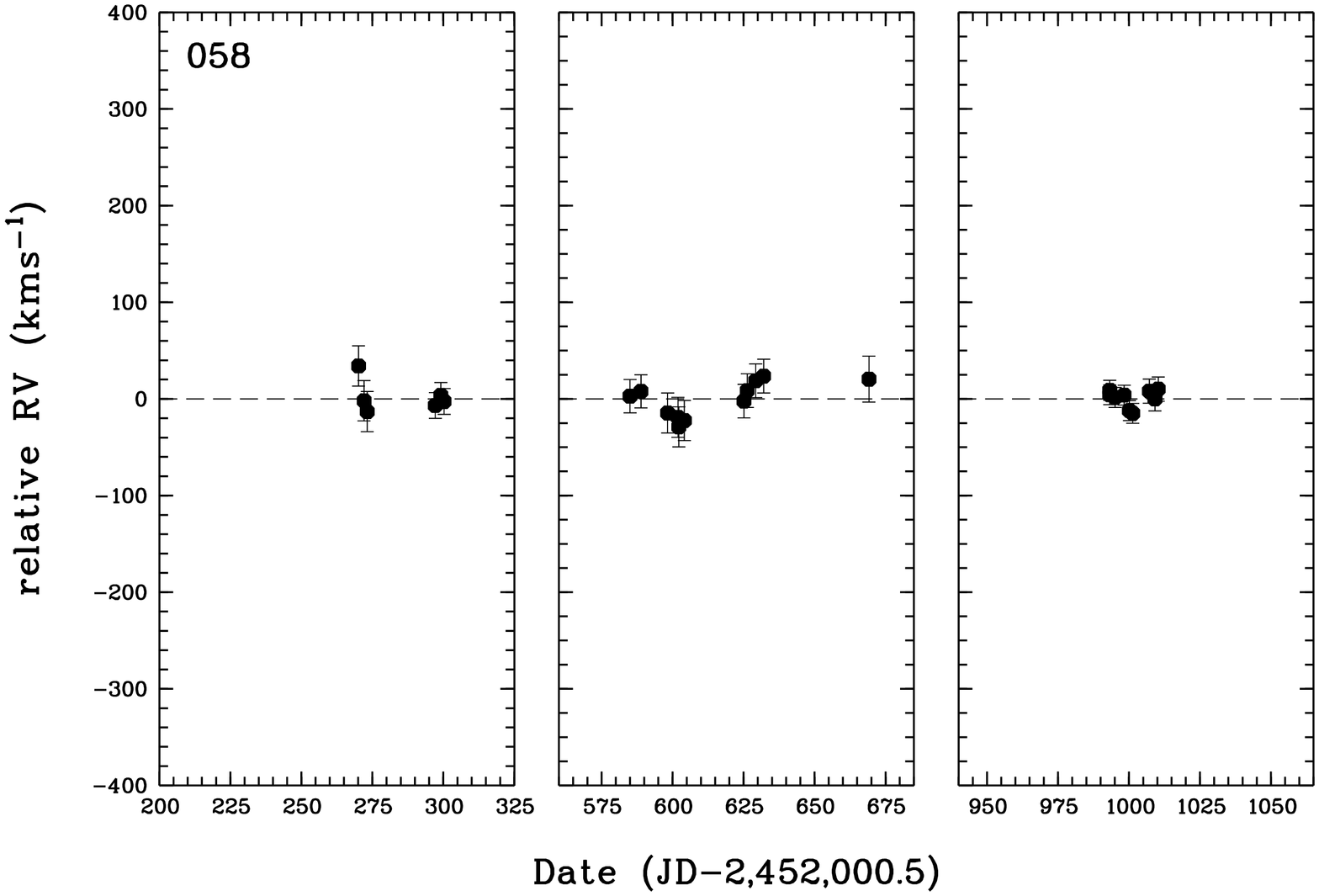}\hfill\\
\includegraphics[width=39mm,angle=-90,trim= 0 60 40 38,clip]{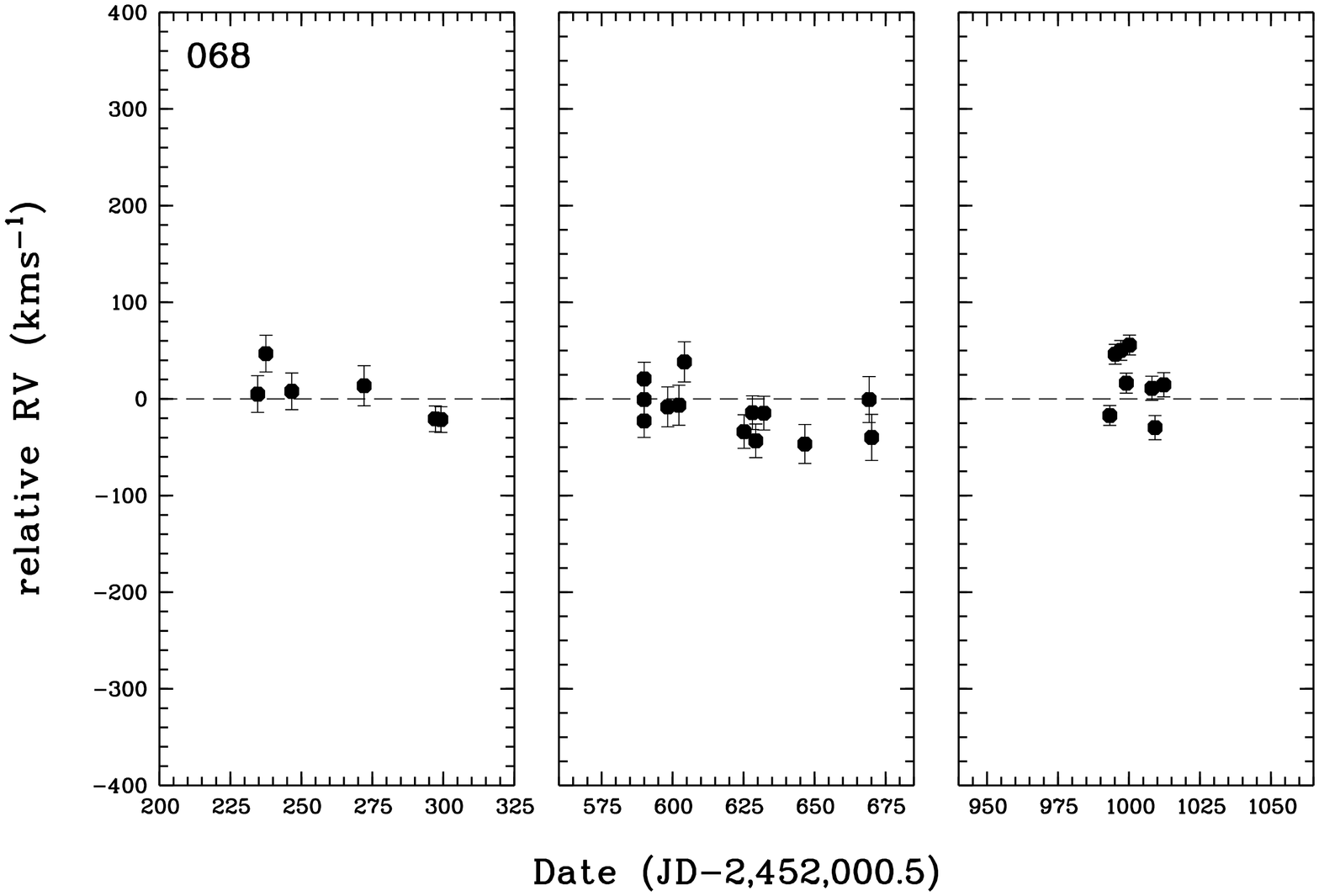}\hfill
\includegraphics[width=39mm,angle=-90,trim= 0 90 40 38,clip]{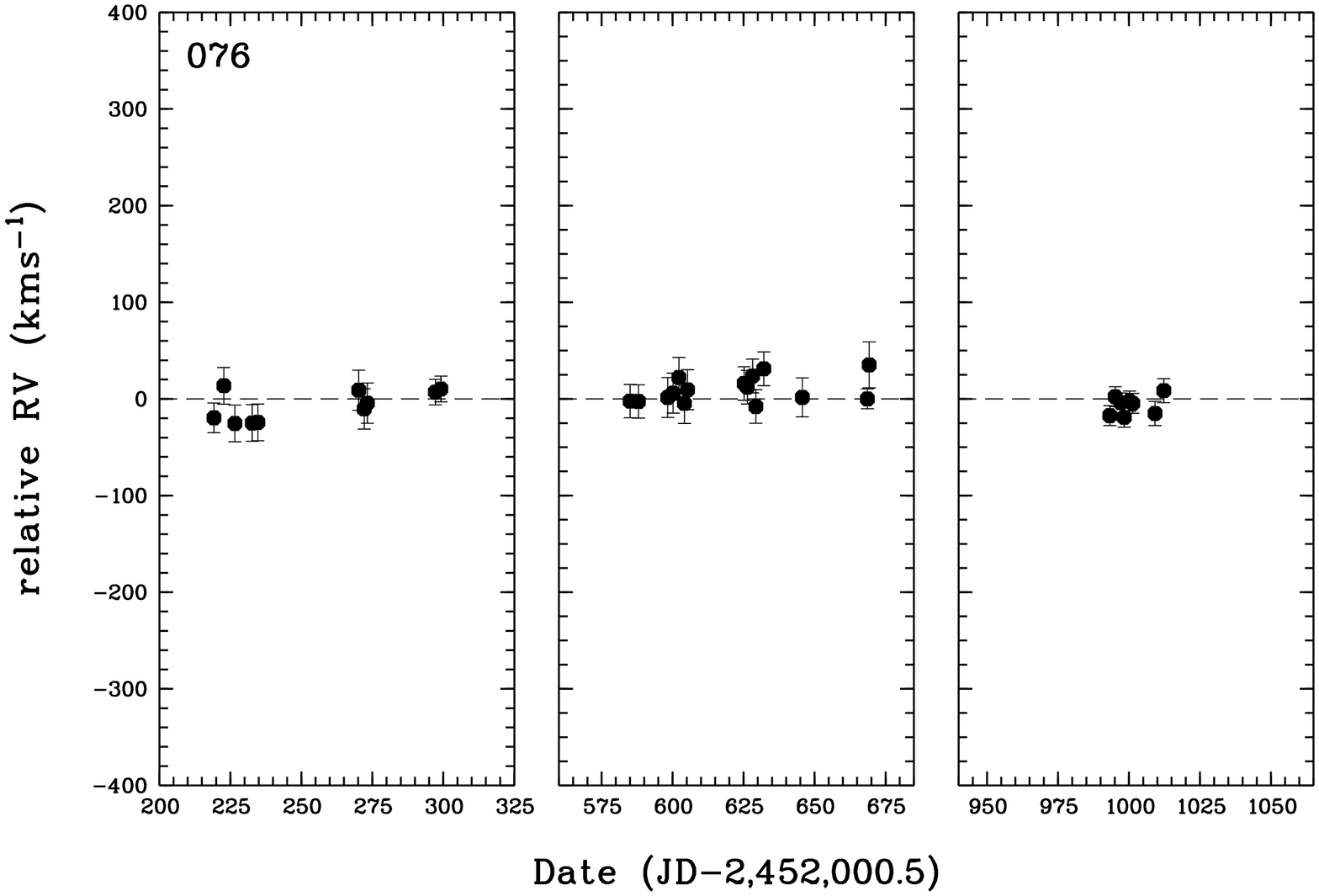}\hfill
\includegraphics[width=39mm,angle=-90,trim= 0 90 40 38,clip]{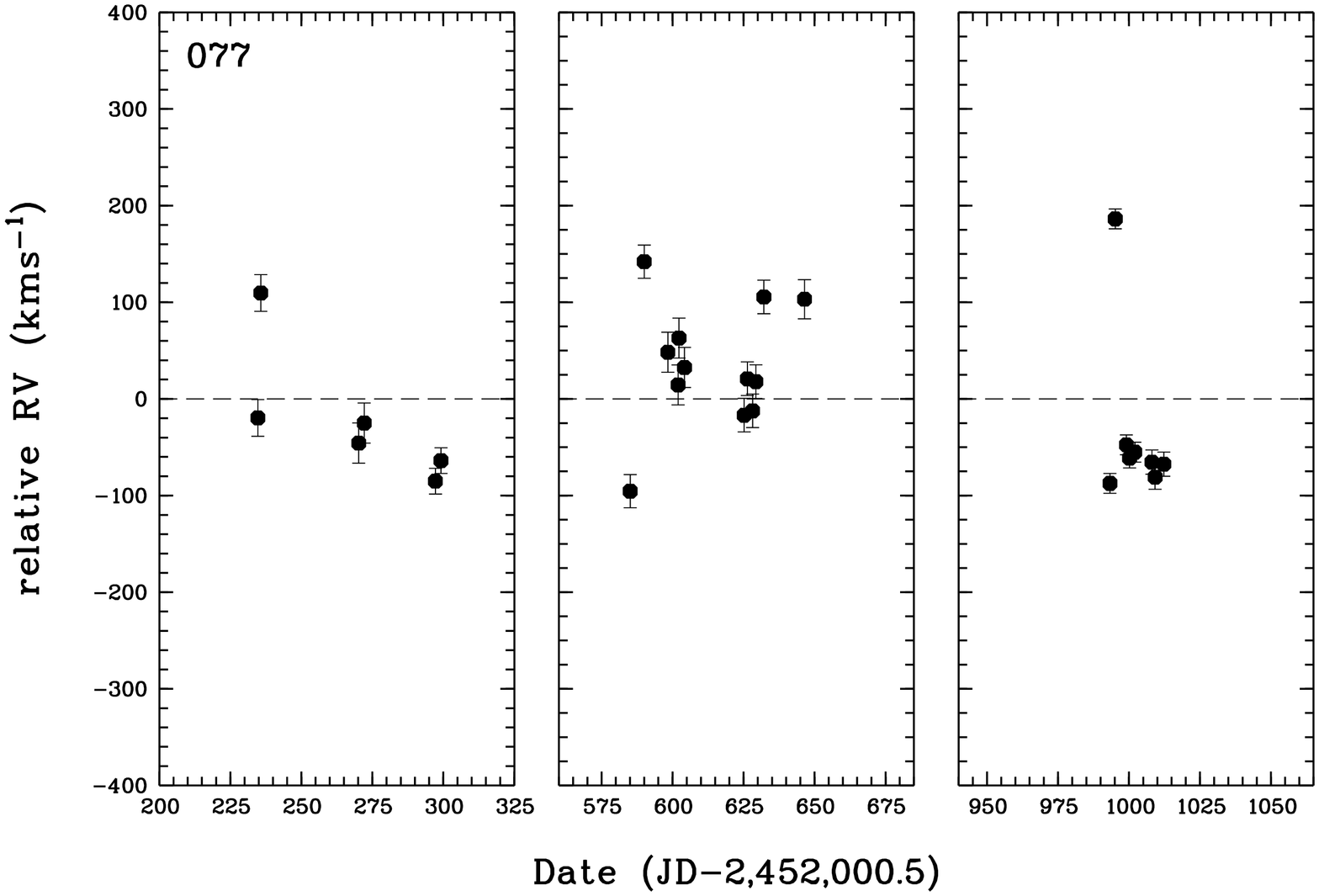}\hfill\\
\includegraphics[width=39mm,angle=-90,trim= 0 60 40 38,clip]{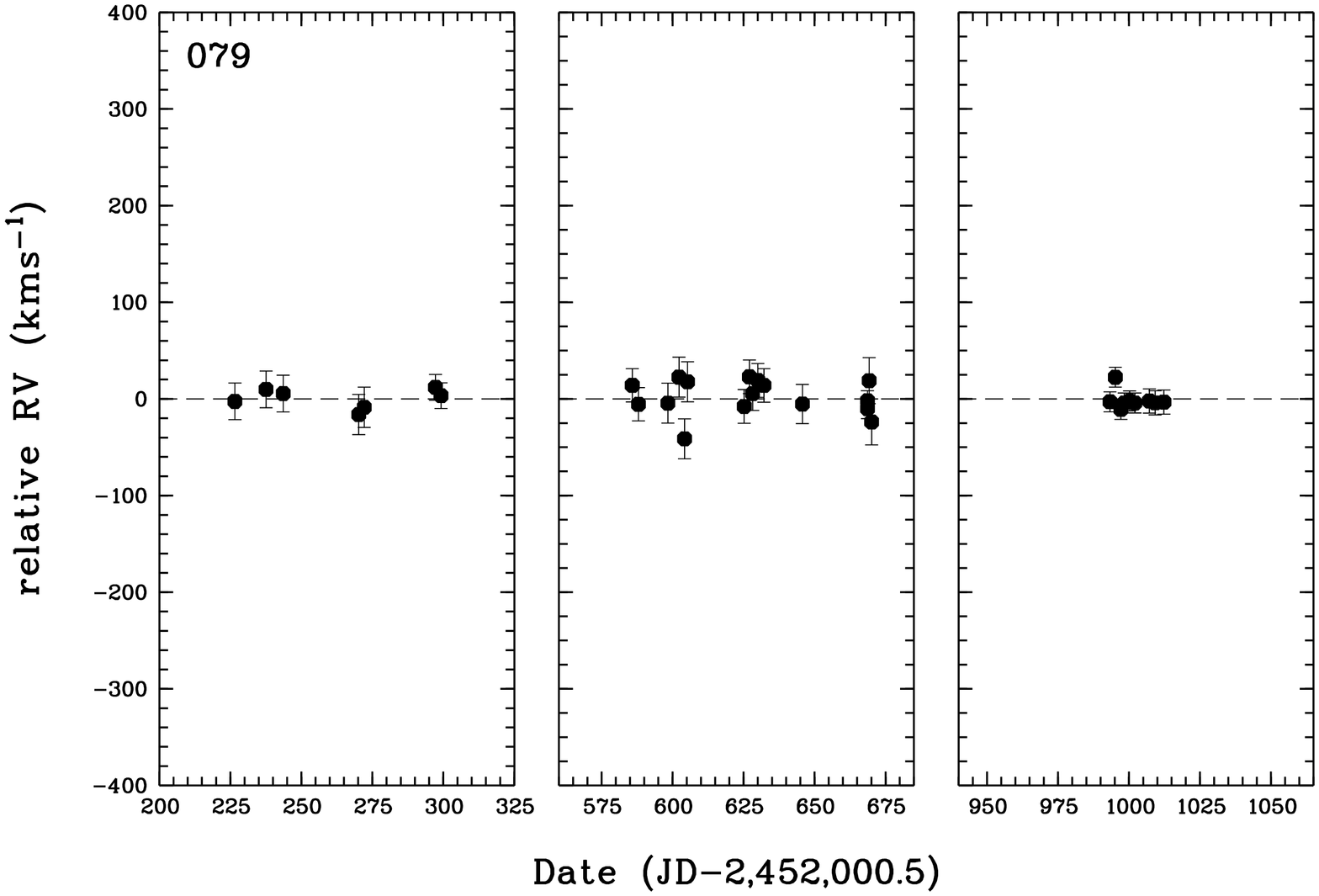}\hfill
\includegraphics[width=39mm,angle=-90,trim= 0 90 40 38,clip]{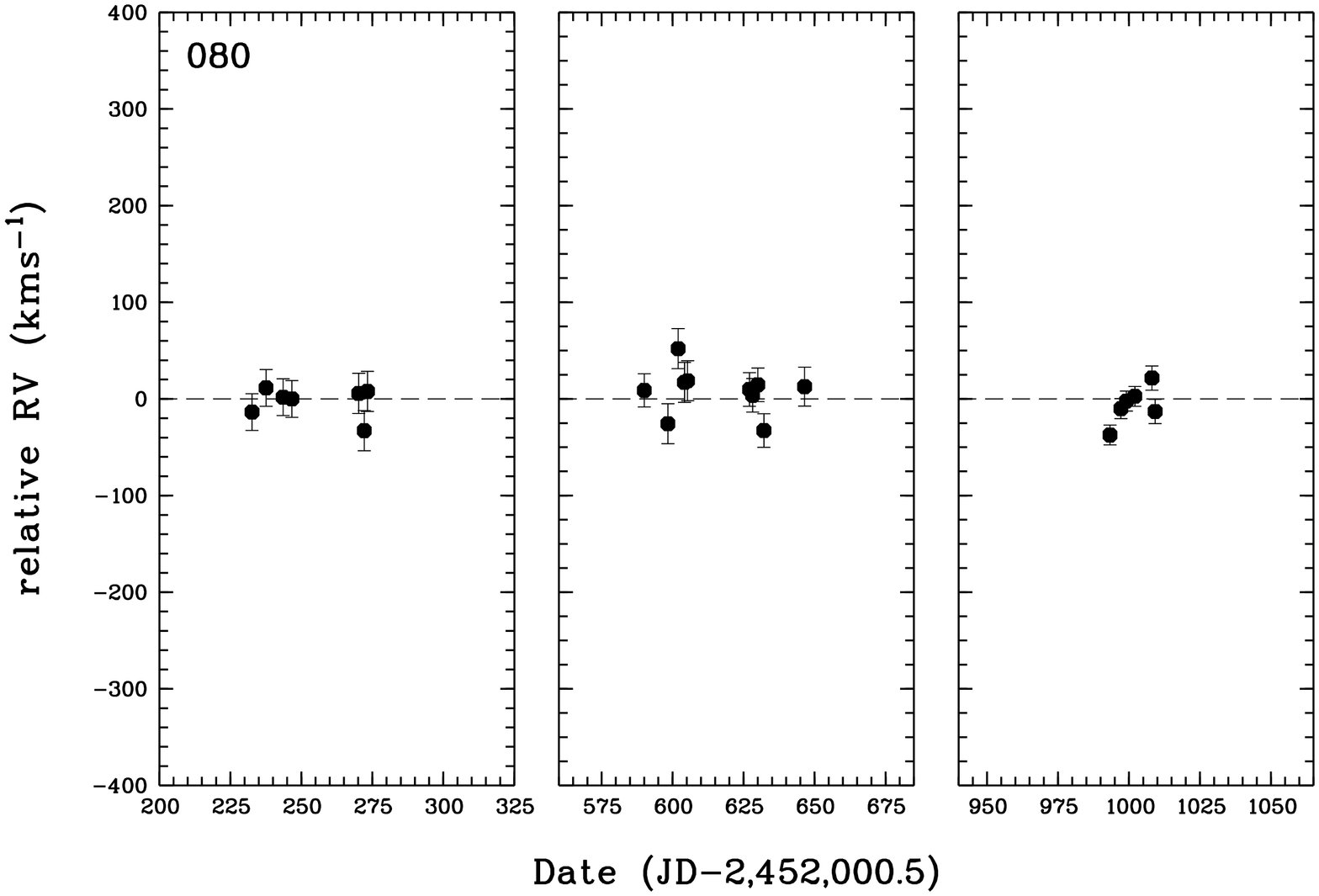}\hfill
\includegraphics[width=39mm,angle=-90,trim= 0 90 40 38,clip]{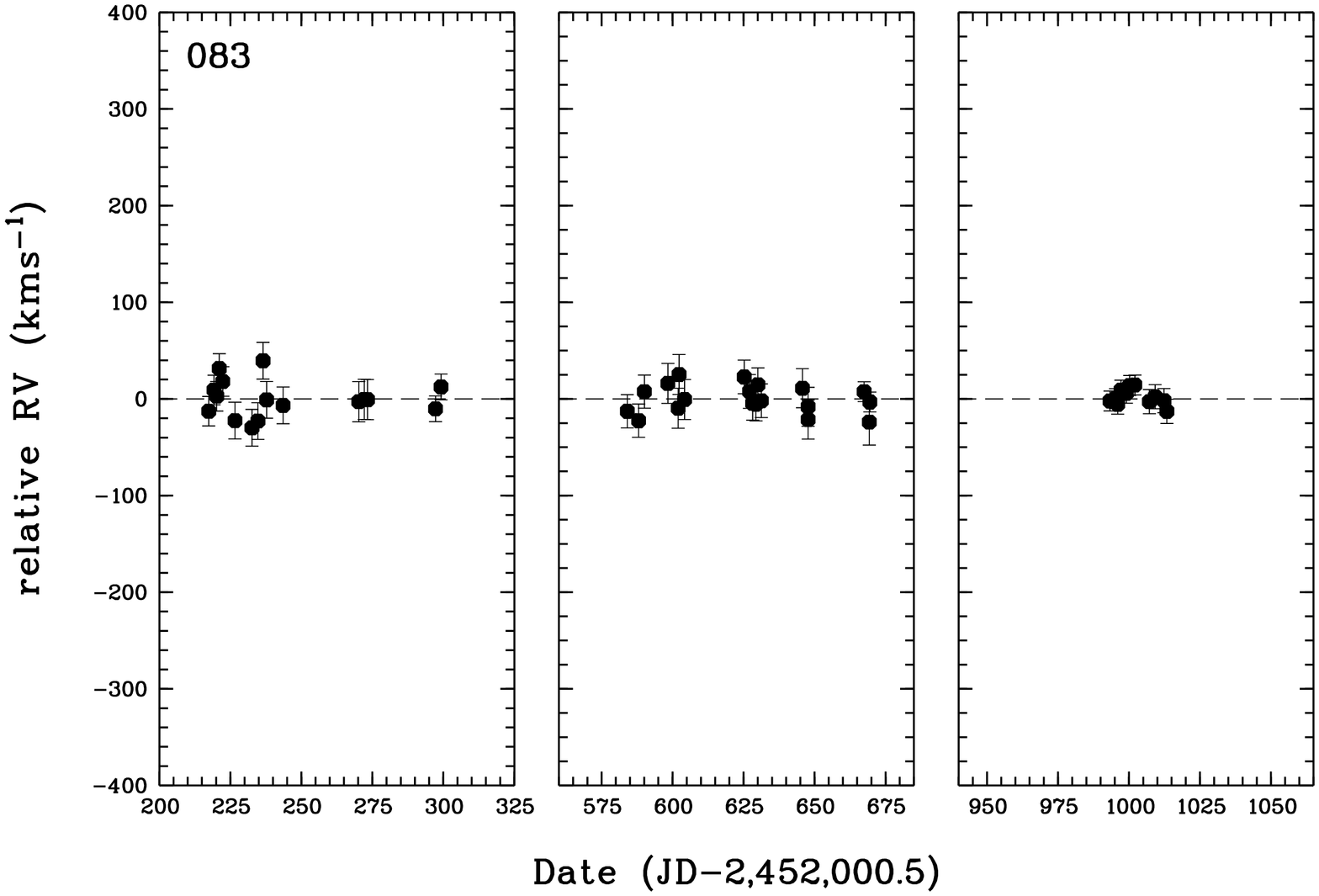}\hfill\\
\end{minipage}
\caption{Relative RVs of our program stars for the three observing seasons.}
\label{allepochs}
\end{figure*}

\begin{figure*}
\begin{minipage}{165mm}
\includegraphics[width=39mm,angle=-90,trim= 0 60 40 38,clip]{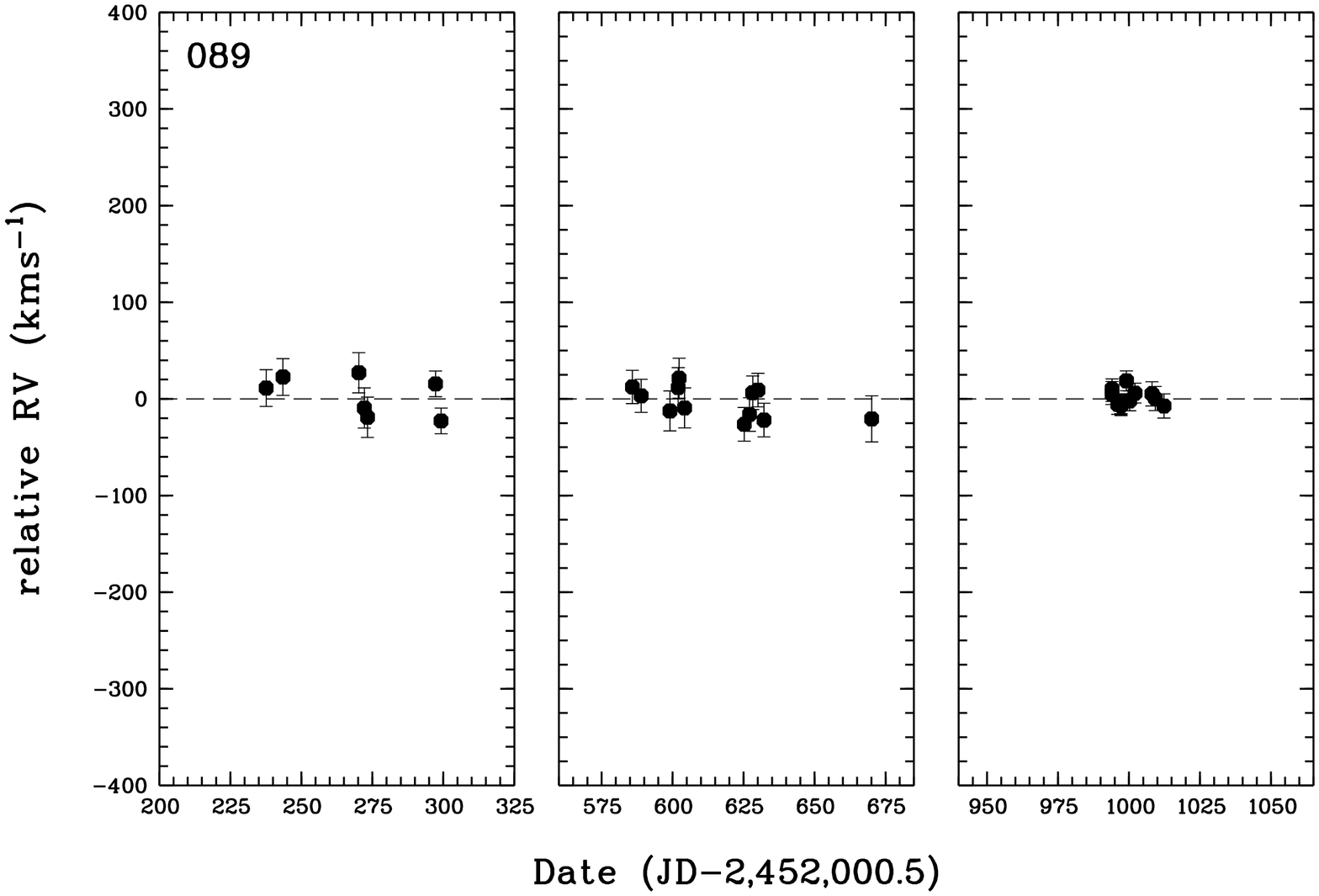}\hfill
\includegraphics[width=39mm,angle=-90,trim= 0 90 40 38,clip]{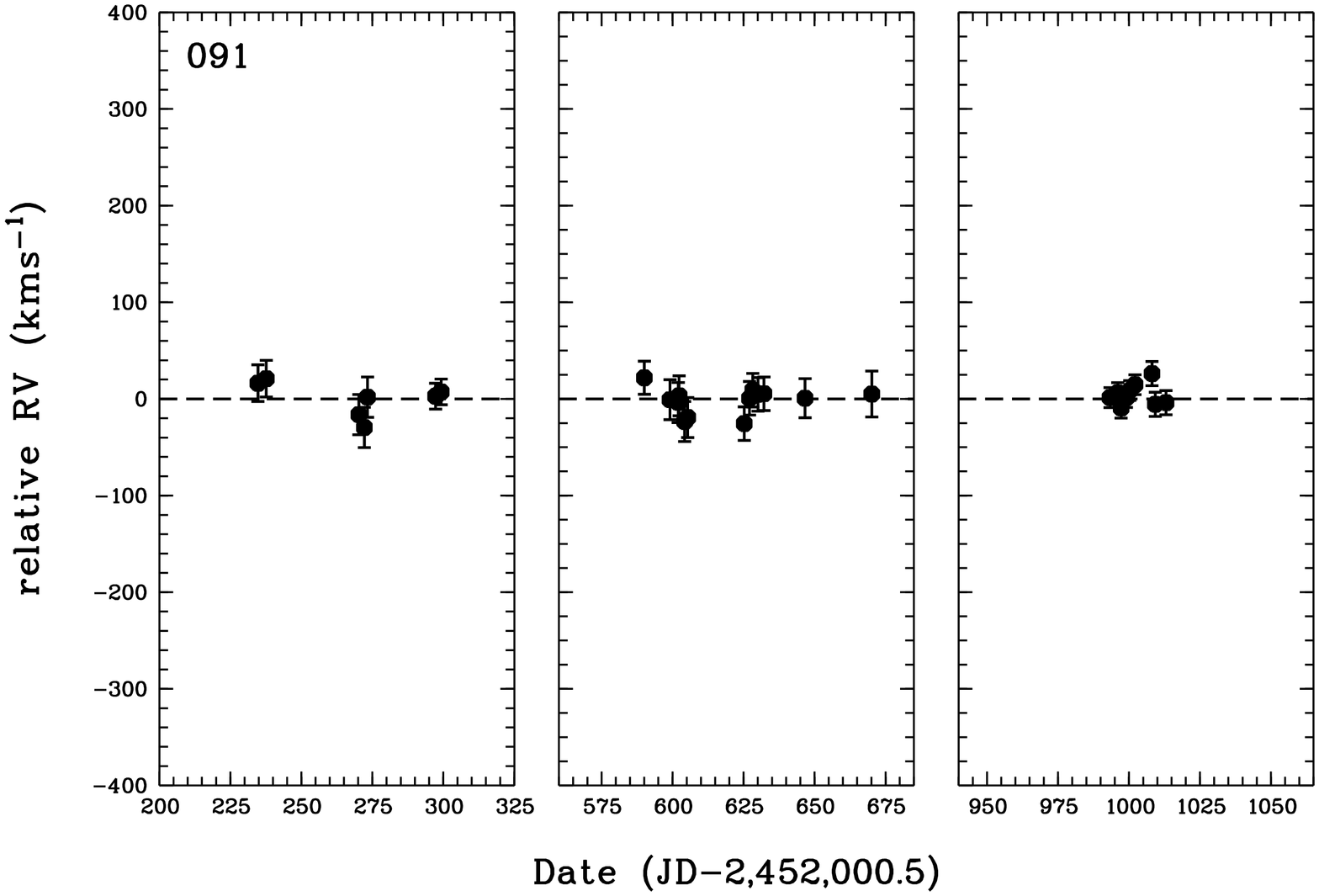}\hfill
\includegraphics[width=39mm,angle=-90,trim= 0 90 40 38,clip]{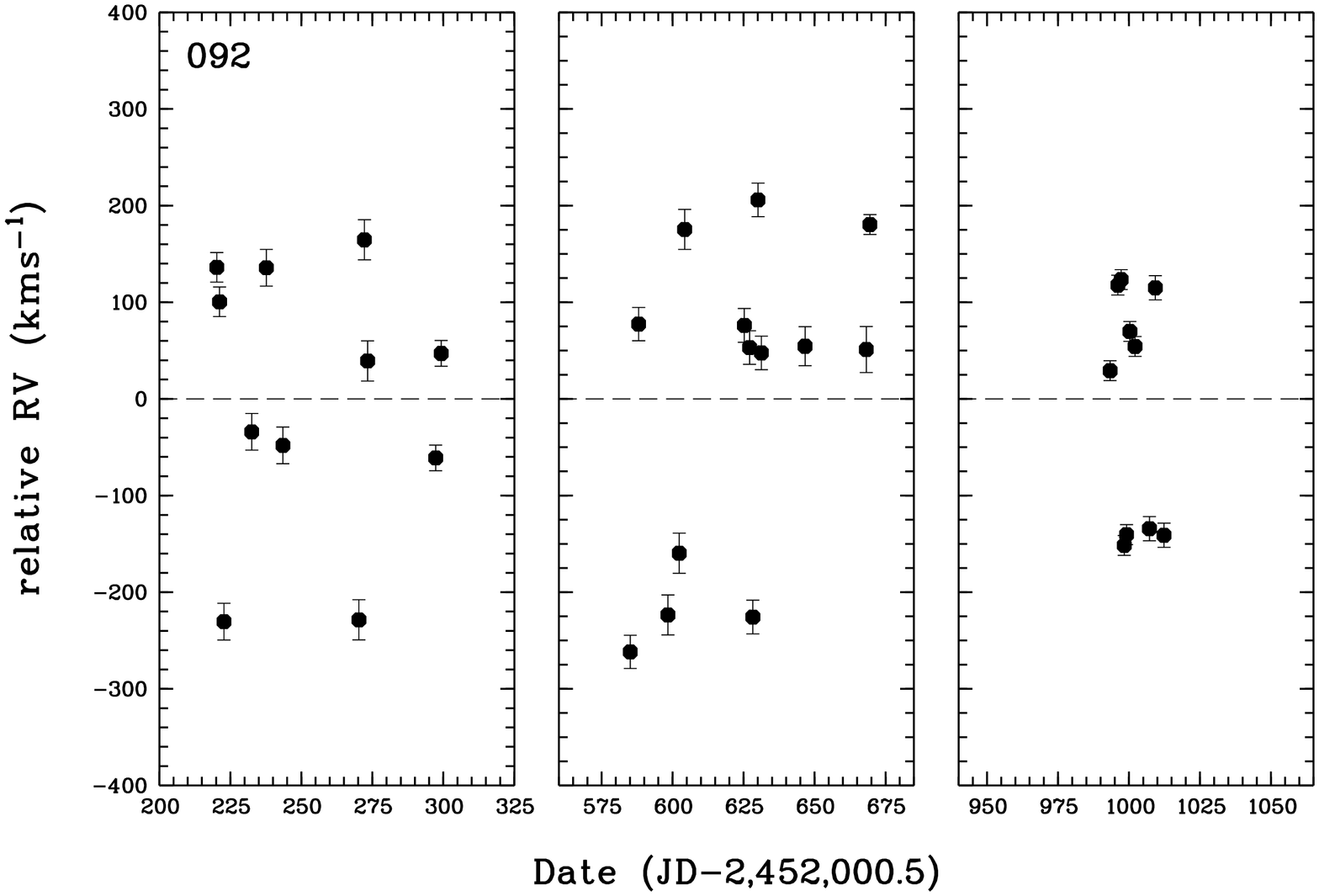}\hfill\\
\includegraphics[width=39mm,angle=-90,trim= 0 60 40 38,clip]{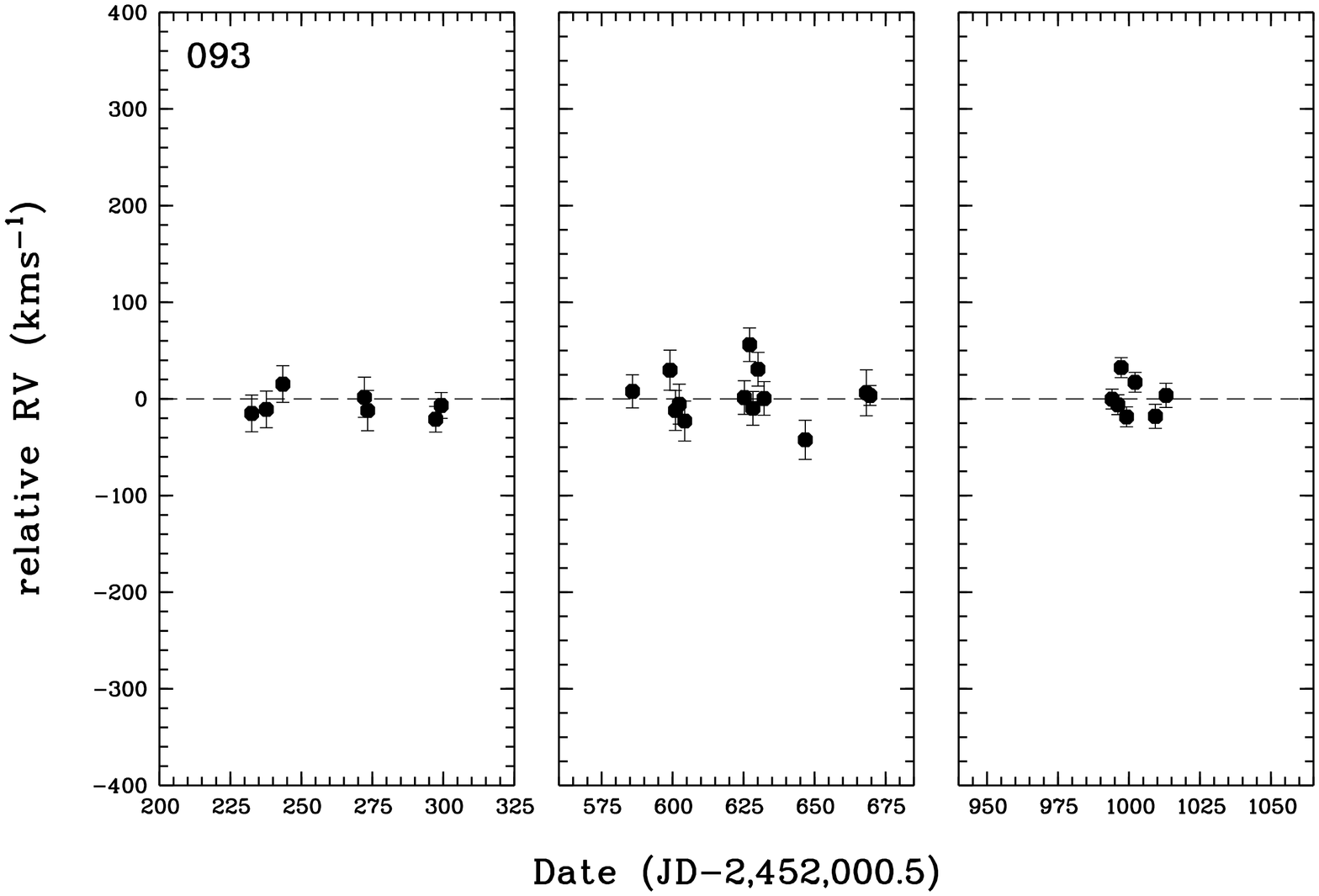}\hfill
\includegraphics[width=39mm,angle=-90,trim= 0 90 40 38,clip]{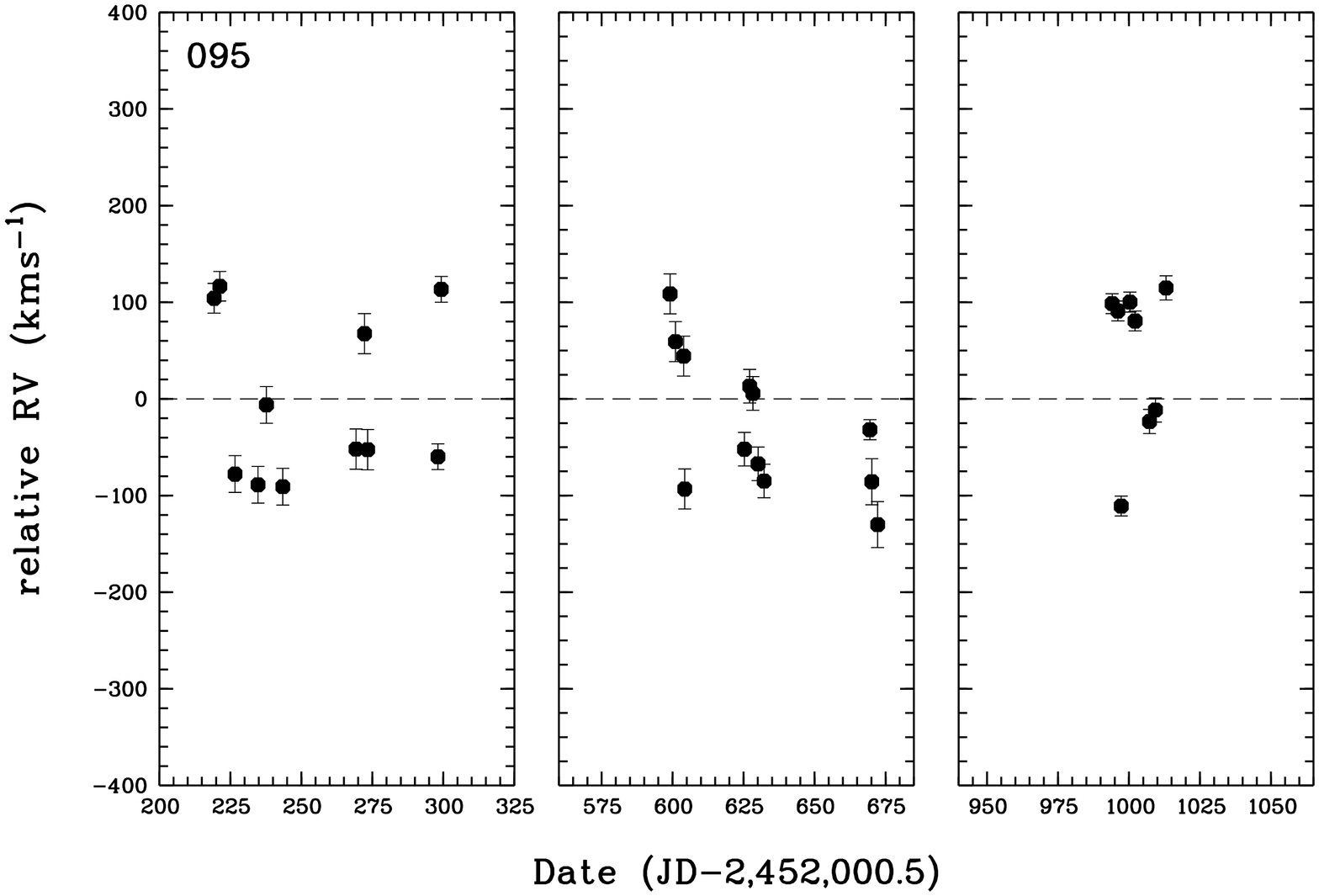}\hfill
\includegraphics[width=39mm,angle=-90,trim= 0 90 40 38,clip]{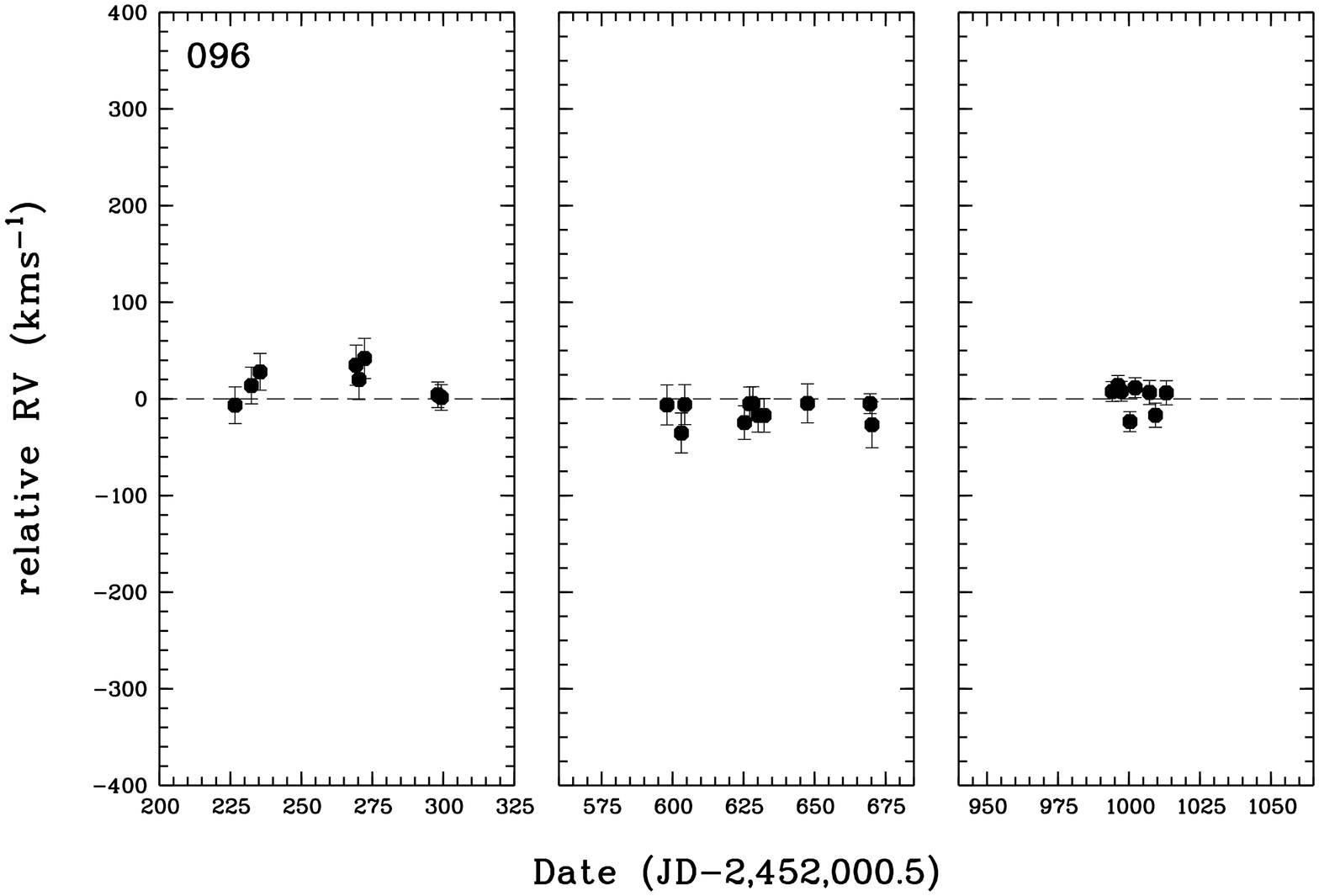}\hfill\\
\includegraphics[width=39mm,angle=-90,trim= 0 60 40 38,clip]{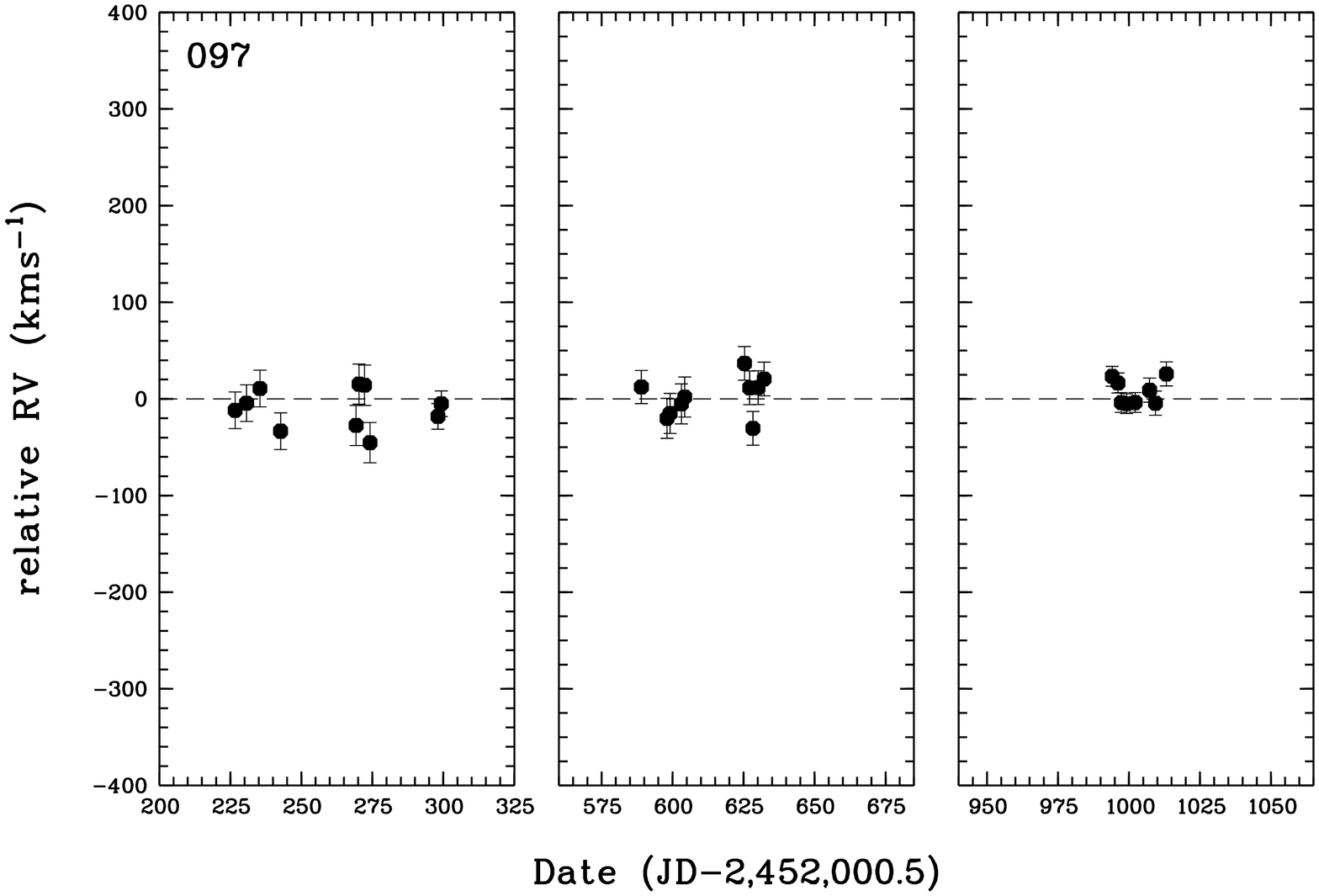}\hfill
\includegraphics[width=39mm,angle=-90,trim= 0 90 40 38,clip]{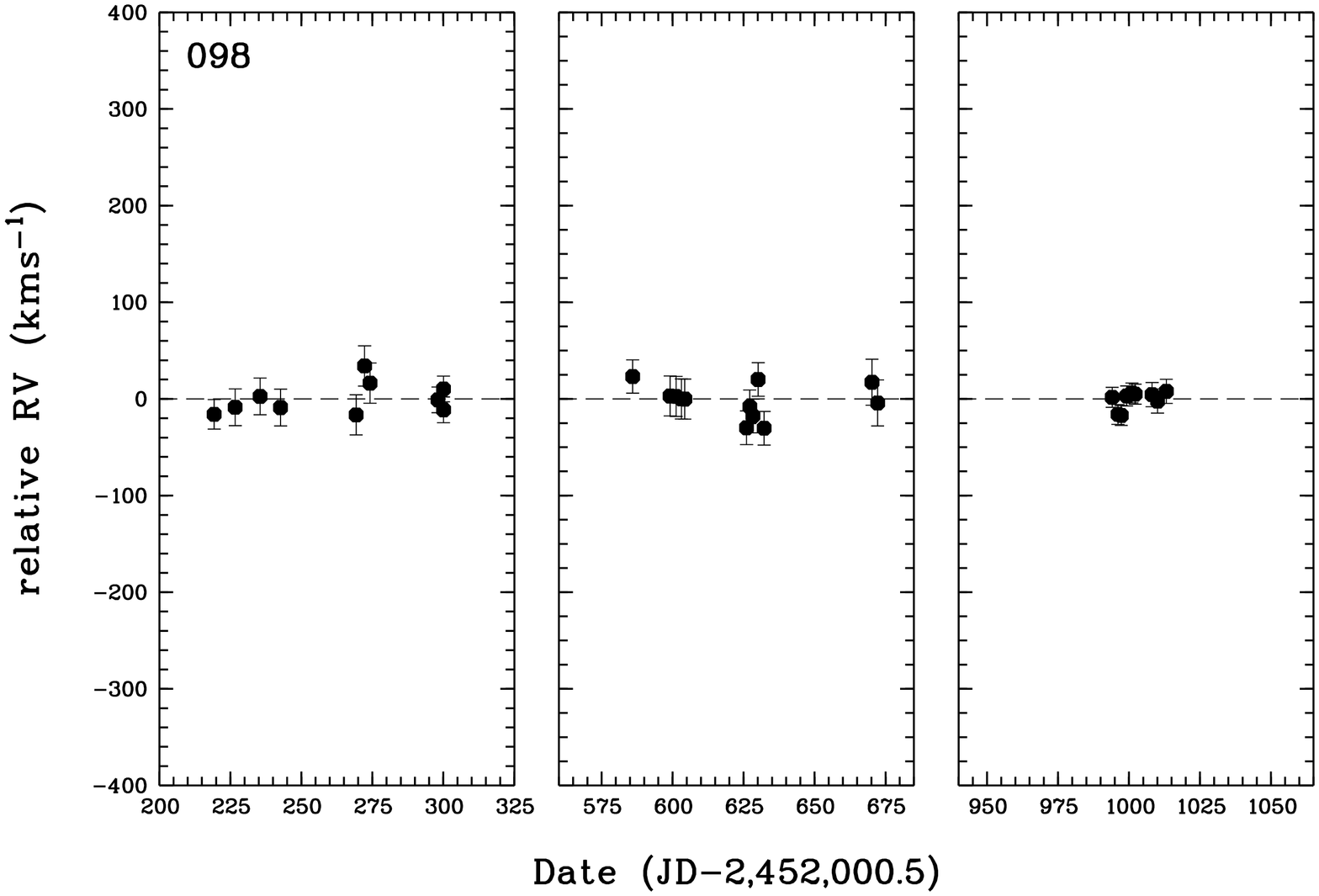}\hfill
\includegraphics[width=39mm,angle=-90,trim= 0 90 40 38,clip]{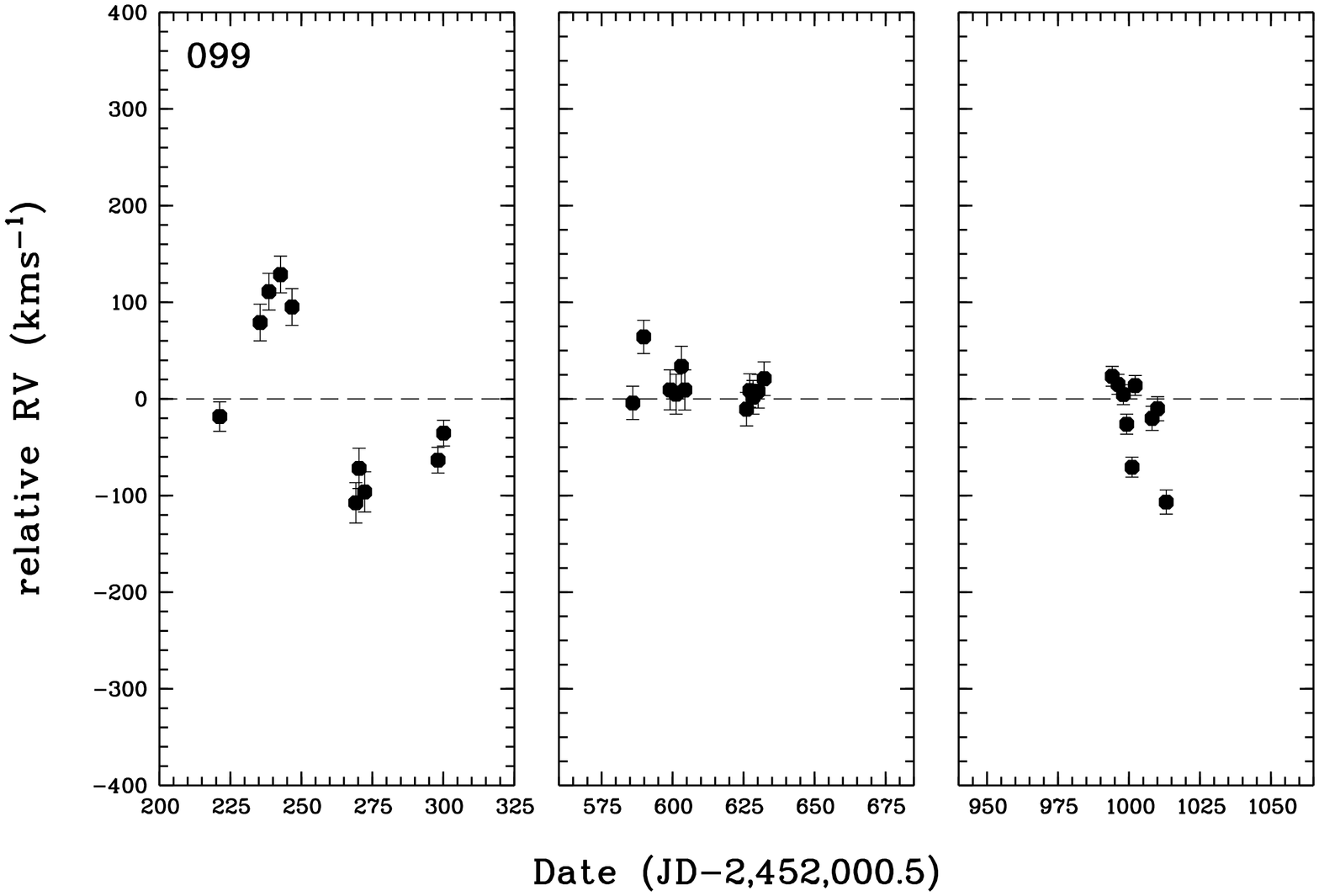}\hfill\\
\includegraphics[width=39mm,angle=-90,trim= 0 60 40 38,clip]{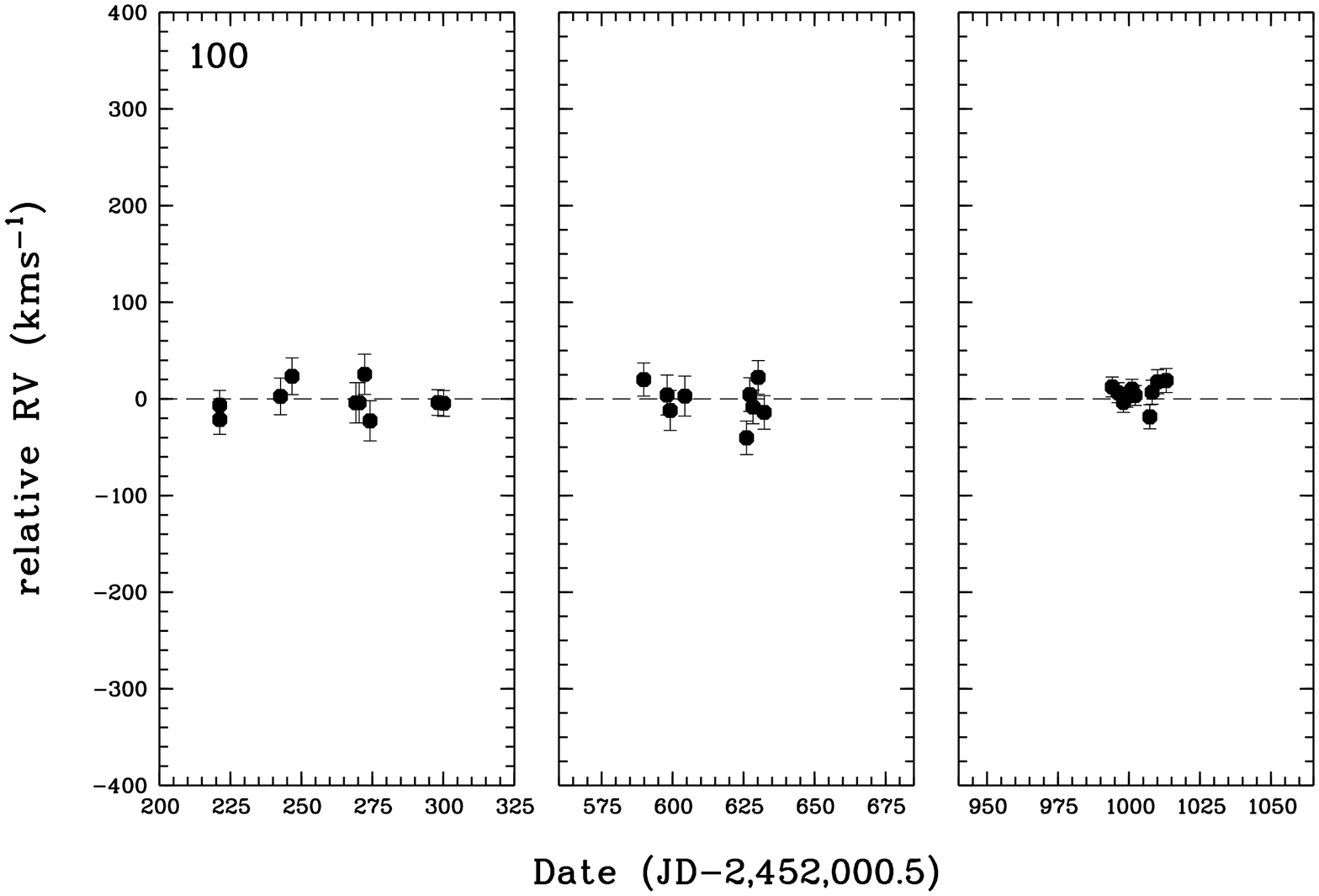}\hfill
\includegraphics[width=39mm,angle=-90,trim= 0 90 40 38,clip]{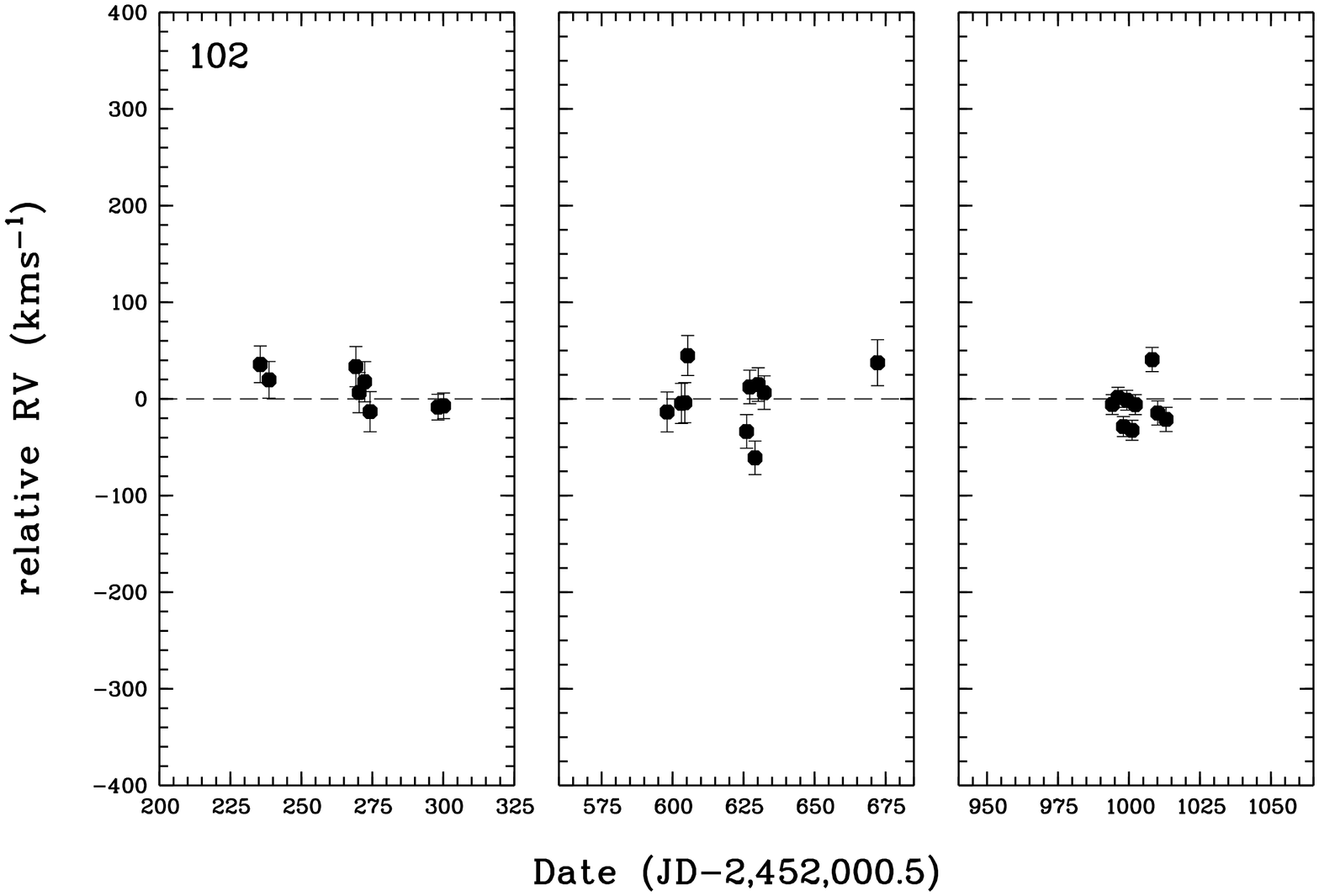}\hfill
\includegraphics[width=39mm,angle=-90,trim= 0 90 40 38,clip]{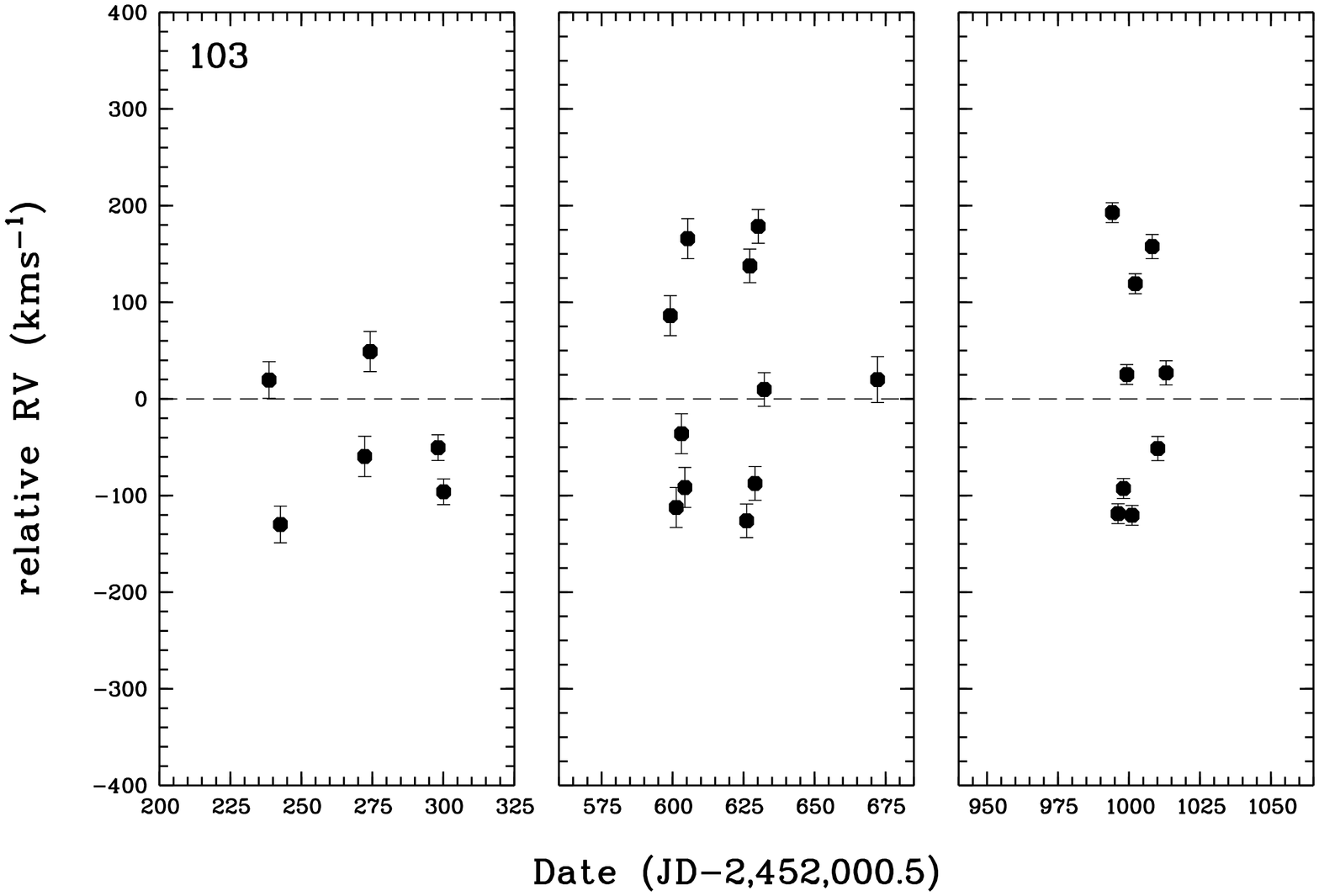}\hfill\\
\includegraphics[width=39mm,angle=-90,trim= 0 60 40 38,clip]{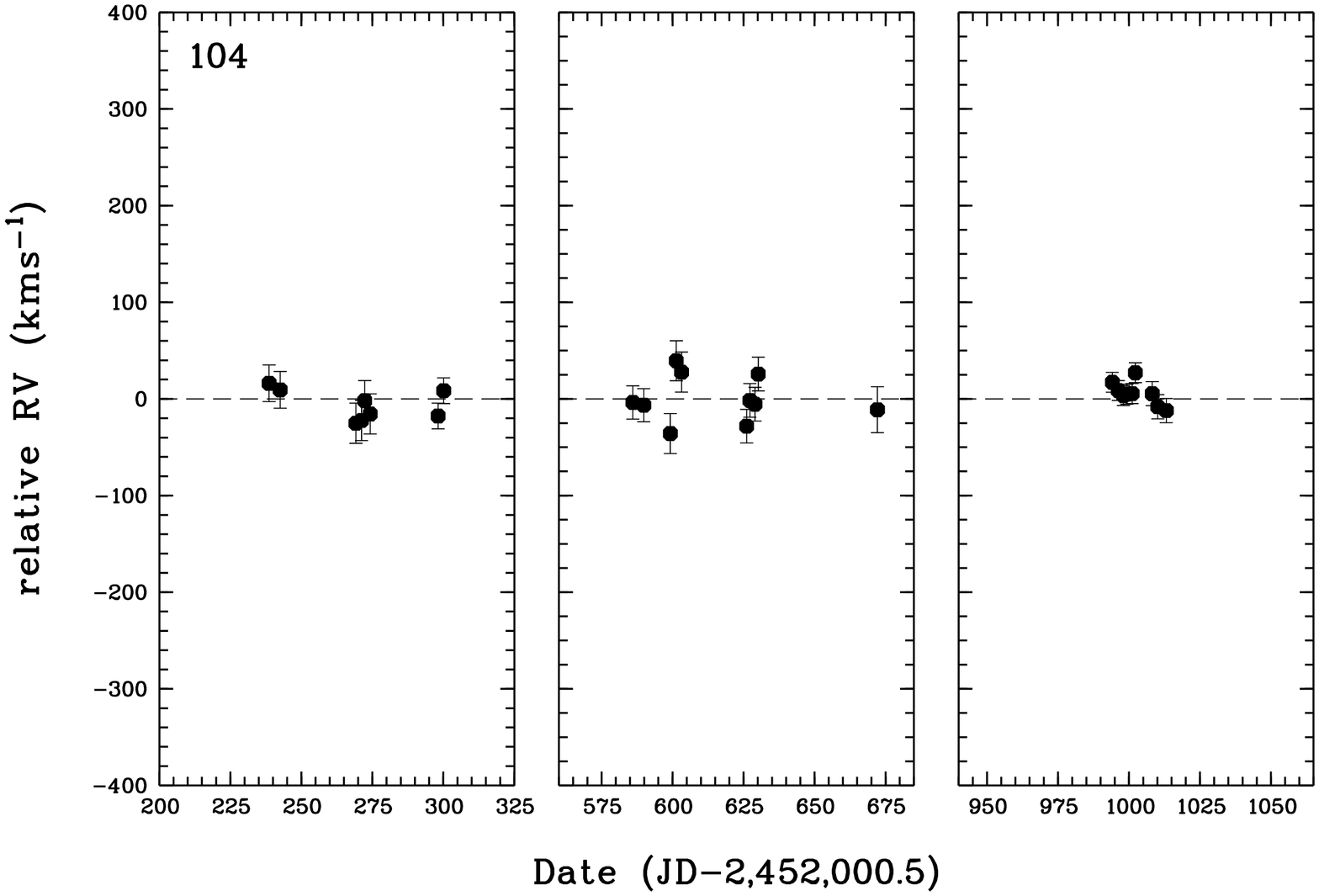}\hfill
\includegraphics[width=39mm,angle=-90,trim= 0 90 40 38,clip]{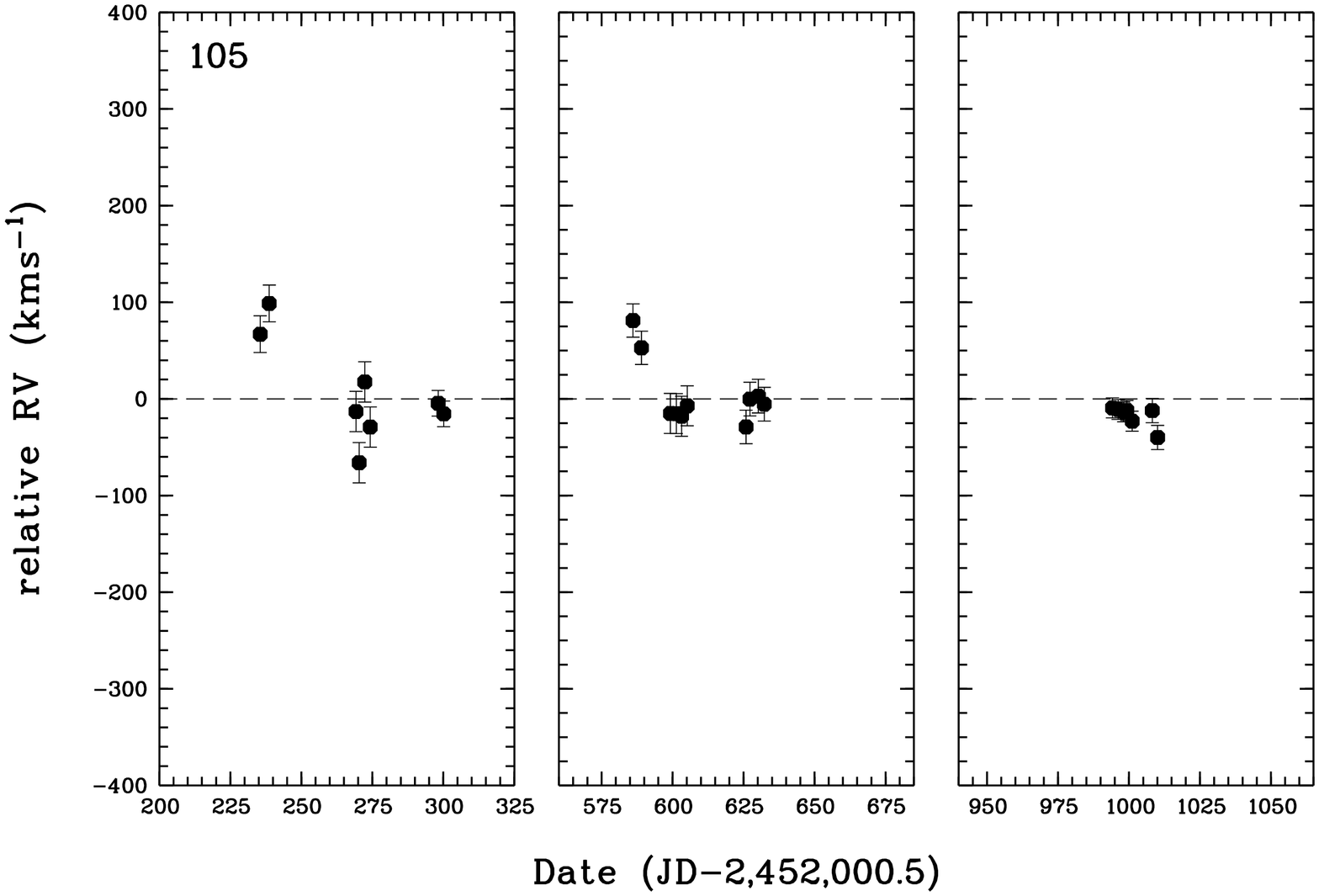}\hfill
\includegraphics[width=39mm,angle=-90,trim= 0 90 40 38,clip]{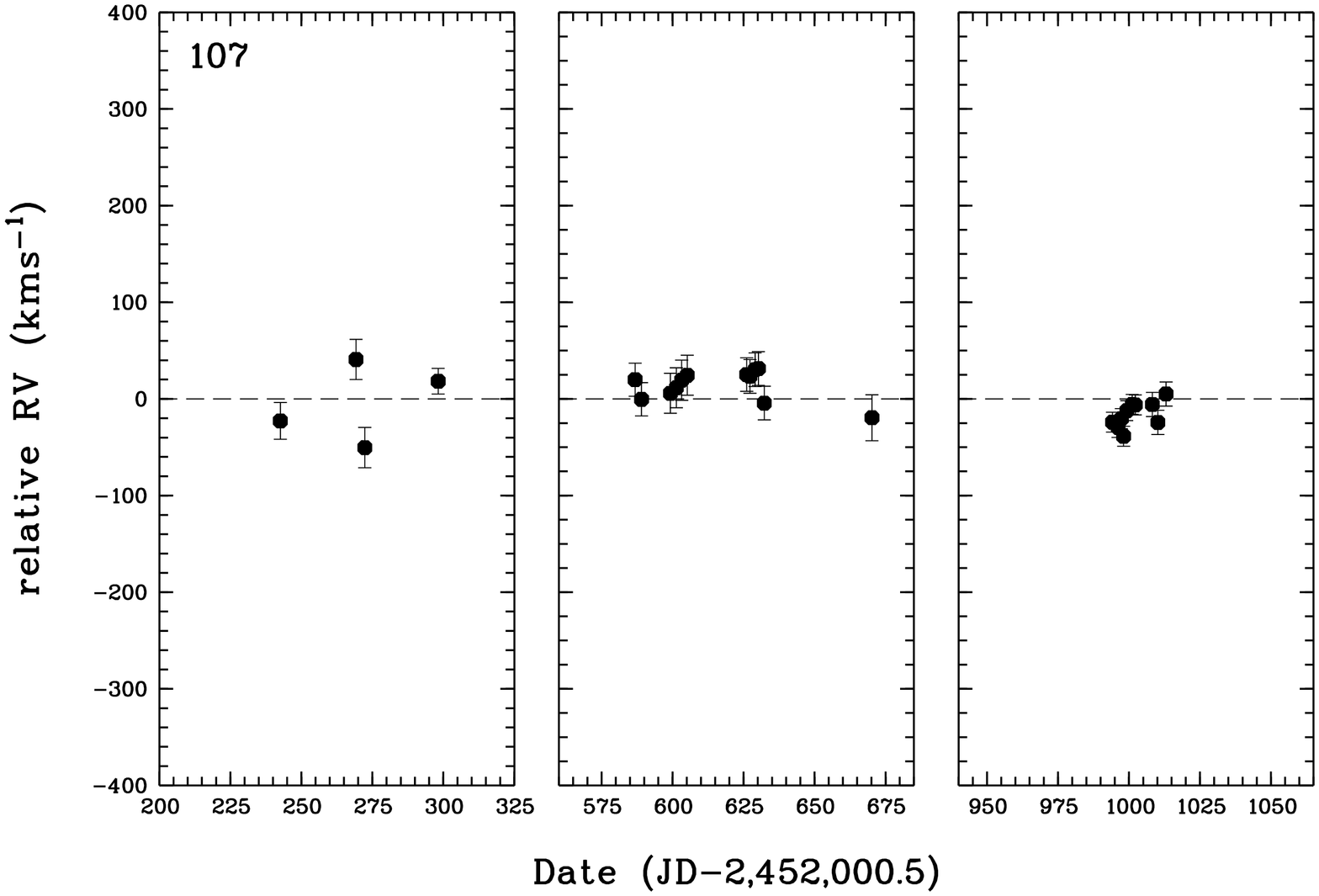}\hfill\\
\includegraphics[width=39mm,angle=-90,trim= 0 60 40 38,clip]{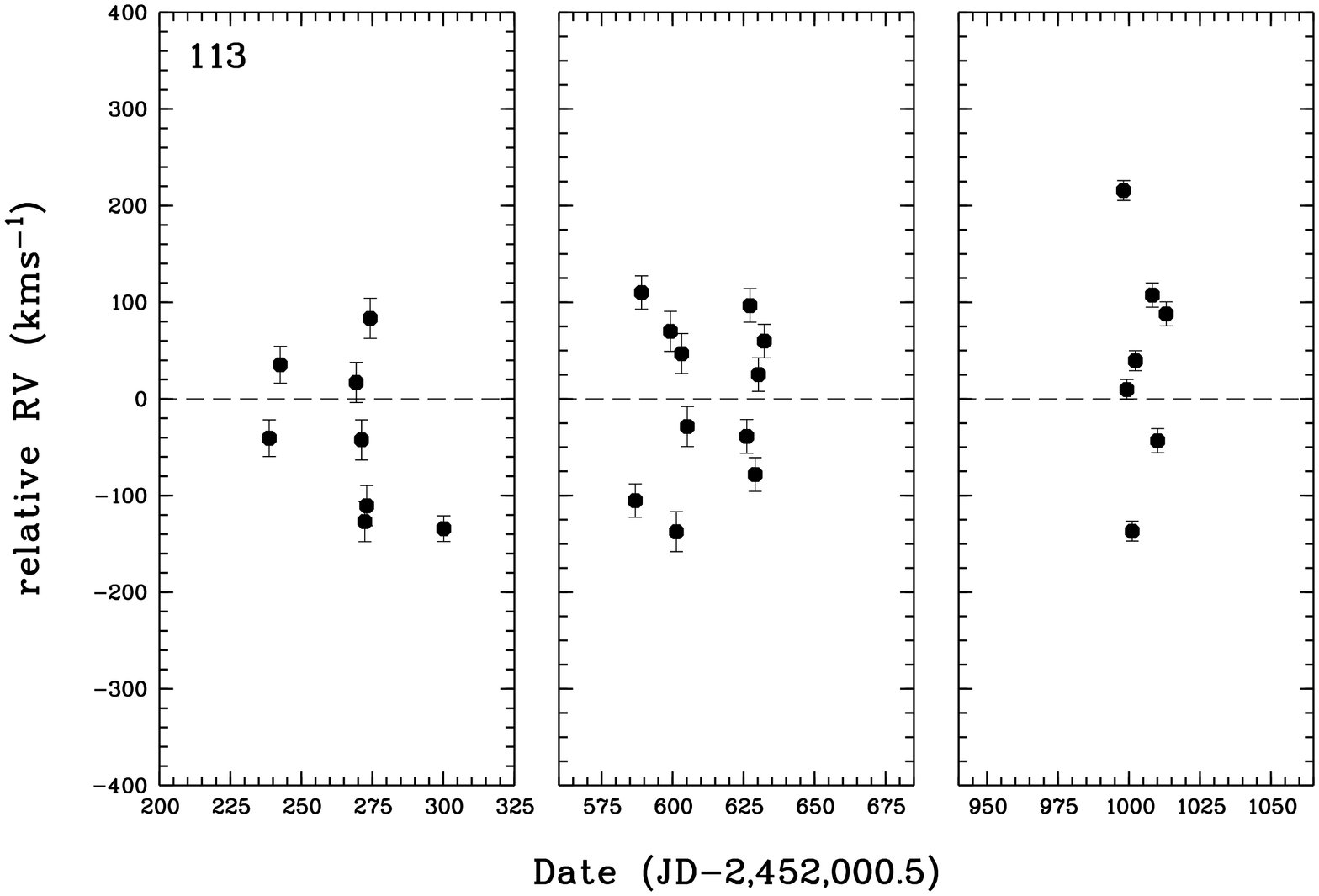}\hfill
\includegraphics[width=39mm,angle=-90,trim= 0 90 40 38,clip]{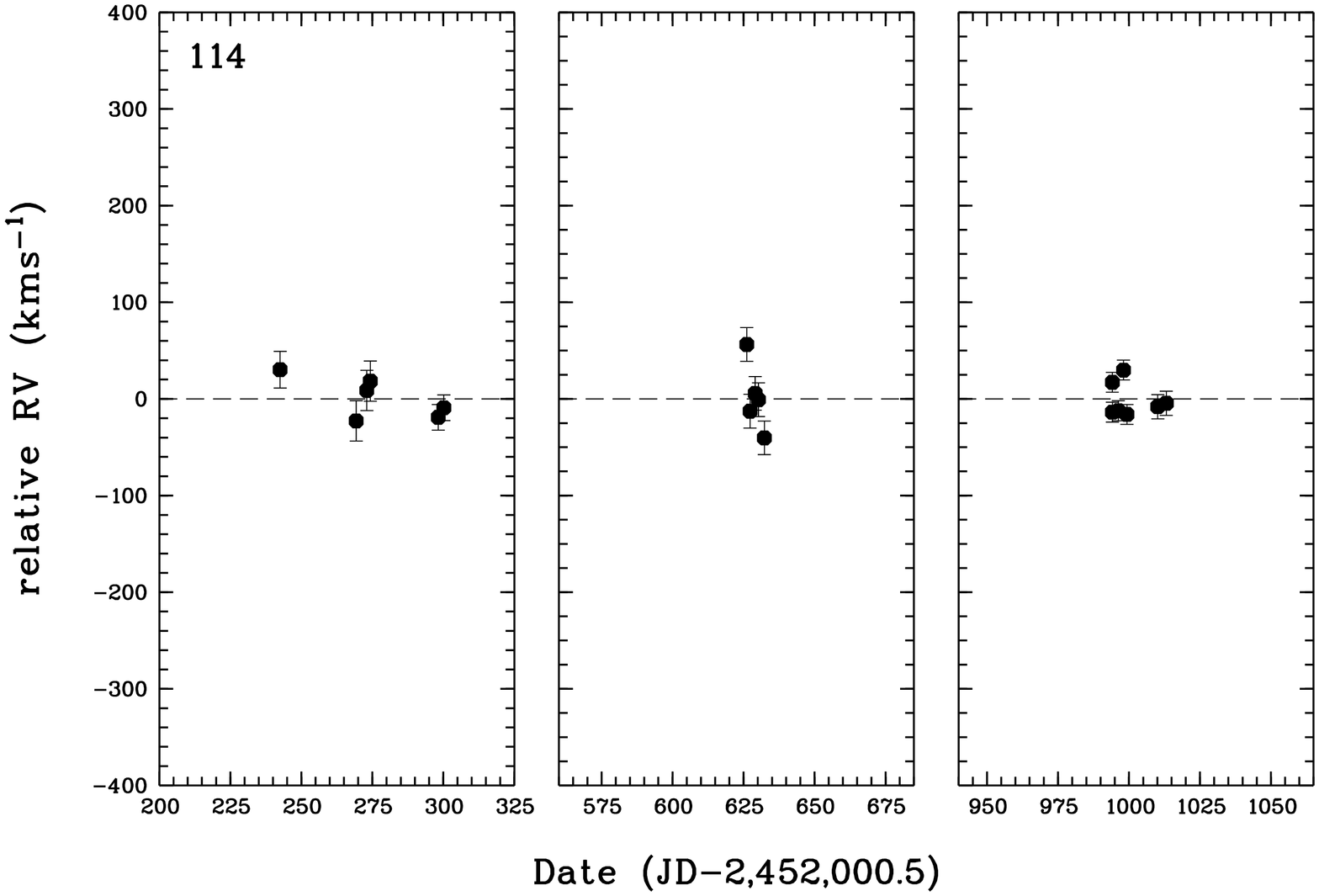}\hfill
\includegraphics[width=39mm,angle=-90,trim= 0 90 40 38,clip]{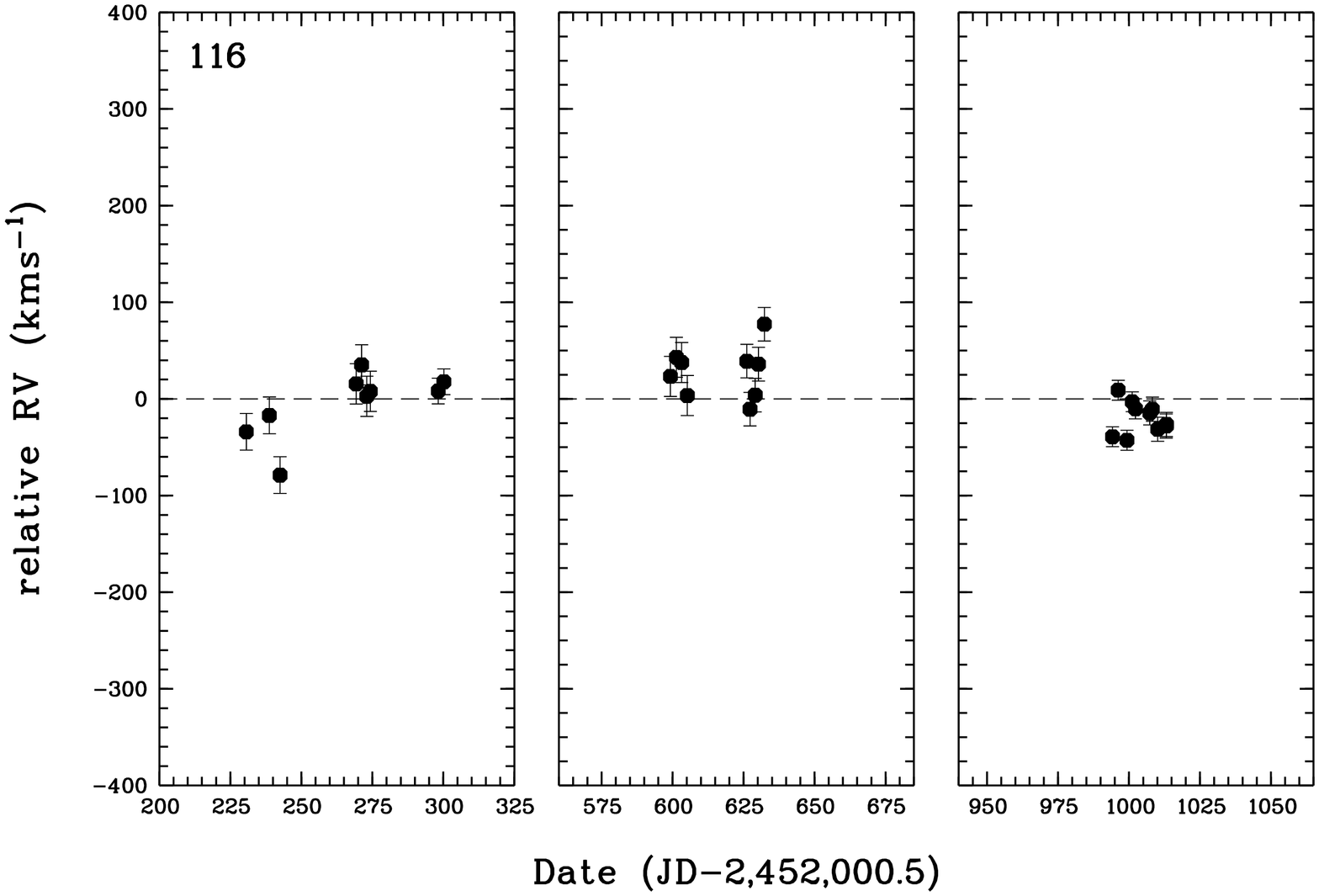}\hfill\\
\end{minipage}
\contcaption{}
\end{figure*}

\begin{figure*}
\begin{minipage}{165mm}
\includegraphics[width=39mm,angle=-90,trim= 0 60 40 38,clip]{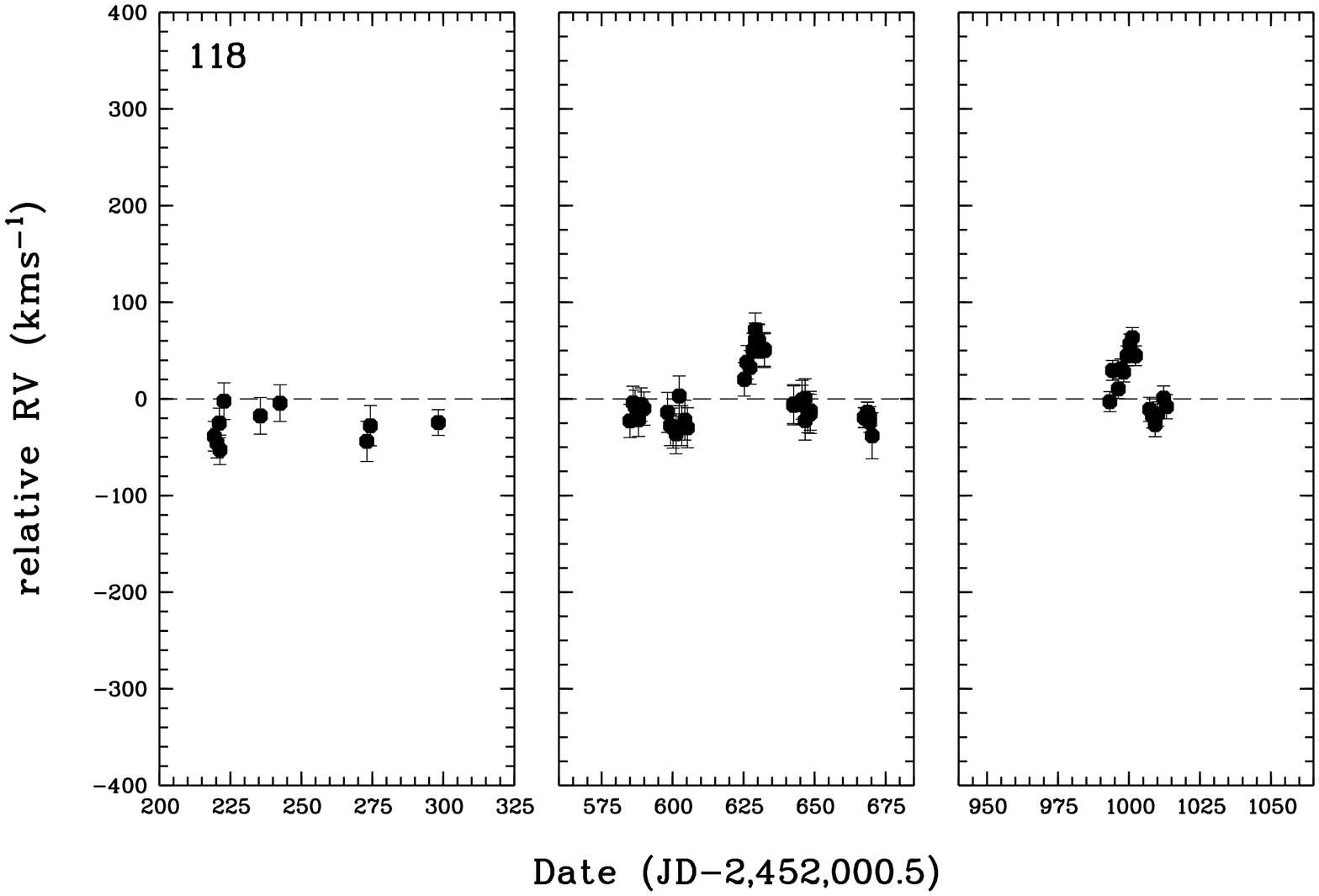}\hfill
\includegraphics[width=39mm,angle=-90,trim= 0 90 40 38,clip]{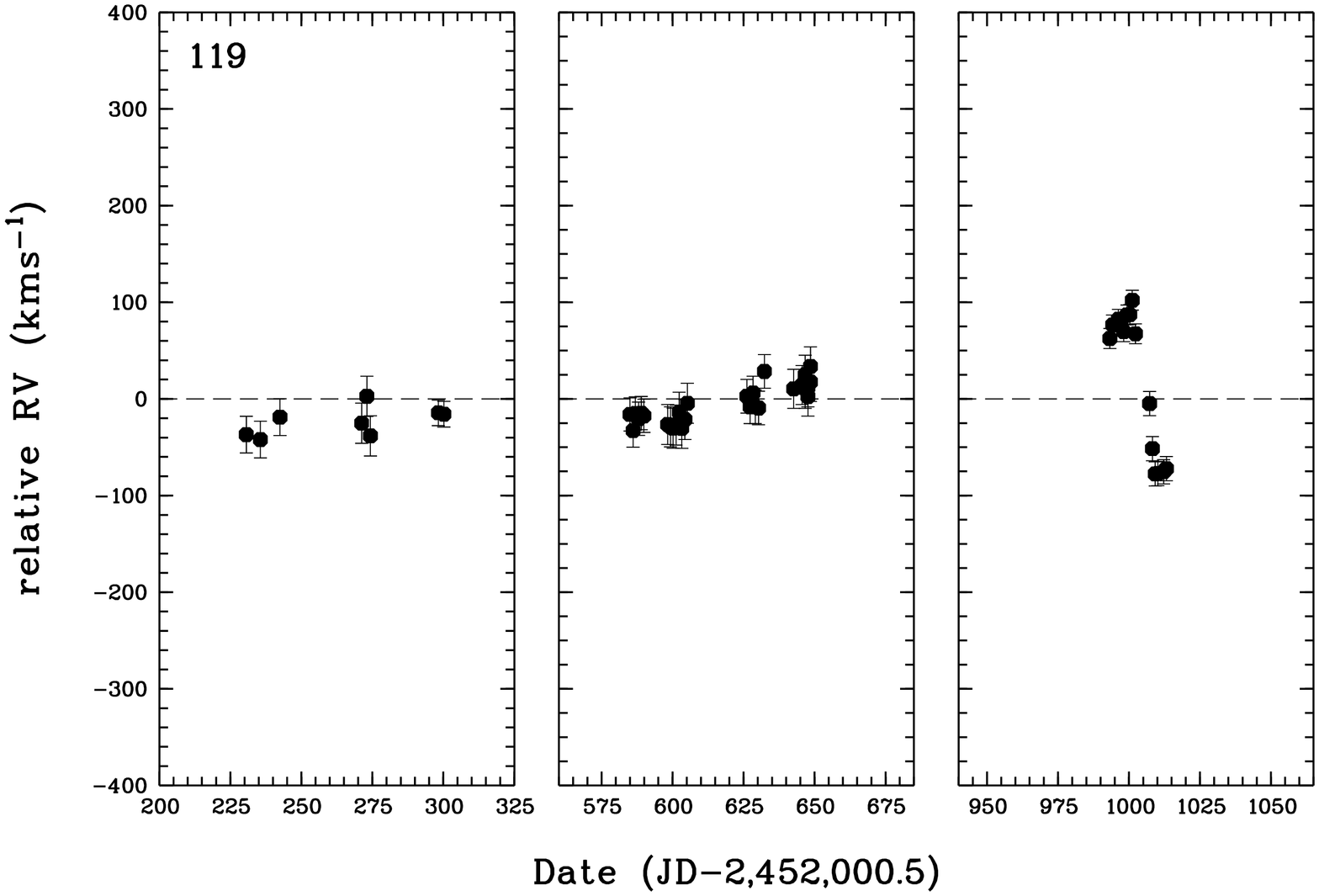}\hfill
\includegraphics[width=39mm,angle=-90,trim= 0 90 40 38,clip]{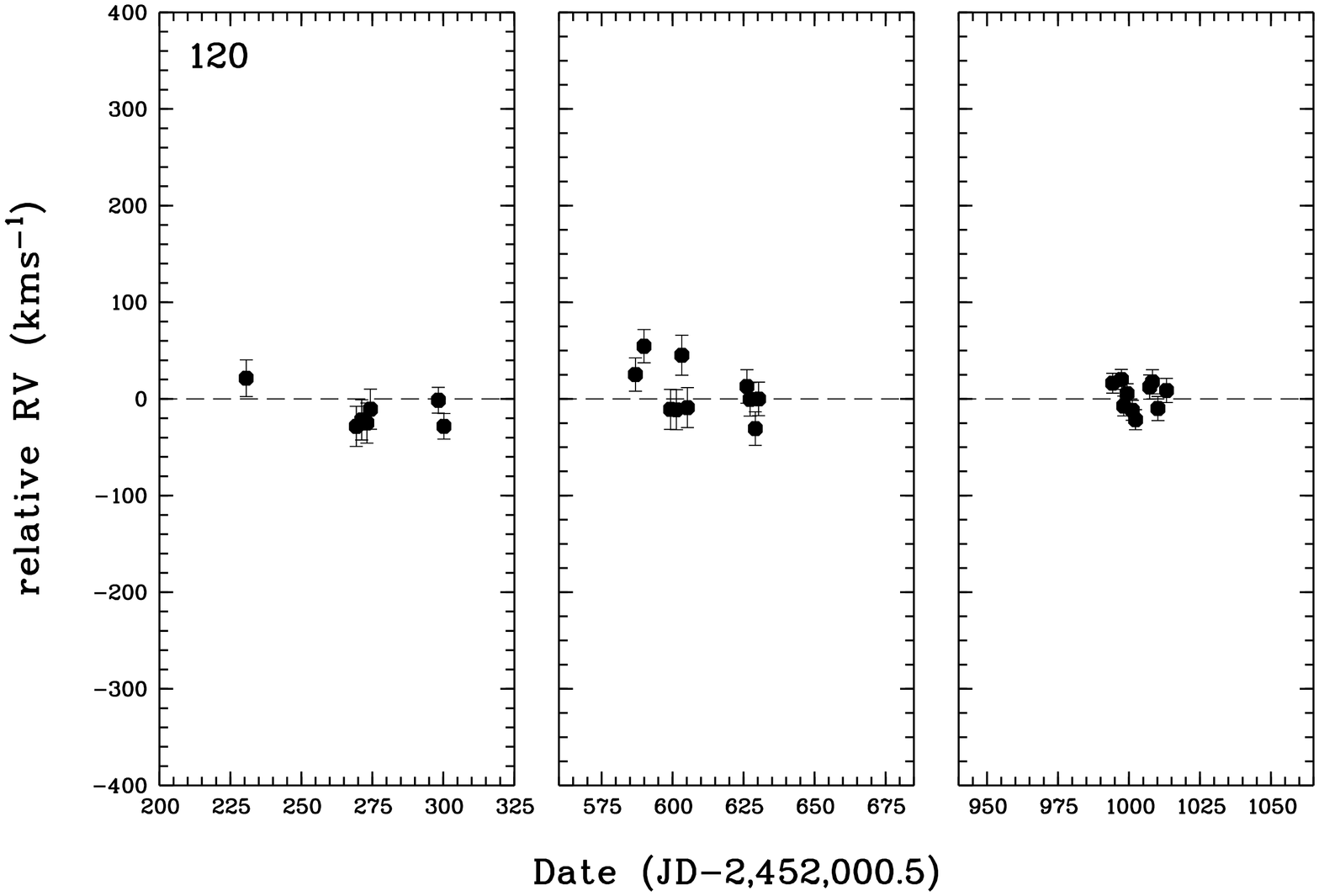}\hfill\\
\includegraphics[width=39mm,angle=-90,trim= 0 60 40 38,clip]{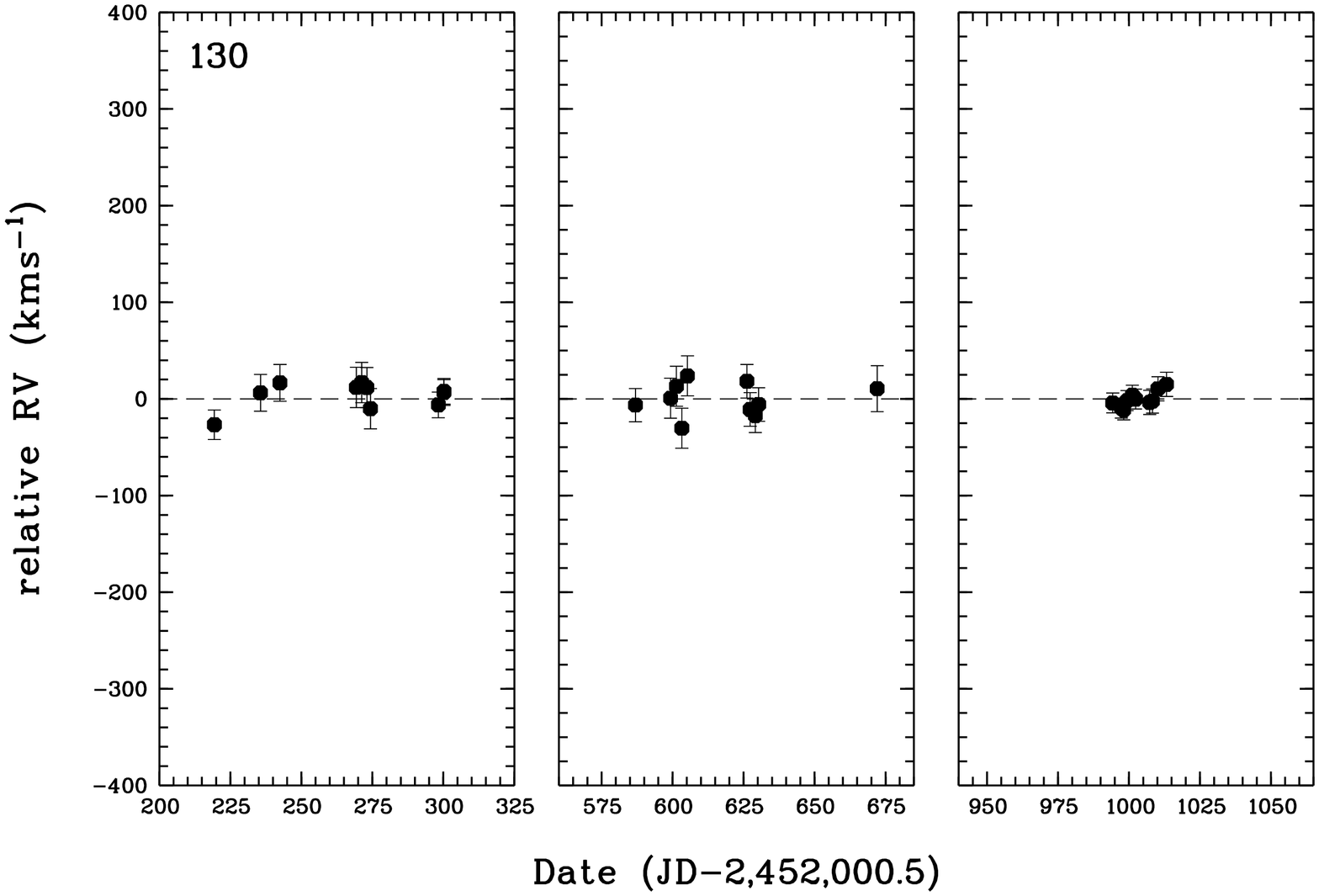}\hfill
\includegraphics[width=39mm,angle=-90,trim= 0 90 40 38,clip]{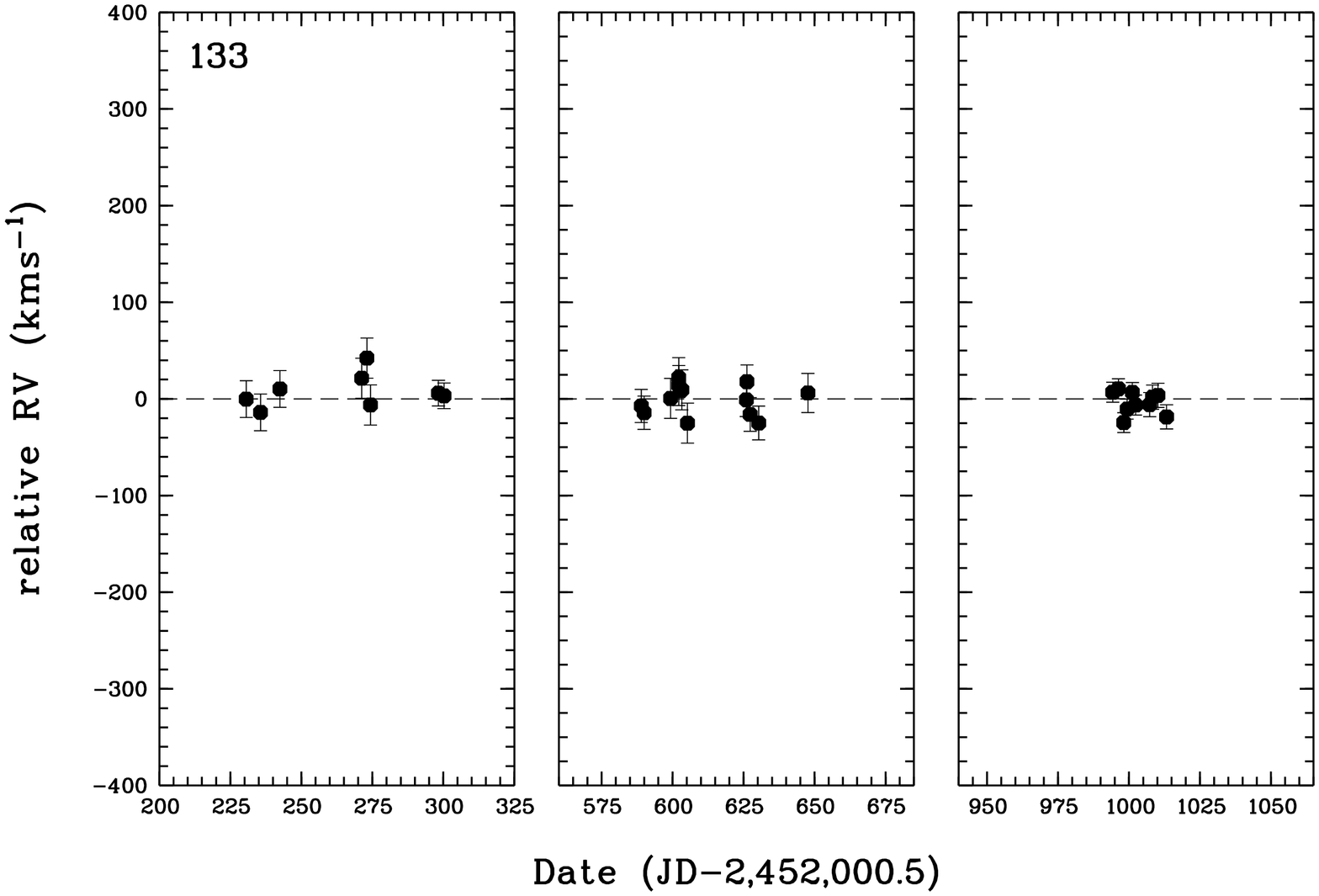}\\
\end{minipage}
\contcaption{}
\end{figure*}

\begin{table}
\caption{Systematic shifts between different runs and standard
deviations of RV per run, used as \emph{a posteriori} error bars. See
text for details.}
\label{shifts}
\centering
\begin{tabular}{llrc}
\hline
Run  & Observatory &   RV shift      & $\sigma_{j}$ \\
 $j$ &     &  kms$^{-1}$  & kms$^{-1}$\\
\hline
1  & CASLEO &  -2.9 & 15.3  \\
2  & MSO    & -16.1 & 19.0  \\
3  & CTIO   &  -3.0 & 20.9  \\
4  & CTIO   & -10.7 & 13.3  \\
5  & SAAO   &  -8.7 & 17.2  \\
6  & CTIO   &   0.2 & 20.7  \\
7  & CTIO   &  -1.9 & 17.4  \\
8  & MSO    &   3.0 & 20.2  \\
9  & SSO    & -24.9 & 10.4  \\
10 & LCO    &  -4.3 & 23.8  \\
11 & CTIO   &   0.0 & 10.3  \\
12 & CTIO   &   3.6 & 12.5  \\
13 & CTIO   &  -0.9 & 12.5  \\
\hline
\end{tabular}
\end{table}

\subsection{Random Variability and Significance Levels}
\label{randomvar}

\begin{figure}
\includegraphics[width=65mm,angle=-90,trim= 0 50 0 0]{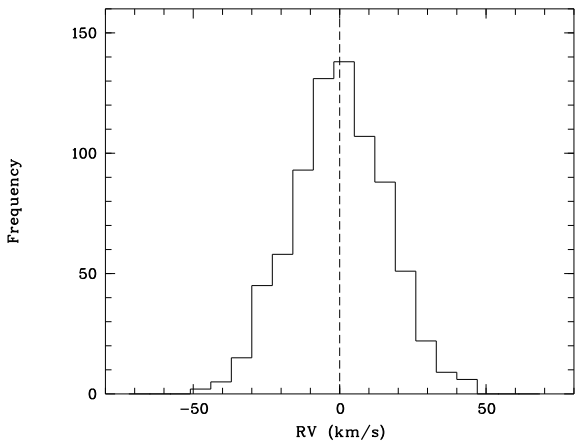}
\caption{Histogram of RV measurements of the 23 reference stars,
obtained by cross-correlation, and after correction for systematic
shifts. In total, 770 data points are used. Binsize is 7
kms$^{-1}$. The unweighted standard deviation around the mean is 16
kms$^{-1}$.}
\label{refhisto}
\end{figure}

Since we are looking for binaries, the property we are initially most
interested in is {\it variability\/}, i.e. the standard deviation
around the mean RV. After all, we expect binaries to show larger RV
scatter than constant stars. But how large a scatter is large enough
for a given star to be identified as variable? To distinguish stars
which are most likely variable from those which, within the
measurement errors, are not, we performed a $\chi^{2}$ test. The idea
is that if {\it all\/} stars in our sample were truly constant, RV
measurements would be scattered around a mean velocity only due to
random measurement errors. In Figure \ref{refhisto}, the 770 RV
measurements of the 23 combined reference stars (the ``super
reference'') are shown in histogram form. A Kolmogorov-Smirnov test
yields that this distribution is normal at a 99\% confidence level,
with $\overline{RV}= 0$ kms$^{-1}$ and $\sigma=16$ kms$^{-1}$, a nice
{\it a posteriori\/} confirmation to use those 23 stars as constant
stars.

We now define the RV amplitude $S$ be the span of RV such that 99.9\%
of the measured RVs fall within this span, i.e. $S = 3 \sigma_{\rm
RV}$. For constant stars, the distribution of the {\it squared\/} RV
amplitudes $S^{2}$ then follows a $\chi^{2}$ distribution with $df$
degrees of freedom, where $df$ is the number of RV measurements per
star (in our case, $df = 33$, because we have 33 data points per star
on average). If a star displays a value of $S^{2}$ exceeding the
99.9\% threshold, its RV scatter is not consistent with the hypothesis
that it is a constant star; hence it will be considered variable.

\begin{figure}
\includegraphics[width=63mm,angle=-90,trim= 0 40 0 0]{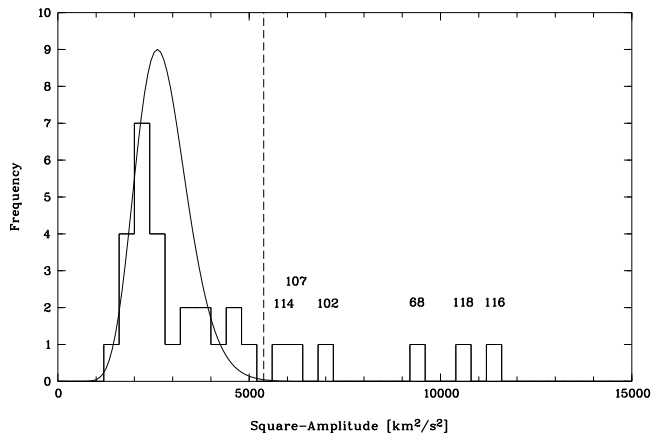}
\caption{Histogram of square-amplitudes $S^{2}$ of our program stars
with the $\chi^{2}$ distribution for $\sigma=16$ kms$^{-1}$
overplotted. Bin-size is 400 km$^{2}$s$^{-2}$. The dashed vertical
line indicates the cut-off value corresponding to $\sigma_{\rm RV}=22.6$
kms$^{-1}$; stars to the right of this cut-off are considered variable
at a 99.9\% confidence level. They are named by their respective BAT99
number. Note that ten stars in our sample have square-amplitudes
larger than 15,000 km$^{2}$s$^{-2}$ (corresponding to RV scatters
$\sigma_{\rm RV}>37.8$ kms$^{-1}$). For reasons of clarity, they are not
shown in this diagram, but listed in Table \ref{exceed}.}
\label{chi2hist}
\end{figure}

A $\chi^{2}(x)$ function with $df=33$ was computed. Values of the
normalized $x$ for $df=30$ and a 99.9\% level were taken from
statistical textbooks (e.g. \citealt{Kreyszig75}) and linearly
interpolated to $N=33$ (which is usually not tabulated); we obtained
$x=64.1$. This corresponds to the cut-off value $S^{2}=5380$
km$^{2}$s$^{-2}$ and thus $\sigma_{\rm cut}=22.6$ kms$^{-1}$. Stars
exceeding this threshold are considered variable at a 99.9\% level
(0.1\% error probability); they are listed in Table \ref{exceed}. The
observed RV square-amplitudes $S^{2}$ are shown in histogram form in
Figure \ref{chi2hist}, with the $\chi^{2}$ function overplotted after
having been adjusted for the chosen bin-size. As we shall see further
below, most, however not all, of the stars exceeding the 99.9
percentile also display periodicities in their RV curves, and indeed
are binaries.

\begin{table}
\caption{Stars whose RV standard deviation $\sigma_{\rm RV}$ exceed the
cut-off value of 22.6 kms$^{-1}$ and which therefore are considered
variable.}
\label{exceed}
\footnotesize
\centering
\begin{tabular}{rr|rr}
\hline
BAT99  & $\sigma_{\rm RV}$ & BAT99 & $\sigma_{\rm RV}$ \\
       &  [kms$^{-1}$]   &       &  [kms$^{-1}$]   \\
\hline
12     &  70.8         &  103  &  106.7 \\
32     &  92.4         &  105  &  37.7  \\
68     &  29.5         &  107  &  23.9  \\
77     &  78.2         &  113  &  93.3  \\
92     &  139.8        &  116  &  32.6  \\
95     &  81.6         &  114  &  23.2  \\
99     &  58.9         &  118  &  31.6  \\
102    &  25.3         &  119  &  44.7  \\
\hline
\end{tabular}
\end{table}
\normalsize

\subsection{Cyclical Variability and Period Analysis}

Variability, i.e. significantly large scatter, of RVs is not a
sufficient criterion for \emph{cyclical\/} variability, let alone for
binarity. There are stars which are erratically variable, i.e. any
measured RV scatter could be of stochastic nature. It is thus
important to verify whether or not any star shows periodic RV
variabilities.

\subsubsection{Period-Search Algorithms}

After having corrected all data sets for systematic shifts, we
re-performed a period search on the unweighted RVs of {\it all\/}
stars in the period range from 1 to 200 days, using a code by
\citet{Kaufer96}. In each step of the iteration, this code computes
C\textsc{lean}ed periodograms and window functions of unevenly
distributed data points according to Deeming (1975; see also
\citealt{Schwarz78}; \citealt{Roberts87}). Significance levels are
calculated for the found periods using Lomb-Scargle statistics
(\citealt{Lomb76}; \citealt{Scargle82}), i.e. by fitting sine waves to
the data points and minimizing the $\chi^{2}$. Of those peaks that
exceed the $3\sigma$ significance level, the one with the highest
power is selected, and its half width at half maximum is adopted as
1-$\sigma$ error on the period.

Our observations are fragmented into 13 runs of different duration and
slightly different typical sampling frequencies between 0.5 and 2 data
points per day and per star. This somewhat reduces aliasing problems,
although $1-\nu$ aliases remain well visible in most periodograms. For
non-circular orbits, where significant power can be contained in the
harmonics of the fundamental frequency, even more side-band peaks are
generated, since the harmonics themselves suffer from
aliasing. However, neither fitting higher-order terms of the Fourier
expansions (up to third order) nor using the analysis-of-variance
(AOV) algorithm (cf. \citealt{Schwarzenberg89}) changed the results
significantly. Thus, only results obtained with Lomb-Scargle
statistics are quoted.

\subsubsection{Stars With Periodicities}
\label{sampling}

\begin{figure*}
\begin{minipage}{165mm}
\includegraphics[width=55mm,angle=-90,trim= 10 10 20 0,clip]{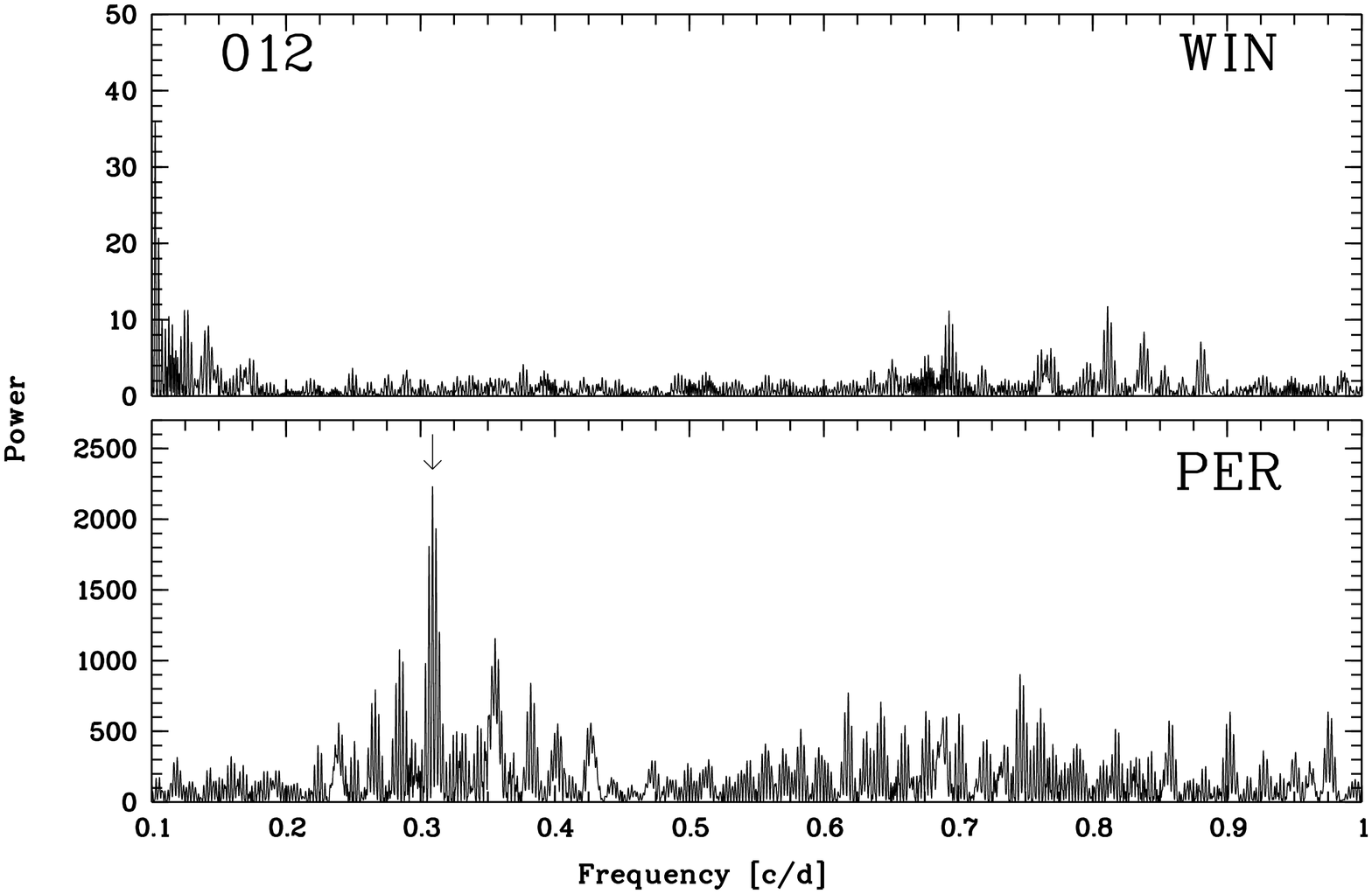}\hfill
\includegraphics[width=55mm,angle=-90,trim= 10 10 20 0,clip]{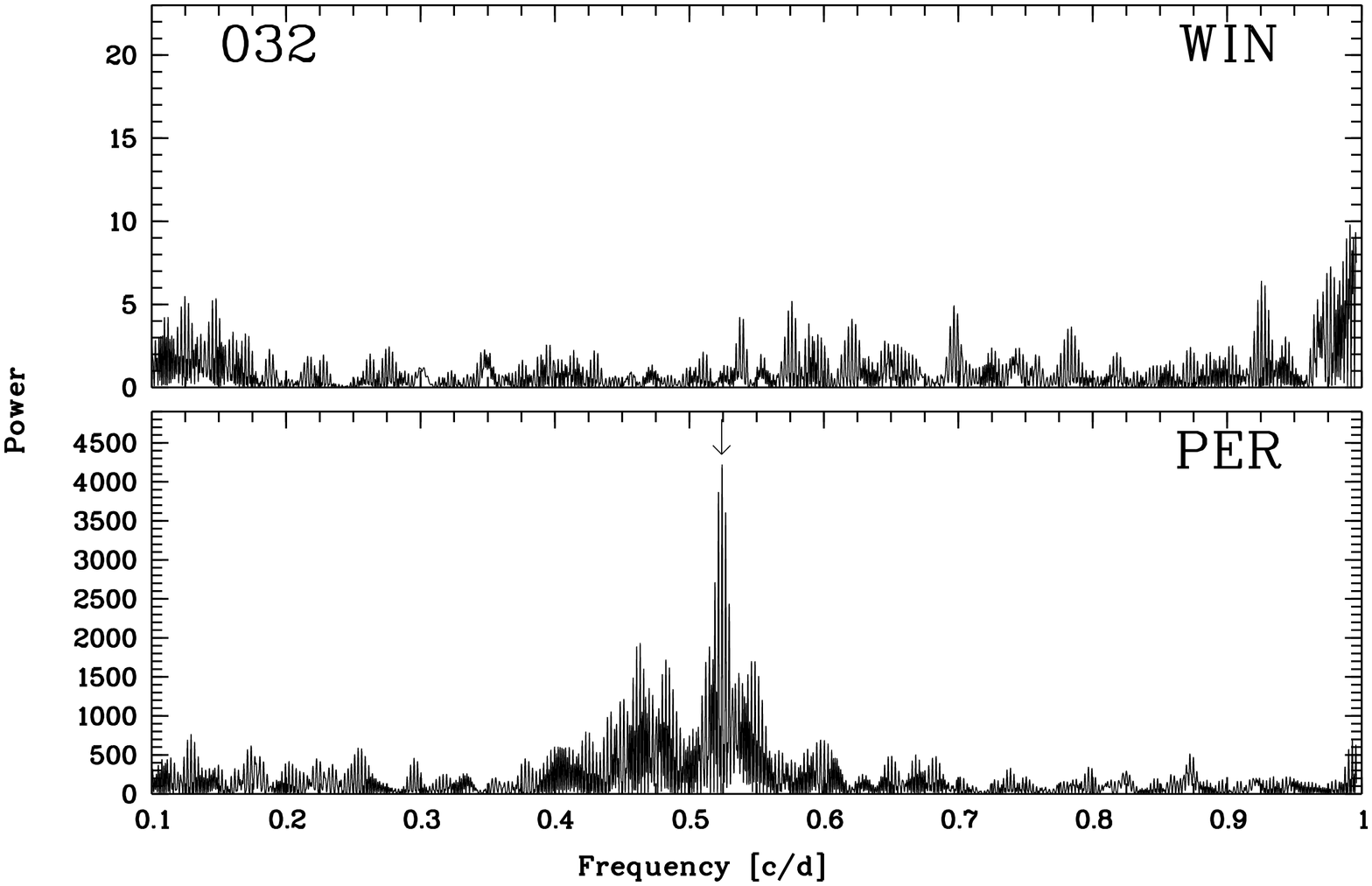}\hfill\\
\includegraphics[width=55mm,angle=-90,trim= 10 10 20 0,clip]{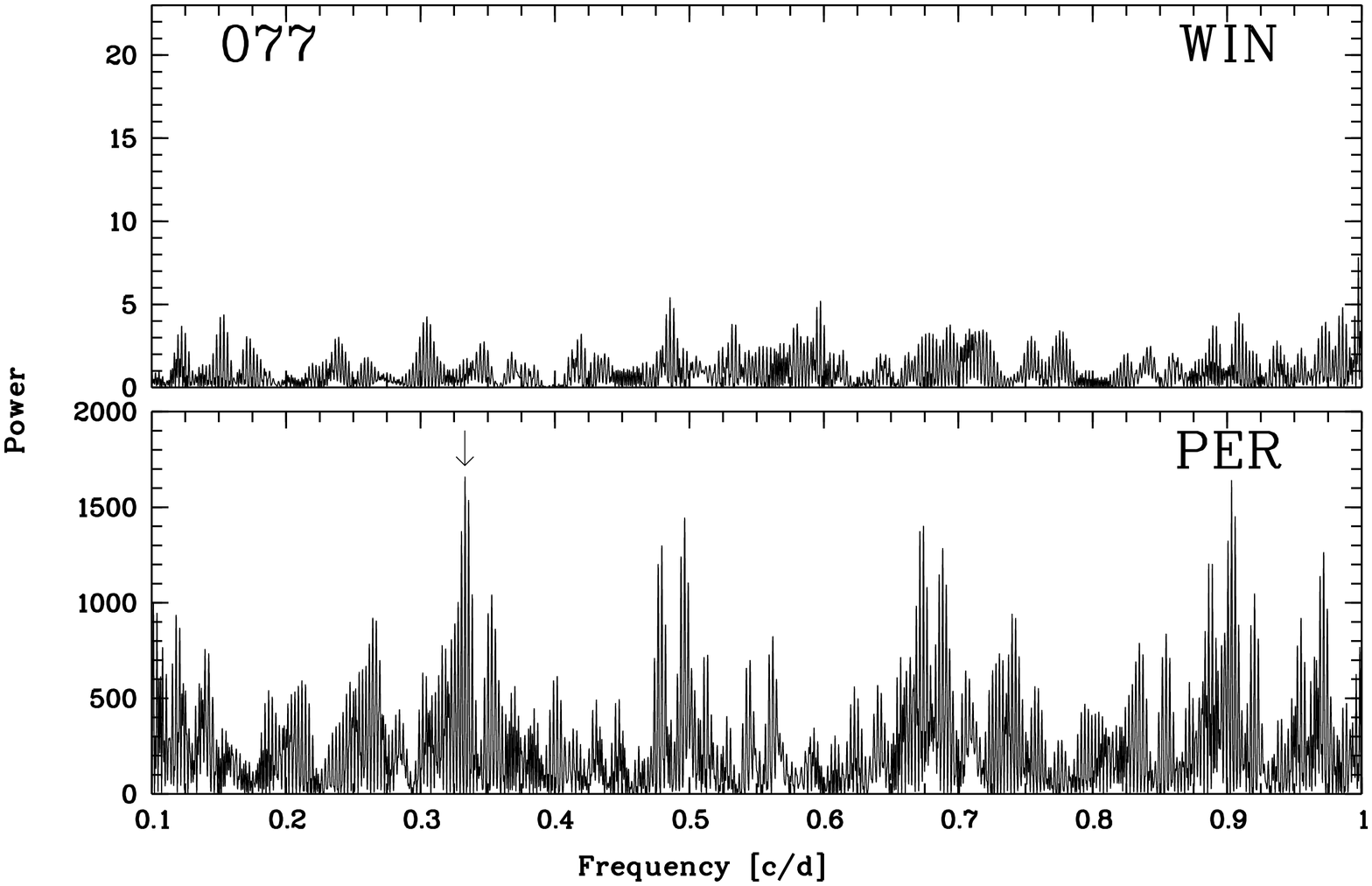}\hfill
\includegraphics[width=55mm,angle=-90,trim= 10 10 20 0,clip]{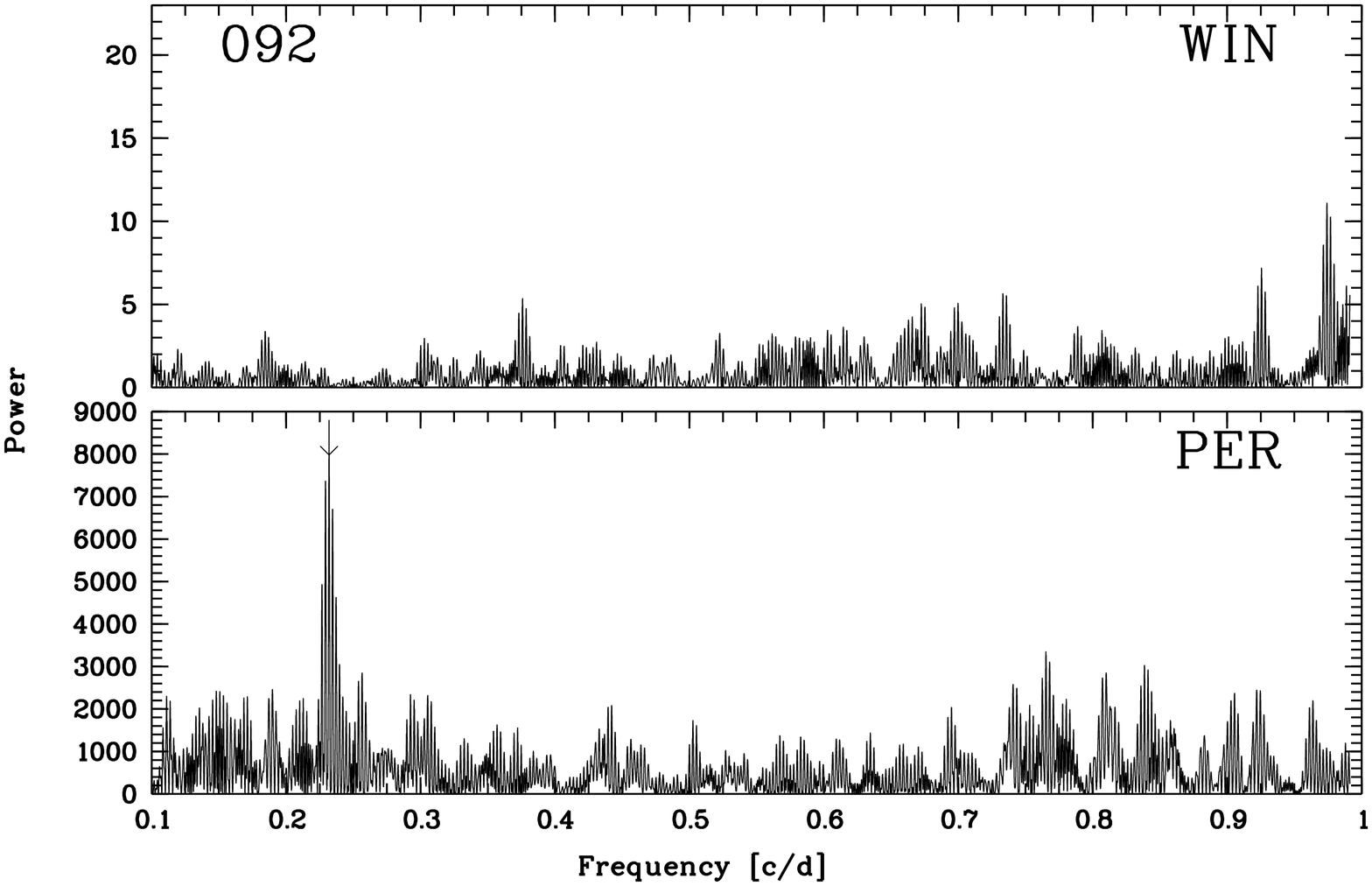}\hfill\\
\includegraphics[width=55mm,angle=-90,trim= 10 10 20 0,clip]{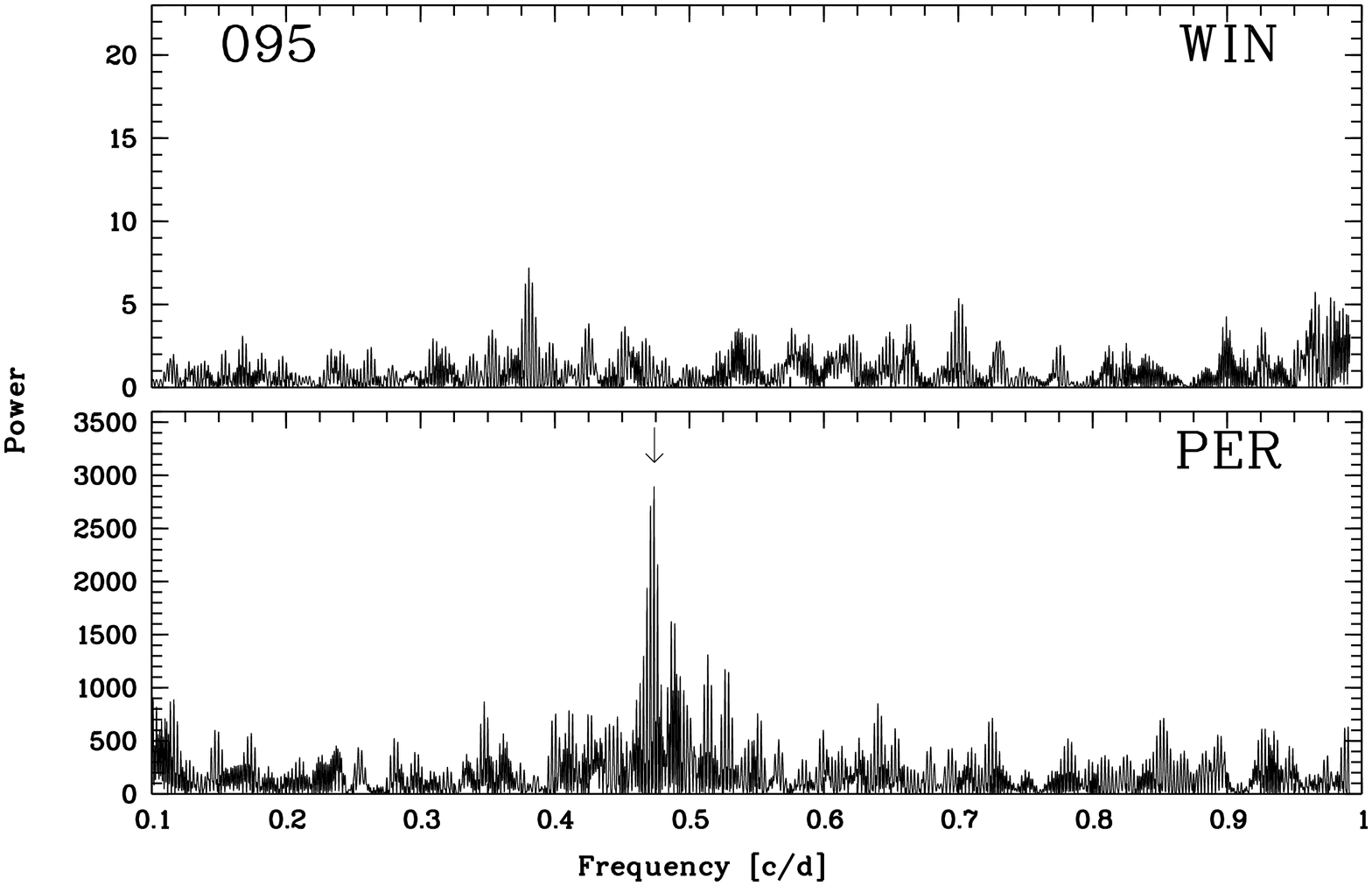}\hfill
\includegraphics[width=55mm,angle=-90,trim= 10 10 20 0,clip]{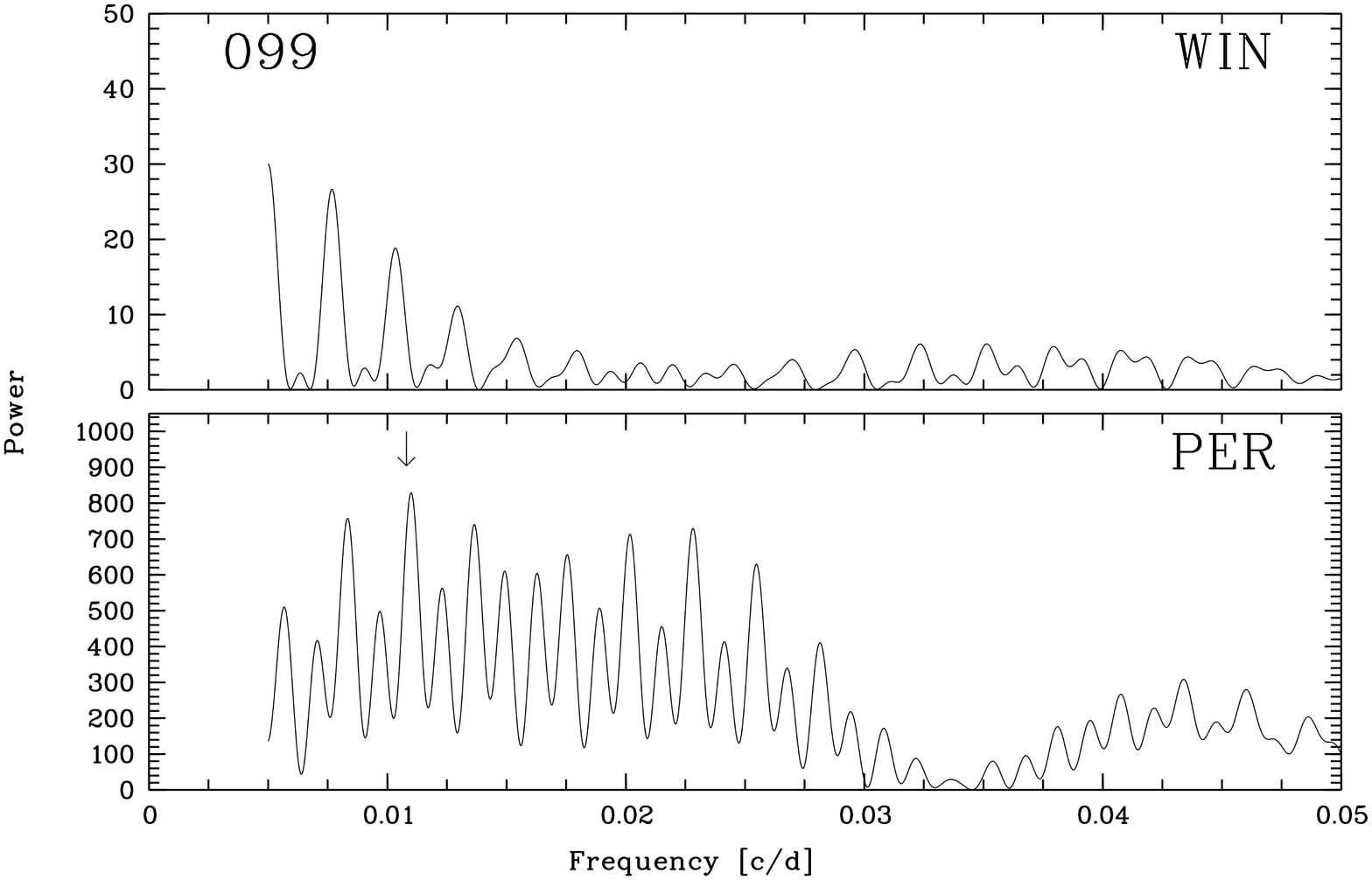}\hfill\\
\includegraphics[width=55mm,angle=-90,trim= 10 10 20 0,clip]{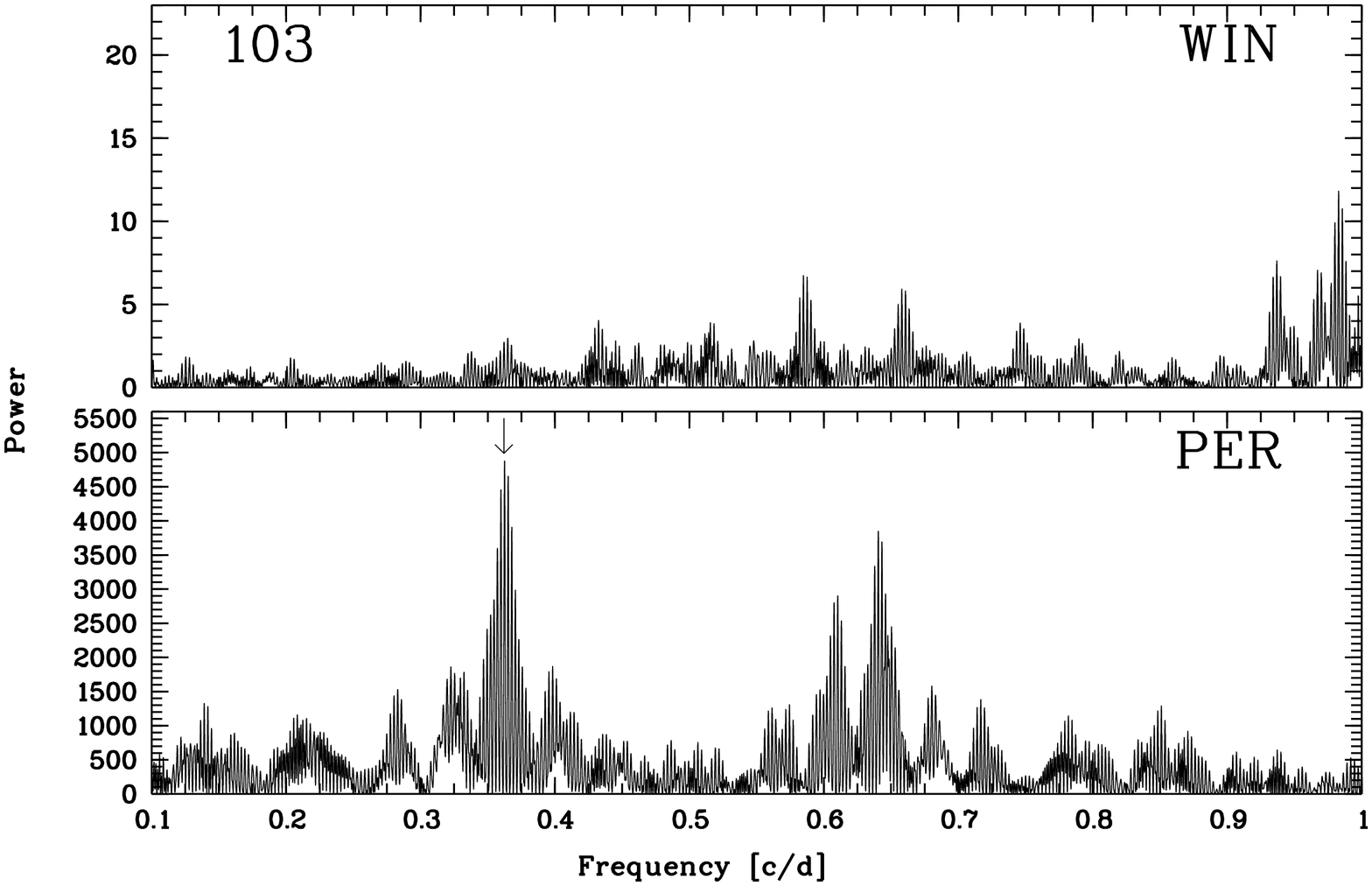}\hfill
\includegraphics[width=55mm,angle=-90,trim= 10 10 20 0,clip]{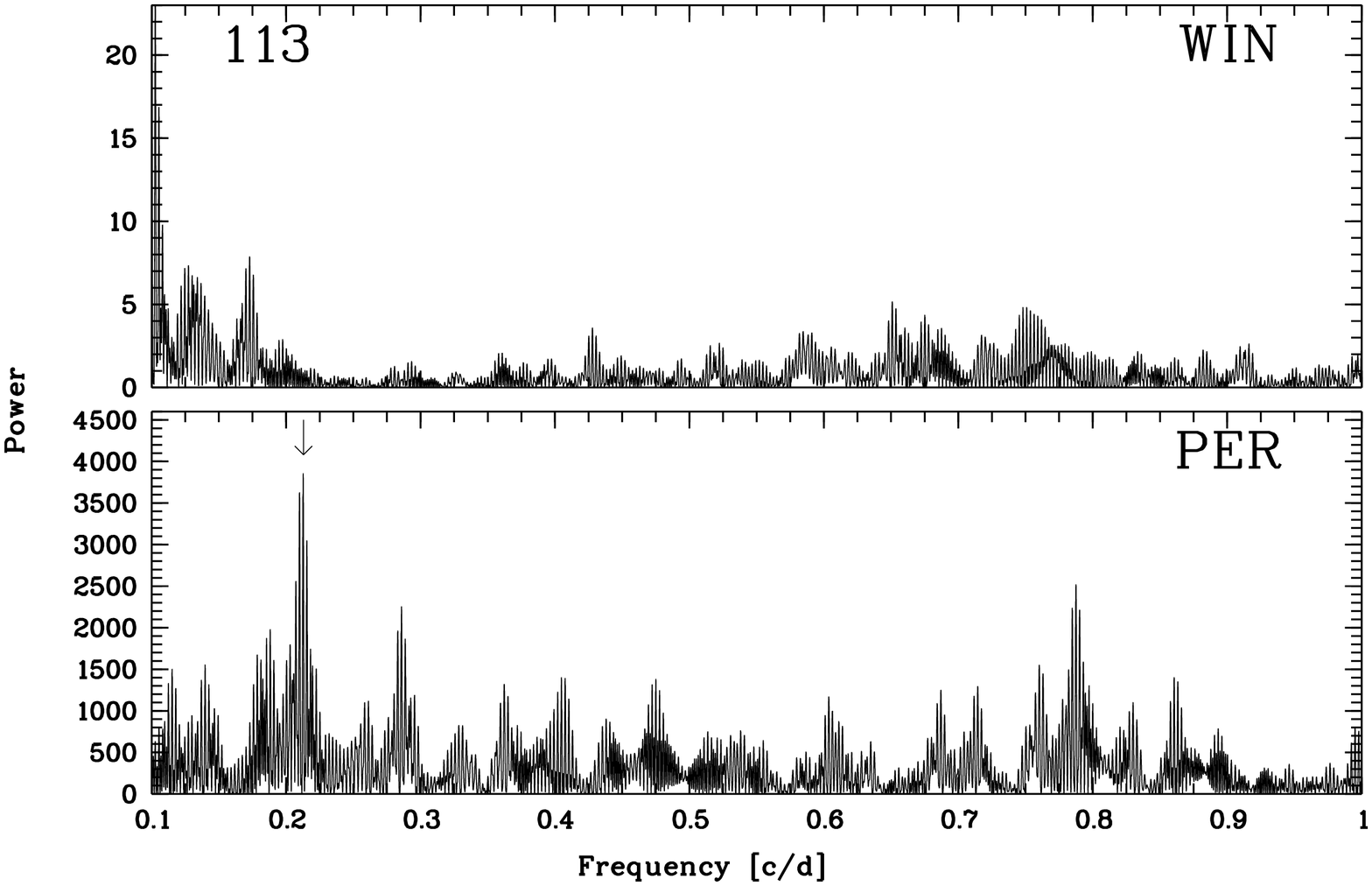}\hfill\\
\end{minipage}
\caption{Window functions (WIN, \emph{upper panels}) and periodograms
(PER, \emph{lower panels}) in the period range of 1 to 100 days (for
BAT99-99: 20 to 200 days), expressed in cycles per day (c/d). BAT
numbers are indicated. The frequency peak corresponding to the period
quoted in this study is indicated by an arrow. Note the strong
differences in $y$-scale between the window functions and the
periodograms.}
\label{periodograms}
\end{figure*}

Periodograms of identified cyclically variable stars are shown in
Figure \ref{periodograms}. Almost all binary periods reported by
previous studies (\citealt{M89}, and references therein) were
reproduced with remarkable similarity, and by combining our data with
published data, we were able to further increase the accuracy of the
periods (see Table \ref{periods}). These revised periods were used in
the further analysis. The confirmed binaries are BAT99-32, 77, and 92.

BAT99-77 is an eccentric binary system with an almost perfect 3-day
period which is why it displays a forest of significant frequency
peaks, harmonics, and aliases. This makes it difficult to determine
which period is the correct one; however, only the 3-day period yields
a coherent RV curve, and a full Keplerian fit to BAT99-77's RVs
confirms the result (see Section \ref{parameters}).

In one curious case, the 2.76-day period reported for BAT99-102
(R140a2) was clearly found, however in its {\it neighboring star\/},
BAT99-103 (R140b). Careful inspection of our logbooks did not reveal
any possible confusion at the telescope, nor is there any indication
of such a mishap in \citet{Moff87}, who first detected the
binary. BAT99-102 forms a visual pair with BAT99-101 (R140a1) which
even under the best seeing conditions at CTIO could not be
separated. Both stars lie very close to BAT99-103, and given our
relatively large slit width, some cross-contamination of emission
lines is not impossible, which could lead to a detection of the same
periodicity in BAT99-102, too. While we cannot propose a solution to
this issue, we will from here on consider BAT99-103 as the 2.76-day
binary.

\begin{table*}
\caption{Periods (in days) found for our program stars in this work
and previous studies. Newly identified, periodically variable stars
are indicated. References are: (1) \citet{MoffSegg86}; (2) \citet{M89};
(3) Moffat et al. (1987; but see text for details).}
\label{periods}
\centering
\begin{tabular}{lcccc}
\hline
BAT99 &  previous studies  &  this work  &  combined data & Ref. \\
\hline
12  & n/a                 & 3.2358 $\pm$ 0.0058 &   n/a                 & new \\
32  & $1.9075 \pm 0.0002$ & 1.9075 $\pm$ 0.0015 & 1.90756 $\pm$ 0.00012 & 1\\
77  & $3.0032 \pm 0.0002$ & 3.0034 $\pm$ 0.0042 & 3.00303 $\pm$ 0.00029 & 2\\ 
92  & $4.3092 \pm 0.0040$ & 4.311  $\pm$ 0.008  & 4.3125  $\pm$ 0.0006  & 2\\
95  & n/a                 & 2.1110 $\pm$ 0.0018 &   n/a                 & new\\
99  & n/a                 & 92.60  $\pm$ 0.31   &   n/a                 & new\\
103 & $2.7596 \pm 0.0001$ & 2.7597 $\pm$ 0.0038 & 2.75975 $\pm$ 0.00027 & 3\\
113 & n/a                 & 4.699  $\pm$ 0.010  &   n/a                 & new \\
\hline
\end{tabular}
\end{table*}

Four new stars were found to display periodic RV curves: BAT99-12, 95
99, and 113. None of these four stars was included in the study of
\citet{M89}, either because it was not listed as a WR star in previous
Catalogue versions, or because it was too faint for Moffat's
magnitude-limited sample ($V \le 13$ mag). Note, however, that
BAT99-99 is at the detection limit, as is clearly illustrated by the
periodogram in which the highest peak is not very clearly pronounced;
also, a full Keplerian fit did not converge (see Section
\ref{parameters}), so that the 92.6-day period has to be considered
very preliminary.

.


\subsubsection{Stars Without Periodicities}

For most stars with a small RV scatter below the variability
threshold, in particular the 23 reference stars, no periods at all
were found, i.e. not even spurious ones. This is also the case for
BAT99-107, for which \citet{M89} had reported a 52.7-day period and
for which no period whatsoever was detected. For some few stars,
marginally significant frequencies were identified, but the data
points yielded incoherent RV curves when folded into the corresponding
phases. Hence, we have discarded these periods as spurious.

For BAT99-119, we did not reproduce the tentative 25.4-day (assumed
circular-orbit) period reported by \citet{M89}, but it was identified
as significantly variable (see Section \ref{randomvar}). Indeed, by
combining our RV data with that of Moffat, we were able to establish
the binary nature of BAT99-119, although it was only after combining
the RV data with unpublished polarimetry that the correct period of
$P=158.8$ days and an orbital solution could be found (Schnurr et al.,
in prep.). Since the system has a long-period \emph{and} a highly
eccentric ($e\sim0.7$) orbit, we are clearly limited by the phase
coverage of our data; had it not been for the polarimetry, we would
have missed the binarity of BAT99-119. On the other hand, we can
reasonably assume that we are just able to detect (circular) binaries
with orbital period up to $\sim100$ days, given that we were able to
identify BAT99-99 as a binary as establish a tentative period of
$\sim92$ days. A detailed discussion of our detection limits is given
in Section \ref{phasefill}).

\subsection{Binaries: The Remaining Orbital Parameters}
\label{parameters}

For variable stars with identified, coherent periods, we attempted to
compute orbital parameters using the program E\textsc{lements}
(cf. \citealt{Marchenko94}). This code allows one to assign weights
$1/\sigma^{2}_{\rm RV}$ to each data point, according to its \emph{a
posteriori} measurement errors determined in Section \ref{standards}
on a per-run basis (Table \ref{shifts}). As input guess for the
orbital period, we used the value which was obtained from the
C\textsc{lean} analysis of the \emph{unweighted} data points, and when
available, we used the periods obtained from the combined data sets
(see above). Since we however did apply weights for the fit of the
orbital solution, we did not fix the period but kept it as a free
fitting parameter. In all cases, the periods found by
E\textsc{lements} were, within the errors, identical to those obtained
from our C\textsc{lean} analysis of either our or the combined data
set. Fixing the value did thus not significantly change the orbital
solution nor the errors. However, we again applied an iterative
approach. For each data point $j$, E\textsc{lements} returns the value
{\it observed minus computed\/}, $(o-c)_{j}$. After the first pass of
E\textsc{lements}, we computed the overall mean $\overline{o-c}$ and
the standard deviation from this mean, i.e. $\sigma_{\rm o-c}$. Any data
point $j$ with $|(o-c)_{j}| > 3 \sigma_{\rm o-c}$ was removed from the
data set, and the orbital solution was fitted again. This procedure
was repeated until all data points were within $3 \sigma_{\rm o-c}$ of the
fitted orbital solution. Usually, this happened after the third
iteration, which means that only few data points had to be
discarded. Naturally, the final orbital period no longer agreed with
the initial guess, because the underlying data set had been
modified. Yet, deviations remained very small.

In all cases where the algorithm converged, E\textsc{lements} returned
an elliptical orbital solution, although $e$ was small in most
cases. Therefore, we repeated the fit and forced a circular solution
by imposing $e=0$, with $\phi = 0$ defined at the time $E_{0}$ of
inferior conjunction, when the WR passes in front of its companion;
thus, $E_{0}$ is different from the time of periastron passage,
$T_{0}$, returned from the elliptical fit. Both sets of parameters are
given in Table \ref{orparamall}; parameters that were fixed for the
circular fit are indicated by the symbol ``@'' after the value.

\begin{figure*}
\begin{minipage}{130mm}
\includegraphics[width=55mm,angle=-90,trim= 0 50 0 0,clip]{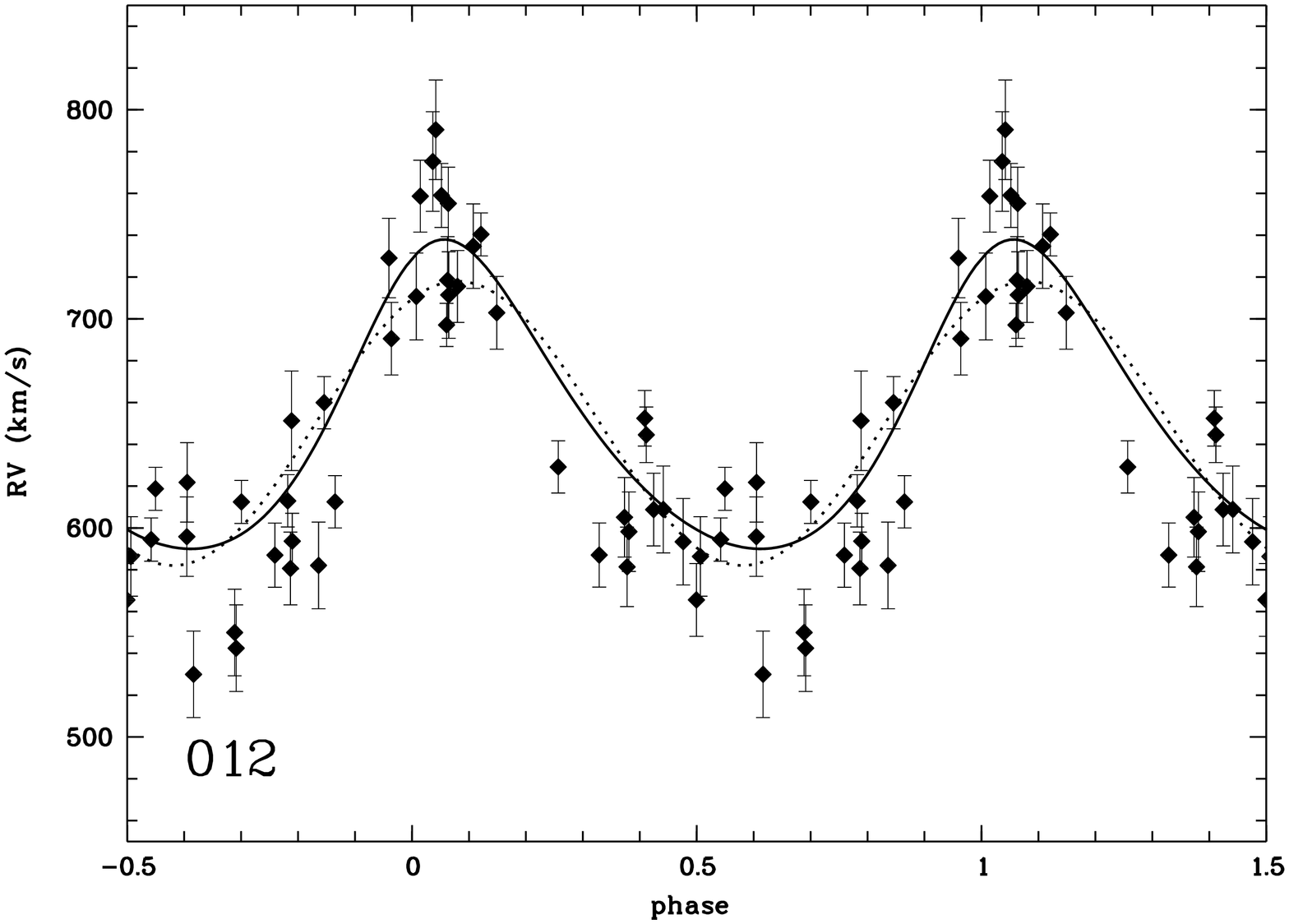}\hfill
\includegraphics[width=55mm,angle=-90,trim= 0 50 0 0,clip]{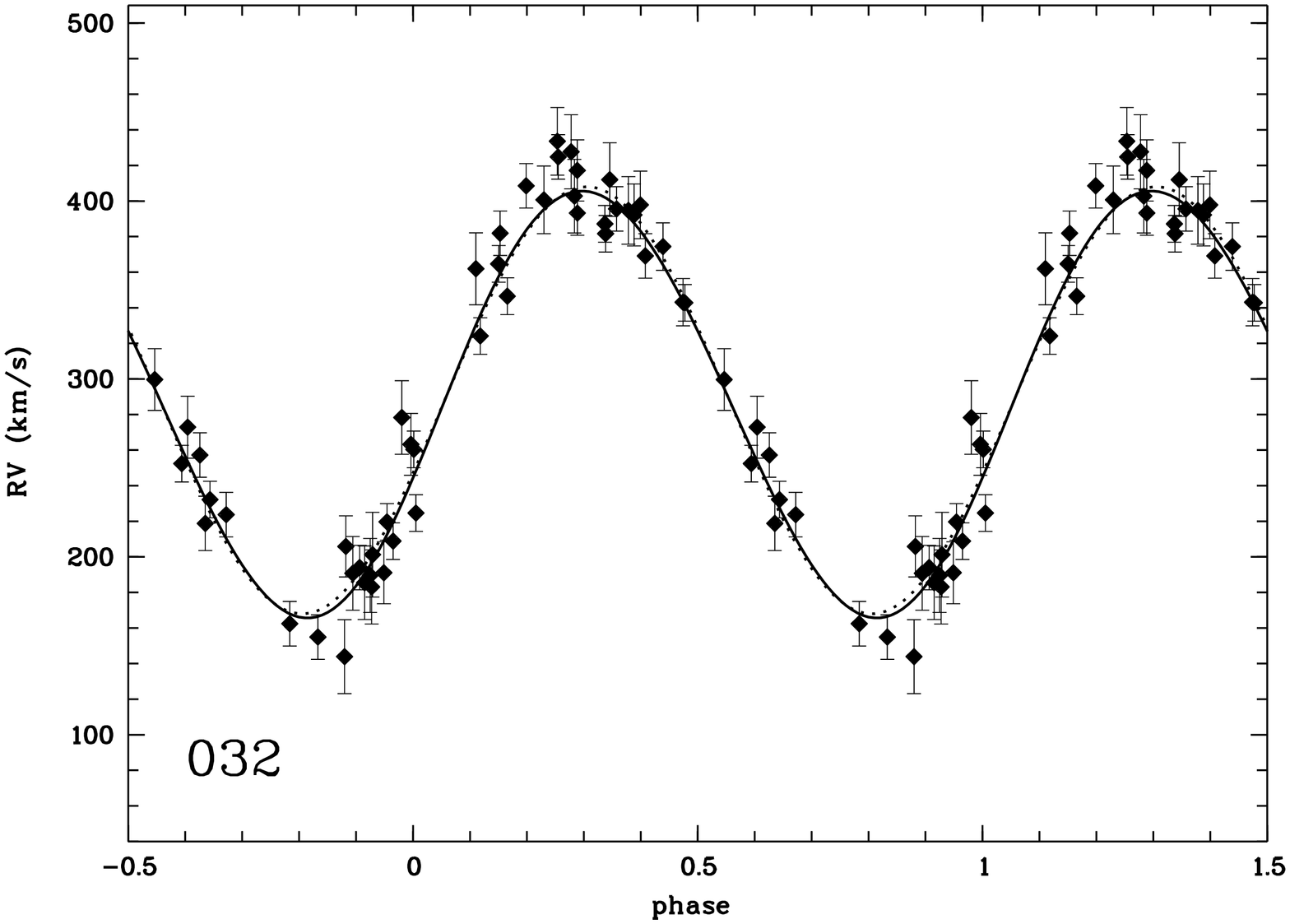}\hfill\\
\includegraphics[width=55mm,angle=-90,trim= 0 50 0 0,clip]{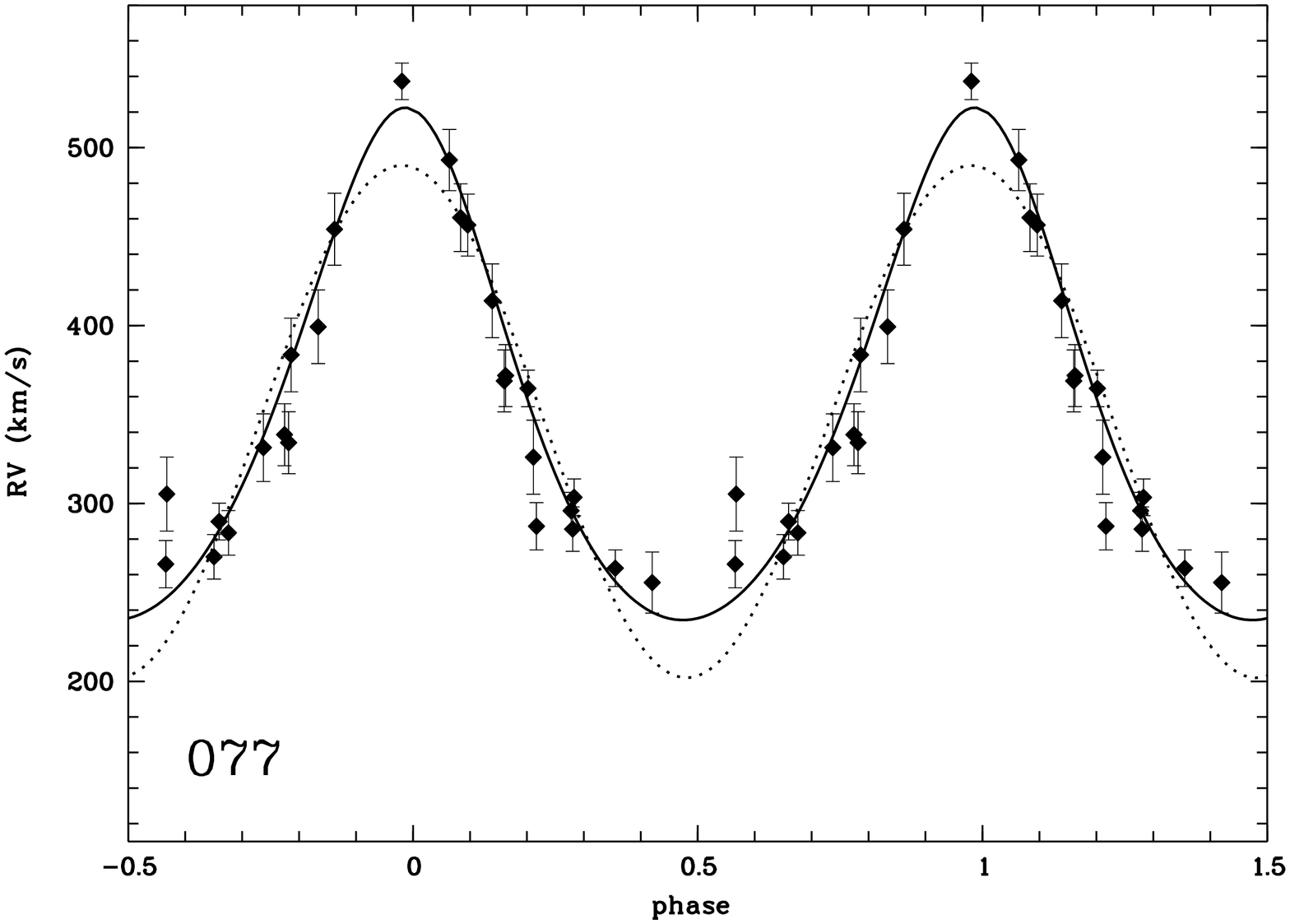}\hfill
\includegraphics[width=55mm,angle=-90,trim= 0 50 0 0,clip]{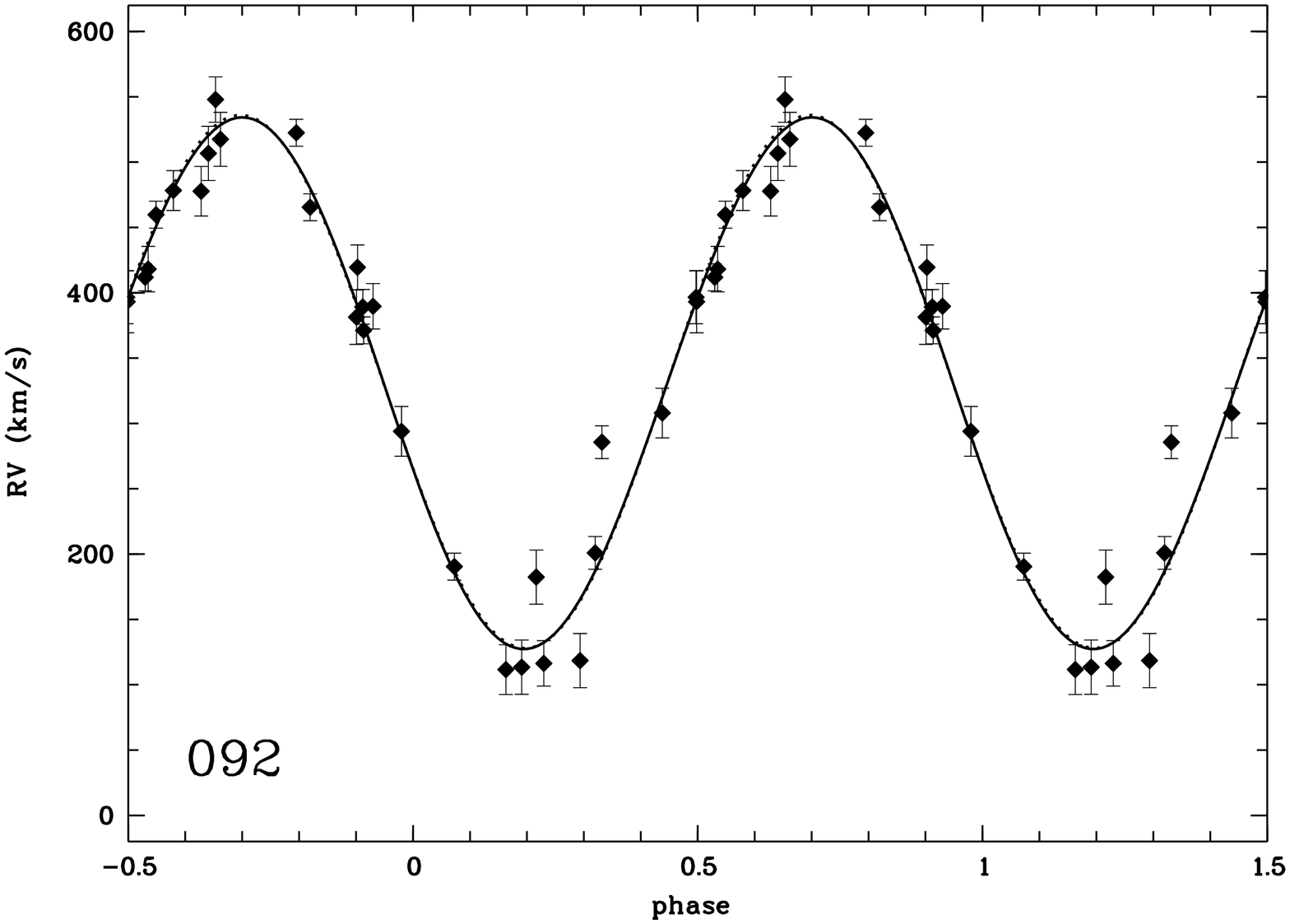}\hfill\\
\includegraphics[width=55mm,angle=-90,trim= 0 50 0 0,clip]{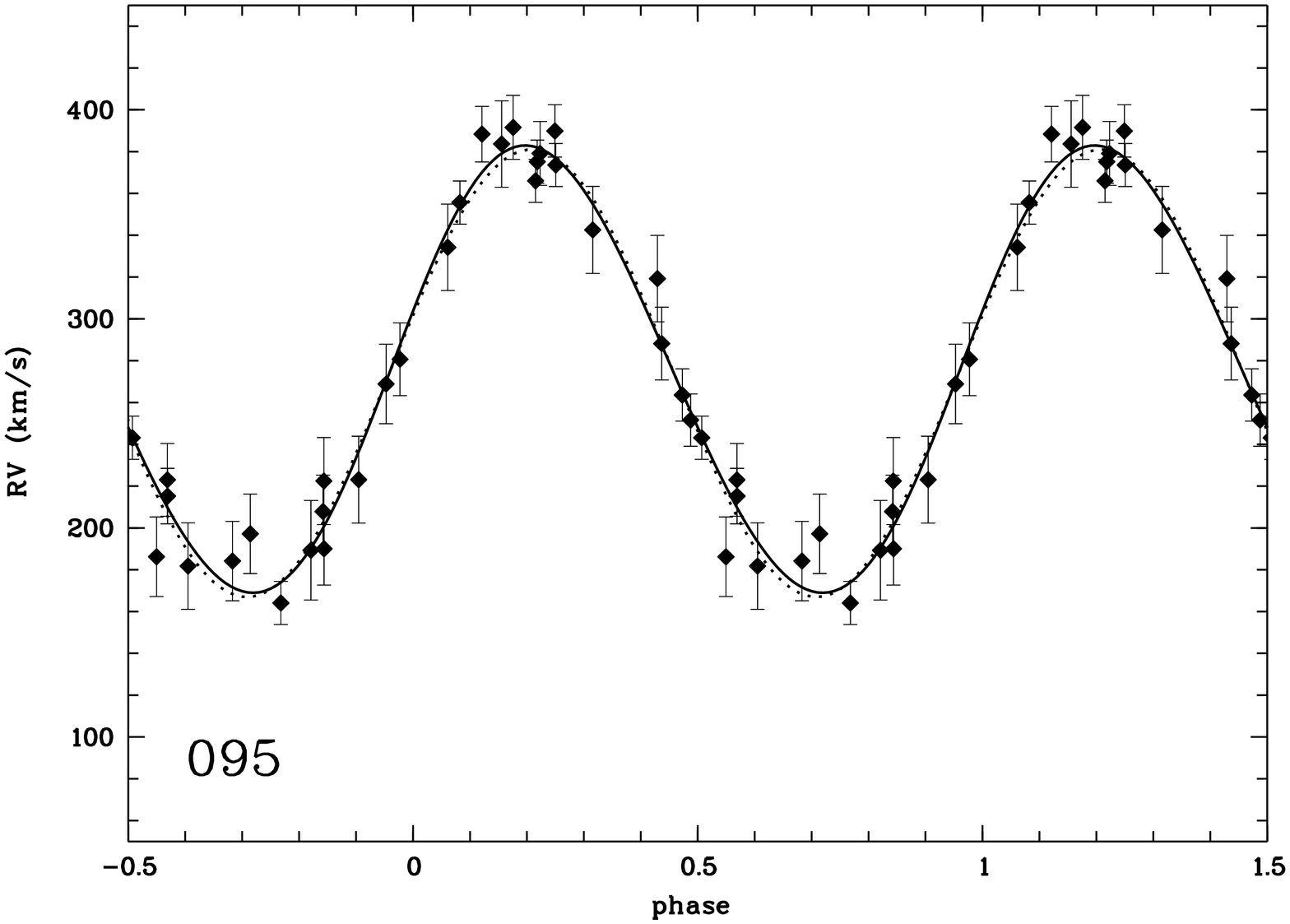}\hfill
\includegraphics[width=55mm,angle=-90,trim= 0 50 0 0,clip]{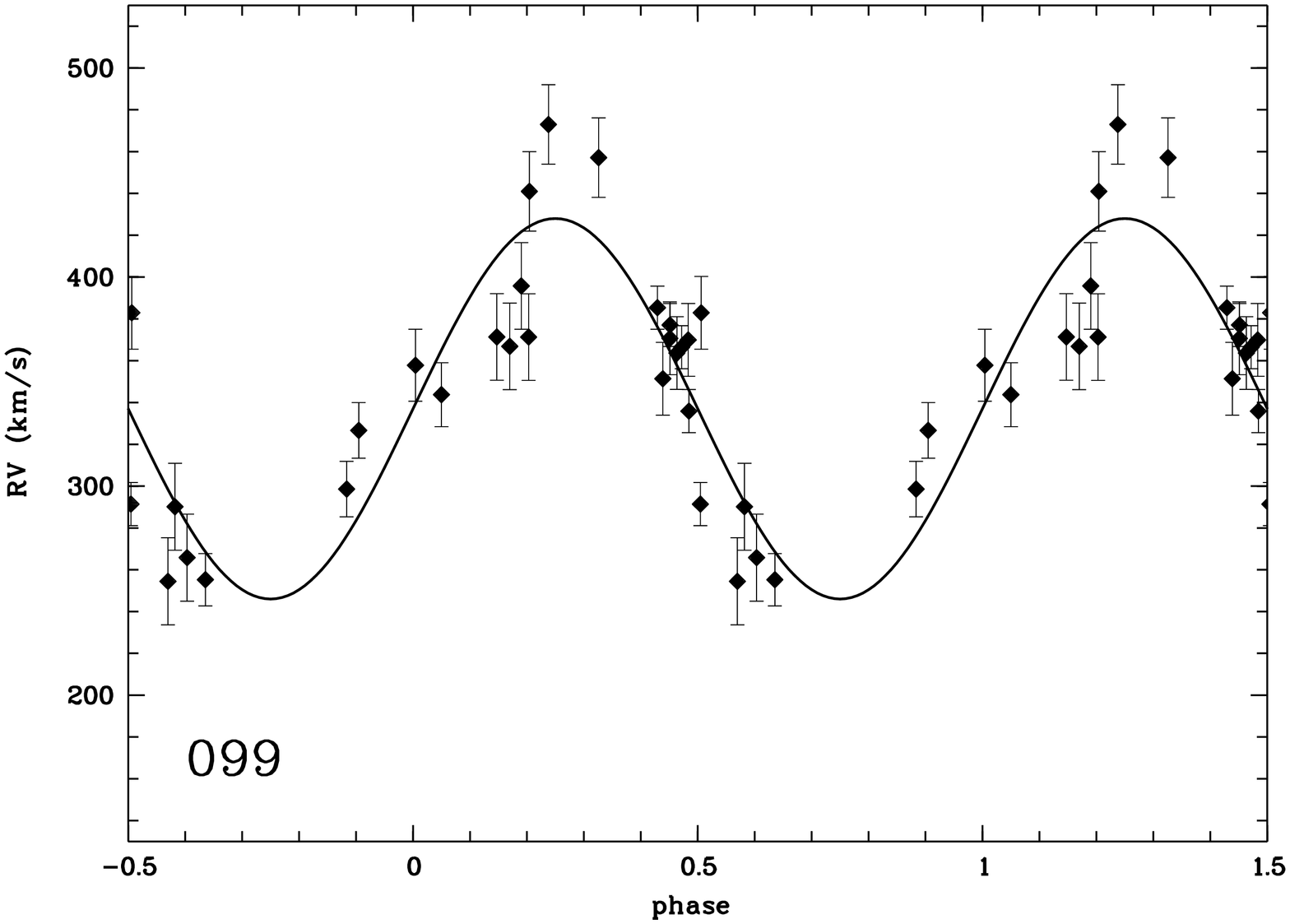}\hfill\\
\includegraphics[width=55mm,angle=-90,trim= 0 50 0 0,clip]{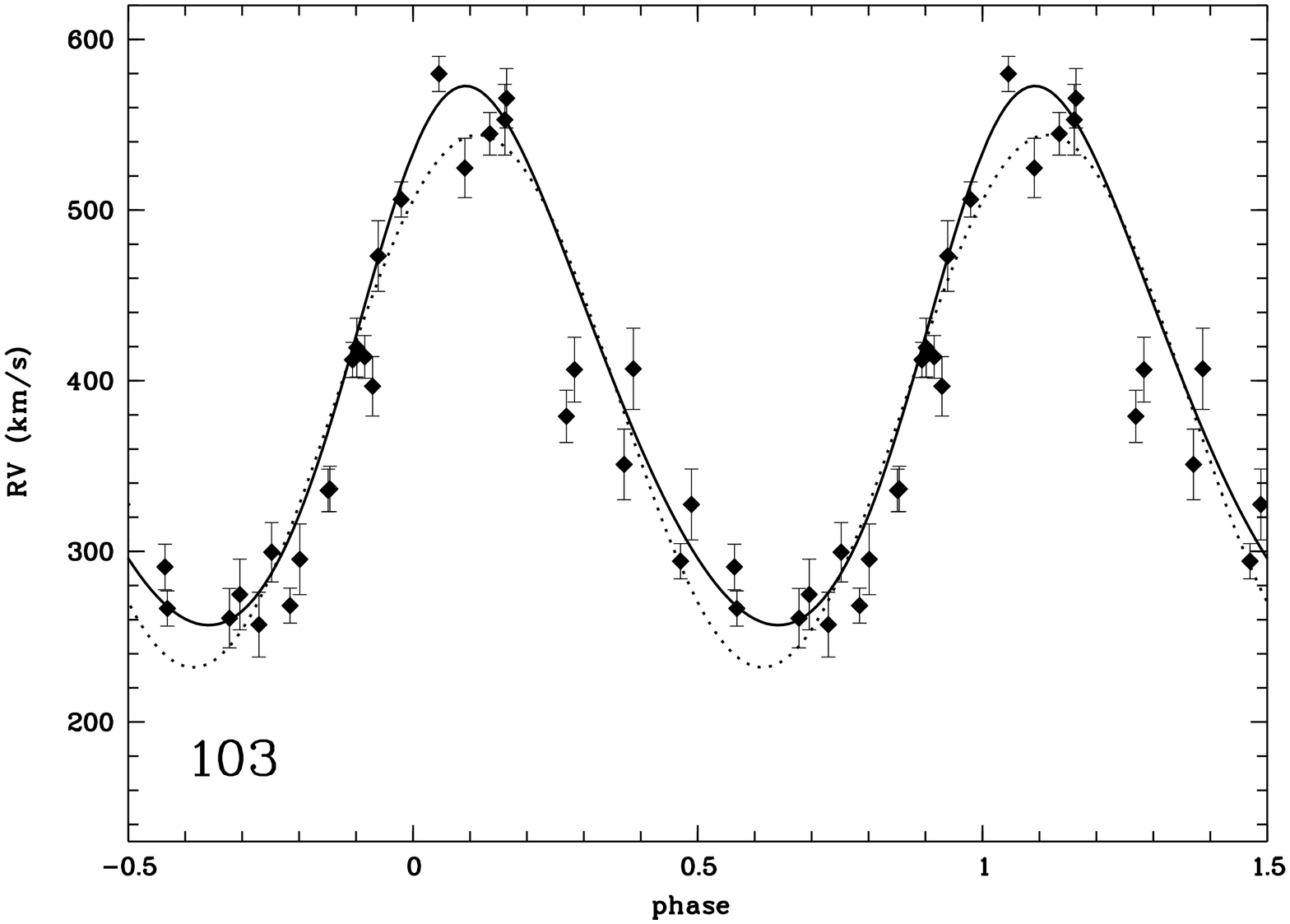}\hfill
\includegraphics[width=55mm,angle=-90,trim= 0 50 0 0,clip]{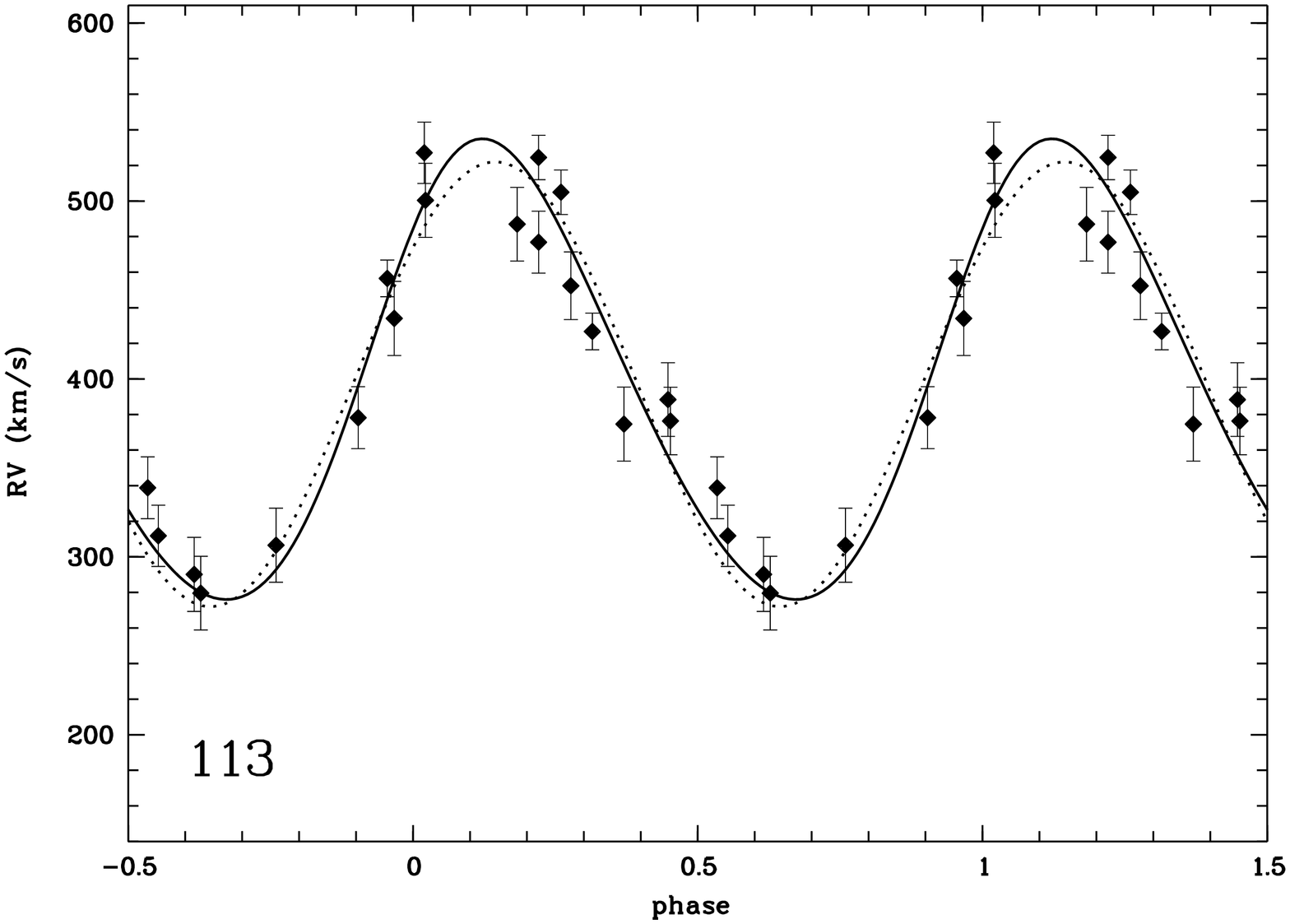}\hfill\\
\end{minipage}
\caption{Orbital solutions for the binary systems, folded into the
respective phases. Shown are both the solution for the ellipctial
(solid) and the circular (dotted) orbit.}
\label{orsolall}
\end{figure*}

In all cases but for BAT99-77, the overall quality of the fit,
expressed by $\sigma_{\rm o-c}$, did not significantly deteriorate. Thus,
for the rest of this study the circular solution was adopted for all
stars, with the exception of BAT99-77. To illustrate the differences
between the free, elliptical solution and the forced, circular one,
both solutions are shown in Figure \ref{orsolall}. Plotted are all data
points which have been retained for the fit after having applied the
iterative $\sigma$ clipping described above. For the purpose of
graphical comparison only, the circular solutions shown were computed
using $E_{0} = T_{0}$ of the given elliptical case, so that their zero
phases coincide for better clarity. Note that in the tables, the dates
for $E_{0}$ and $T_{0}$ are different.

In the following, we will discuss the binary status of individual
stars which display cyclical RV variability.

--- {\bf BAT99-12:\/} This newly identified variable shows a clear
    periodicity of 3.2358 days in its RV curve obtained from the
    He\textsc{ii} $\lambda$4686 emission lines. Although the RV curve
    is somewhat noisy when the RV points are folded into the
    corresponding phase, the orbital fit converges. The obtained
    eccentricity is surprisingly large for such a short-period
    orbit. This might be because the He\textsc{ii} $\lambda$4686
    emission is subject to severe distortions due to WWC. Forcing a
    circular fit deteriorates the overall error $\sigma_{\rm o-c}$ only
    slightly, so that a circular motion with slightly changed orbital
    parameters was adopted for the rest of the paper. The systemic
    velocity $\gamma$ obtained from the orbital fit is very large,
    $\gamma = 650 \pm 8$ kms$^{-1}$, and qualifies this system for a
    fast runaway binary (see also \citealt{Massey05}). This is very
    remarkable given the fact that such close and massive binary
    systems are very ``hard'' with respect to gravitational
    interactions with other cluster members. This object clearly
    deserves more attention; we have therefore obtained
    higher-quality, follow-up observations which will be the subject
    of a future study.

\begin{table*}
\centering
\caption{Orbital parameters for the WR component of our binary
systems. Results are given for both the free, elliptical fit and the
forced, circular solution. For the latter, $\omega$ is not defined
(n/d). Depending on the eccentricity of the orbit, zero-phase
$\phi_{0}$ is either the time of periastron passage, $T_{0}$, or the
time of inferior conjunction (i.e., WR star in front), $E_{0}$.}
\label{orparamall}
\begin{tabular}{r r@{}@{ }p{3.5mm}@{}l c r@{}@{ }p{3.5mm}@{}r r@{}@{ }p{3.5mm}@{}r r@{}@{ }p{3.5mm}@{}r r@{}@{ }p{3.5mm}@{}r  c }
\hline
BAT99 &         & $P$ &              &   $e$      &     & $K$ &                       && $\omega$ &                        &  &$\gamma$&                        &     $\phi_{0}$& &      &  $\sigma_{\rm o-c}$  \\
      &  \multicolumn{3}{c}{(days)}  &            &  \multicolumn{3}{c}{(kms$^{-1}$)} & \multicolumn{3}{c}{($^{\circ}$)}   & \multicolumn{3}{c}{(kms$^{-1}$)}   &\multicolumn{3}{c}{(HJD-2,400,000.5)}    &  [km/s]\\
\hline
12    & 3.2358  &$\pm$& 0.0058  & $0.34 \pm 0.06$ &  74 &$\pm$& 5                     & -29 &$\pm$& 11                     & 642 &$\pm$& 13                     & 52269.84 &$\pm$& 0.09 & 24.8 \\
      &         &     &         & 0               &  68 &$\pm$& 8                     &     & n/d &                        & 650 &$\pm$& 8                      & 52272.58 &$\pm$& 0.10 & 27.1 \\[1mm]
32    & 1.90756 &$\pm$& 0.00012 & $0.06 \pm 0.02$ & 120 &$\pm$& 3                     & 250 &$\pm$& 22                     & 288 &$\pm$& 6                      & 53011.57 &$\pm$& 0.12 & 13.5 \\
      &         &     &         & 0               & 120 &$\pm$& 3                     &     &  n/d&                        & 288 &$\pm$& 6                      & 53011.68 &$\pm$& 0.12 & 14.2 \\[1mm]
77    & 3.00303 &$\pm$& 0.00029 & $0.32 \pm 0.02$ & 144 &$\pm$& 4                     &   7 &$\pm$& 4                      & 333 &$\pm$& 8                      & 52631.87 &$\pm$& 0.04 & 10.9 \\
      &         &     &         & 0               & 144 &$\pm$& 15                    &     & n/d &                        & 346 &$\pm$& 10                     & 52637.05 &$\pm$& 0.06 & 16.9 \\[1mm]
92    & 4.3125  &$\pm$& 0.0006  & $0.02 \pm 0.02$ & 204 &$\pm$& 5                     & 109 &$\pm$& 66                     & 332 &$\pm$& 7                      & 52998.03 &$\pm$& 0.04 & 16.8 \\
      &         &     &         & 0               & 204 &$\pm$& 5                     &     & n/d &                        & 332 &$\pm$& 7                      & 52999.96 &$\pm$& 0.04 & 17.0 \\[1mm]
95    & 2.1110  &$\pm$& 0.0018  & $0.07 \pm 0.03$ & 107 &$\pm$& 3                     & 285 &$\pm$& 18                     & 274 &$\pm$& 9                      & 52999.87 &$\pm$& 0.10 & 10.6 \\
      &         &     &         & 0               & 107 &$\pm$& 3                     &     & n/d &                        & 274 &$\pm$& 9                      & 52999.78 &$\pm$& 0.09 & 10.9 \\[1mm]
99    &92.60    &$\pm$& 0.31    & 0               &  91 &$\pm$& 19                    &     & n/d &                        & 337 &$\pm$& 16                     & 53045.90 &$\pm$& 1.30 & 30.5 \\[1mm]
103   & 2.75975 &$\pm$& 0.00027 & $0.23 \pm 0.03$ & 158 &$\pm$& 4                     & -41 &$\pm$& 7                      & 388 &$\pm$& 8                      & 53007.80 &$\pm$& 0.05 & 19.9 \\
      &         &     &         & 0               & 156 &$\pm$& 10                    &     & n/d &                        & 388 &$\pm$& 8                      & 53010.22 &$\pm$& 0.05 & 23.8 \\[1mm]
113   & 4.699   &$\pm$& 0.010   & $0.20 \pm 0.05$ & 130 &$\pm$& 8                     & 308 &$\pm$& 16                     & 390 &$\pm$& 10                     & 52993.07 &$\pm$& 0.13 & 18.1 \\
      &         &     &         & 0               & 125 &$\pm$& 15                    &     & n/d &                        & 397 &$\pm$& 11                     & 52992.53 &$\pm$& 0.14 & 20.5 \\
\hline
\end{tabular}
\end{table*}

--- {\bf BAT99-32:\/} This star is a known binary with an orbital
    period of 1.9076 days, the shortest known among the WN binaries in
    the LMC; yet it does not display the largest RV amplitude (see
    BAT99-92). This might indicate that the system is seen under a low
    inclination angle. Within the errors, the system has a circular
    orbit, as might be expected for such short systems.

--- {\bf BAT99-77:\/} This star is a known binary with an almost
    integer-day period which makes it hard to determine. However, the
    orbital fit confirms our period, and yields quite a large
    eccentricity for such a short system. Consequently, the overall
    error increases significantly when a circular solution is
    imposed. Presently, we do not have enough data to confirm whether
    this is due to WWC-induced distortions of the He\textsc{ii}
    $\lambda$4686 emission or whether the orbit is indeed
    non-circular, but for the rest of this paper, we adopt the
    elliptical solution.

--- {\bf BAT99-92:\/} This star is a known binary. The orbital fit
    yields a remarkably small eccentricity, so that the circular
    solution is adopted. This system shows the largest RV amplitude of
    our program stars, although it does not have the shortest
    period. This might indicate that it is seen under a large
    inclination angle, which renders BAT99-92 potentially interesting
    for a photometric campaign to obtain the inclination angle from a
    light curve.

--- {\bf BAT99-95:\/} This is a newly discovered binary in the 30 Dor
    region. Despite its short orbital period, it has a remarkably
    small RV amplitude. A circular solution is adopted for this
    system.

--- {\bf BAT99-99:\/} This is a newly discovered binary system located
    in the periphery of the R136 cluster in the center of the 30 Dor
    region. \citet{Massey05} has reported this star to be a binary,
    but not given an orbital period. From our analysis, we find a
    rather long period of 92.6 days, close to our detection limits
    (see above). Unfortunately, folding the data points into the
    corresponding phase yields a rather noisy RV curve. A free
    (elliptic) fit does not converge, so that we have forced a
    circular solution; however, $\sigma_{\rm o-c}$ is uncomfortably large,
    probably because the true orbit is elliptical indeed. We thus
    report a tentative set of orbital parameters for BAT99-99. Errors
    on individual parameters are not stated, rather we quote the
    $\sigma_{\rm o-c}$ of the total solution.

--- {\bf BAT99-103:\/} This star is a known binary system, located in
    the periphery of the R136 cluster. Again, we are unable to verify
    whether or not the non-zero, albeit mild eccentricity is real, but
    the circular fit is only moderately worse. Since the RV curve is
    not very clean, both WWC-induced distortions and the weakness of
    the He\textsc{ii} $\lambda$4686 emission line are the most likely
    causes for this result.

--- {\bf BAT99-113:\/} This is a newly discovered binary system in the
    periphery of R136. Regarding the non-zero eccentricity, the same
    remarks apply as for the previous system.

\subsection{Systemic Velocities and Runaway Stars}
\label{runaway}

\begin{figure}
\includegraphics[width=65mm,angle=-90,trim= 0 50 0 0,clip]{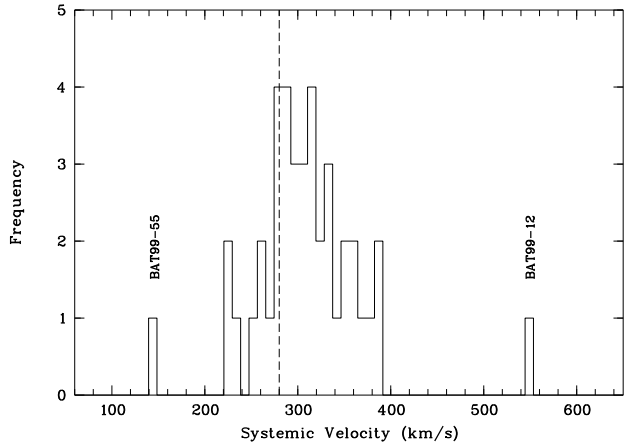}
\caption{Histogram of mean systemic velocities of our program stars as
measured by Gauss fits to the He\textsc{ii} $\lambda$4686 emission
line. The dashed vertical line indicates the expected systemic
velocity of the LMC, 280 kms$^{-1}$. Note the two outliers, BAT99-12
and BAT99-55; these are most likely runaway stars. See text for more
details.}
\label{sysvel_hist}
\end{figure}

For single program star, both mean $\overline{RV}_{\rm sys}$ and
standard deviations were computed from RVs obtained through Gauss fits
to the He\textsc{ii} $\lambda$4686 emission line (see Section
\ref{rvmeasures}). For binary stars, the systemic velocity $\gamma$ and
its error returned from the orbital fit were used. Resulting
$\overline{RV}_{\rm sys}$ are listed in Table \ref{sysveltable_all}
and shown in histogram form in Figure \ref{sysvel_hist}.

\begin{table}
\centering
\footnotesize
\caption{Systemic velocities and RV scatter for our program stars.}
\label{sysveltable_all}
\begin{tabular}{llrcc}
\hline
BAT99& $\overline{RV_{\rm sys}}$ & $\sigma_{\rm RV}$ & Binary? & remarks\\
     &  [kms$^{-1}$]   & [kms$^{-1}$]      &         &\\ 
\hline
 12 &    650  &   70.8 &   yes     & runaway \\
 13 &    277  &   12.5 &   ...     & ...\\
 16 &    330  &   13.3 &   ...     & ...\\
 22 &    255  &   13.3 &   ...     & ...\\
 30 &    345  &   17.3 &   ...     & ...\\
 32 &    288  &   92.4 &   yes     & ...\\
 33 &    293  &   16.2 &   ...     & ...\\
 44 &    398  &   15.8 &   ...     & ...\\
 45 &    245  &   11.4 &   ...     & ...\\
 54 &    316  &   18.2 &   ...     & ...\\
 55 &    140  &   14.6 &   ...     & runaway? \\
 58 &    311  &   14.7 &   ...     & ...\\
 68 &    321  &   29.5 &   ...     & ...\\
 76 &    274  &   15.5 &   ...     & ...\\
 77 &    333  &   78.2 &   yes     & ...\\
 79 &    288  &   14.3 &   ...     & ...\\
 80 &    374  &   20.5 &   ...     & ...\\
 83 &    ...  &   14.9 &   ...     & ...\\
 89 &    303  &   14.3 &   ...     & ...\\
 91 &    324  &   13.7 &   ...     & ...\\
 92 &    332  &  139.8 &   yes     & ...\\
 93 &    373  &   20.5 &   ...     & ...\\
 95 &    274  &   81.6 &   yes     & ...\\
 96 &    274  &   18.7 &   ...     & ...\\
 97 &    305  &   19.6 &   ...     & ...\\
 98 &    321  &   14.7 &   ...     & ...\\
 99 &    337  &   58.9 &   yes     & ...\\
100 &    307  &   15.4 &   ...     & ...\\
102 &    317  &   25.3 &   ?       & ...\\
103 &    388  &  106.7 &   yes     & ...\\
104 &    342  &   18.4 &   ...     & ...\\
105 &    273  &   37.7 &   ...     & ...\\
107 &    303  &   23.9 &   ...     & ...\\
113 &    397  &   93.3 &   yes     & ...\\
114 &    393  &   23.2 &   ...     & ...\\
116 &    373  &   32.6 &   ...     & ...\\
118 &    301  &   31.6 &   ?       & ...\\
119 &    332  &   44.7 &   yes     & ...\\
120 &    282  &   21.8 &   ...     & ...\\
130 &    287  &   13.2 &   ...     & ...\\
133 &    280  &   15.4 &   ...     & ...\\
\hline
\end{tabular}
\end{table}

Averaged over our complete sample, we find $314 \pm 10$ kms$^{-1}$,
somewhat more than the systemic velocity of the LMC, which is $280 \pm
20$ kms$^{-1}$ (e.g. \citealt{Kim98}), but consistent with the value
\citet{Foell03b} reported for their WNE sample stars, $324 \pm 6$
kms$^{-1}$. However, we re-analyzed Foellmi et al.'s data and found
$345 \pm 6$ kms$^{-1}$, which is then incompatible with our results
for WNL stars. The origin of this systematic difference is not
entirely clear, but most likely stems from the way the measurement
were carried out; we used line-fitting, while Foellmi et al. used
bisectors on selected parts of the He\textsc{ii} $\lambda$4686
emission line. For consistency, we will use our values in the
following discussion. In Table \ref{sysvel_table}, the systemic
velocities of the different subsamples are shown; all errors quoted
are the error of the mean ($eom = \sigma_{\rm RV}/\sqrt{N}$).

The measured redshift does not come unexpectedly; it is well known
among observers that some emission lines of WR stars yield redshifted
systemic velocities. Some authors have attributed this phenomenon to a
prominent electron-scattering wing on the red flank of the emission
line (e.g. \citealt{AuBlerk72}), others to the presence of a
blue-shifted P Cygni absorption, diminishing the blue flank of the
emission profile (e.g. \citealt{Bartz01}). \citet{Hillier89}, on the
other hand, explains the observed redshift by radiative-transfer
effects in optically thick lines\footnote{We are indebted to the
anonymous referee for pointing this out to us.}.

\begin{figure}
\includegraphics[width=62mm,angle=-90,trim= 0 15 0 0,clip]{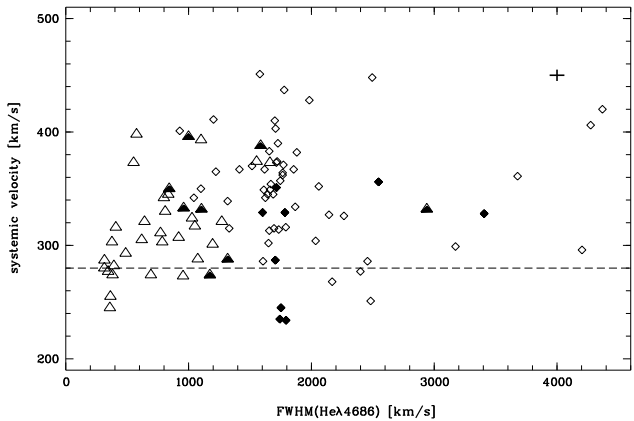}
\caption{Mean systemic velocities plotted against the FWHM (in
velocity units) of He\textsc{ii} $\lambda$4686 emission line for the
WN stars in the LMC. Triangles denote the 41 WNL stars, lozenges
denote the 61 WNE stars. Filled symbols are identified binaries. The
dashed line indicates the systemic velocity of the LMC ($\sim$280
kms$^{-1}$). For clarity, error crosses have been omitted, but a
typical error cross ($\pm eom = \pm \sigma/\sqrt{N}$) is shown in the
upper right-hand corner. Most stars have larger systemic velocities
than the LMC, and the star-to-star scatter is significant. The two
runaway candidates, BAT99-12 and 55, have been omitted for this graph
so that the $y$ axis is not unduly compressed. See text for more
details.}
\label{sysvel_vfwhm}
\end{figure}

Both Foellmi et al. (2003b, their Figure 8) and Bartzakos et
al. (2001, their Figure 8) reported a negative correlation between the
line width (FWHM) and the measured systemic velocities, showing that
broader lines better reflected the true systemic (i.e. LMC)
velocity. However, after combining the data sets for the WNL and WNE
stars, we found that to the contrary, there is a a \emph{positive}
correlation between the linewidth and the measured systemic
velocities; the combined data are shown in Figure \ref{sysvel_vfwhm}.
Furthermore, we have binned stars to certain FWHM ranges, and
calculated the bin mean systemic velocity and its error. The results
are listed in Table \ref{groupsysvel}. Indeed it seems that
\emph{only} the most narrow-lined WNL stars (FWHM $\lse 500$
kms$^{-1}$) do correctly reflect the systemic velocity of the LMC,
whereas broader-lined stars, i.e. those with optically thicker winds,
\emph{on average} yield redder systemic velocities, in line with
Hillier's (1989) explanation. Note that contrary to what
\citet{Foell03b} reported, the measured systemic velocities for stars
with FWHMs broader than 500 kms$^{-1}$ remain essentially constant at
$\sim330$ kms$^{-1}$, rather than to converge again towards the true
LMC value.

Two interesting cases, BAT99-55, a constant star, and BAT99-12, a
binary, are marked in Figure \ref{sysvel_hist} with their names. These
stars have systemic velocities which are quite different from the
systemic velocity of the LMC, 146 kms$^{-1}$ (BAT99-55) and 650
kms$^{-1}$ (BAT99-12; circular orbit assumed), respectively. Thus,
these stars are excellent runaway candidates (for a definition of
runaway stars, see \citealt{Blaauw61}). Moreover, \citet{Massey05}
report a systemic velocity of 430 kms$^{-1}$ for BAT99-12, obtained
from absorption lines in their better-resolved spectra, confirming its
high systemic velocity; our much larger velocity is most likely an
artefact due to the mentioned P Cygni absorption in the blue flank of
the He\textsc{ii} $\lambda$4686 emission line. The runaway nature is
another property that BAT99-12 shares with its Galactic counterpart,
the O4Inf star $\zeta$ Pup.

\begin{table}
\centering
\caption{Average systemic velocities for different WN samples in the
LMC, as observed by this study (WNL, including the O3f/WN6 stars) and
by Foellmi et al. (2003b; WNE). Values are in kms$^{-1}$.}
\label{sysvel_table}
\begin{tabular}{ll}
\hline
Subsample & $\overline{RV}_{\rm sys}$\\
\hline
WNL, all                   &  314 $\pm$ 10 \\
WNL, no binaries, no runways  &  305 $\pm$ 7  \\
WNE, all       &  345 $\pm$ 6  \\
WNE, no binaries, no runways  &  353 $\pm$ 6  \\
WN, all                           &  334 $\pm$ 5  \\
WN, no binaries, no runaways  &  338 $\pm$ 5  \\
\hline
LMC                  &  280 $\pm$ 20 \\
\hline
\end{tabular}
\end{table}
\normalsize

\begin{table*}
\small
\centering
\caption{Mean systemic velocities of stars binned into different FWHM
ranges. $N$ is the number of stars per FWHM bin, $\overline{RV}$
the mean systemic velocity, $\sigma$ the RV scatter, and $eom =
\sigma/\sqrt{N}$ the error of the mean systemic velocity in the
respective bin.}
\label{groupsysvel}
\begin{tabular}{l|c|c|c|c|c}
\hline
FWHM [kms$^{-1}$]     & 0-500 & 500-1000 & 1000-1500 & 1500-2000 & 2000+\\
$N$                   & 10 & 16 & 16 & 40 & 17 \\
$\overline{RV}$ [kms$^{-1}$] & 281 & 335 & 333 & 335 & 332 \\
$\sigma$ [kms$^{-1}$] & 21 & 41 & 38 & 49 & 55\\
$eom$ [kms$^{-1}$]    & 6.6 & 10.3 & 9.5 & 7.7 & 13.3 \\
\hline
\end{tabular}
\end{table*}
\normalsize


\subsection{Mean Spectra and Spectral Types}
\label{reclass}

Spectra from individual observatories were concatenated to obtain an
average, high-S/N spectrum. To verify the spectral types listed in the
BAT99 catalogue, the classification criteria of Smith et al. (1996;
hereafter SSM96) were applied. In most cases, the change in spectral
type was only very minor; results are listed in Table
\ref{spectable}. The most important change is BAT99-92, which is also
a confirmed binary. This star is listed as WN6
(cf. \citealt{MoffSegg86}), but closer inspection of the spectrum
reveals that BAT99-92 is hotter. However, its true spectral type is
very difficult to determine because of the dominating B supergiant
spectrum. We tentatively assign a WN3:b(+O)+B1Ia type to this system.

We have also measured both the full-width at half maximum (FWHM) and
the equivalent width (EW) of the He\textsc{ii} $\lambda$4686 emission
line (see Table \ref{spectable}), with their standard deviations of
the time series $\sigma_{\rm EW}$ as errors. Both for crowded stars,
whose dilution depends on seeing or slit-positioning, and for
weaker-lined stars such as O3If/WN6 stars, $\sigma_{\rm EW}$ can be
rather large. FWHMs, however, remain unaffected by dilution effects,
with measurement errors of the order of 10 to 15\% in all cases. Of
course, some stars might be intrinsically variable, but an in-depth
study of the line-variability phenomenon and its underlying causes is
a very complex task and beyond the scope of the present paper. Results
of such a study are planned for a future paper.

Many of our program stars still contain residual hydrogen, and some
are even expected to be still in the CHB phase. The main indicator for
the presence of H in the WR spectrum is the alternating He\textsc{ii}
Pickering decrement. If not stated otherwise, nebular lines did not
hamper the spectral classification. In the 30 Dor region, however, the
alternating Pickering decrement of helium lines suffers considerably
from nebular pollution, and the determination of the H content remains
(very) uncertain; such subtypes feature a ``:'' (or even ``::'') to
indicate this uncertainty. Note also that the unsubtracted nebular
emission leads to partially filled-up absorptions. Together with our
relatively low spectral resolution, which affects narrower absorption
lines more than broader emission lines, some O3If/WN6 stars, whose
very weak He\textsc{ii} $\lambda$4686 emission lines (small EWs)
clearly distinguishes them from genuine WN stars
(\citealt{Walborn86}), display an artificially attenuated
absorption-line spectrum, thus favoring a (diluted) WN- rather than an
O-star classification. Indeed, Massey et al. (2004, 2005) proposed the
O2If classification for some of the O3If/WN6 stars using
higher-resolution data and applying the criteria of
\citet{Walborn02}. Nevertheless, two hot slash-stars were reclassified
into WN types, BAT99-80 (now WN5ha:) and BAT99-105 (now WN7), based on
the relatively large EW and the presence of a He\textsc{ii}
$\lambda$5411 emission. Their spectrum resembles that of genuine, yet
diluted WN stars. In turn, BAT99-68 is listed as WN5-6 in the BAT99
catalogue, although it was classified as Of by SSM96. This star is
located in a cluster, hence its spectrum is severely diluted. However,
because of both the weak EW and the absence of a He\textsc{ii}
$\lambda$5411 emission line, we re-classify this star as Of/WN7, the
WN part of the spectral type being based on the emission-line
spectrum.

Regarding the Ofpe/WN9 or ``cool slash-stars'', we did not re-classify
them, since most of them have already been re-classified into WN10-11
by \citet{CroSmi97}, and we maintain this classification. The
exception is BAT99-33 (R99), which has already been reported to show a
peculiar spectrum that is inconsistent with a WR classification (see
\citealt{CroSmi97} for more details). We thus maintain the Ofpe/WN in
our Table \ref{spectable}, but add a ``?''. In the case of BAT99-107,
which is listed as WNL/Of, we change the spectral type to WN9h::a.

A montage of the (re-)classified, mean spectra of each of our program
stars are displayed in Figure \ref{spectypes}. Note that for stars in
the 30 Dor region, nebular lines have been clipped; thus, in some
spectra, the H$\beta$ and H$\gamma$ lines are truncated.

\begin{table}
\small
\caption{Spectral types, and equivalent width (EW) and full-width at
half maximum (FWHM) of the He\textsc{ii} $\lambda$4686 emission line
of our program stars. Spectral types that have been changed by us are
also given.}
\label{spectable}
\begin{tabular}{lllrr}
\hline
BAT & Spec. Type & Spec. Type     &   EW            &    FWHM        \\
 99 & (BAT99)    & (this work)    &  [\AA]          &   [\AA]        \\ 
\hline
 12   & O3If*/WN6   &             & $-12.3 \pm 1.6$ & $17.9 \pm 1.9$ \\
 13   & WN10        &             & $ -2.7 \pm 0.3$ & $ 5.4 \pm 0.7$ \\
 16   & WN8h        & WN7h        & $-40.2 \pm 1.3$ & $12.7 \pm 0.3$ \\
 22   & WN9h        &             & $ -4.6 \pm 0.5$ & $ 5.7 \pm 0.7$ \\
 30   & WN6h        &             & $-50.0 \pm 1.5$ & $13.1 \pm 0.3$ \\
 32   & WN6(h)      &             & $-74.0 \pm 3.8$ & $20.6 \pm 2.0$ \\
 33   & Ofpe/WN9    & Ofpe/WN9?   & $ -2.8 \pm 0.7$ & $ 7.6 \pm 0.7$ \\
 44   & WN8h        & WN8ha       & $-26.8 \pm 1.0$ & $ 9.0 \pm 0.3$ \\
 45   & WN10h       &             &                 &                \\
 54   & WN9h        & WN8ha       & $ -7.2 \pm 0.8$ & $ 6.4 \pm 0.4$ \\
 55   & WN11h       &             & $ -0.5 \pm 0.2$ & $ 4.7 \pm 1.5$ \\
 58   & WN6h        & WN7h        & $-38.8 \pm 1.3$ & $12.1 \pm 0.2$ \\
 68   & WN5-6       & Of/WN7      & $ -6.1 \pm 1.5$ & $10.0 \pm 1.0$ \\
 76   & WN9h        & WN9ha       & $ -8.4 \pm 0.6$ & $ 6.0 \pm 0.5$ \\
 77   & WN7         & WN7ha       & $-11.4 \pm 3.1$ & $15.0 \pm 1.6$ \\
 79   & WN7+OB      & WN7ha+OB    & $-26.0 \pm 0.8$ & $16.8 \pm 0.8$ \\
 80   & O4If/WN6    & WN5h:a      & $ -9.2 \pm 1.5$ & $24.3 \pm 3.3$ \\
 83   & Ofpe/WN9    & LBV         &                 &                \\
 89   & WN7h        &             & $-65.8 \pm 2.1$ & $12.3 \pm 0.7$ \\
 91   & WN7         & WN6h:a      & $-10.5 \pm 3.0$ & $16.1 \pm 0.5$ \\
 92   & WN6+B1Ia    & WN3:b(+O)   &                 &                \\
      &             &       +B1Ia & $-13.5 \pm 1.2$ & $46.0 \pm 3.6$ \\
 93   & O3If/WN6    &             & $ -4.8 \pm 0.5$ & $ 8.6 \pm 1.3$ \\
 95   & WN7h        & WN7         & $-84.6 \pm 9.8$ & $18.4 \pm 1.1$ \\
 96   & WN8(h)      & WN8         & $-26.4 \pm 6.0$ & $10.8 \pm 0.8$ \\
 97   & O3If*/WN7 &             & $ -6.2 \pm 1.6$ & $ 9.7 \pm 1.1$ \\
 98   & WN6(h)      & WN6         & $-19.4 \pm 6.2$ & $19.9 \pm 1.7$ \\
 99   & O3If*/WN6 &             & $ -4.1 \pm 0.6$ & $13.2 \pm 1.5$ \\
100   & WN6h        & WN7         & $-26.5 \pm 3.6$ & $14.4 \pm 1.7$ \\
102   & WN6+O       & WN6         & $-21.5 \pm 4.3$ & $16.4 \pm 2.0$ \\
103   & WN6         &             & $-27.1 \pm 5.7$ & $24.8 \pm 3.5$ \\
104   & O3If*/WN6 &             & $ -4.2 \pm 0.6$ & $12.5 \pm 1.5$ \\
105   & O3If*/WN6 & WN7         & $-18.7 \pm 5.2$ & $14.9 \pm 1.8$ \\
107   & WNL/Of      & WN9h::a     & $ -1.7 \pm 0.4$ & $ 5.9 \pm 0.7$ \\
113   & O3If*/WN6 &             & $ -5.1 \pm 1.6$ & $15.6 \pm 1.2$ \\
114   & O3If*/WN6 &             & $ -3.5 \pm 0.6$ & $17.2 \pm 1.5$ \\
116   & WN5h        & WN5h:a      & $-29.5 \pm 6.5$ & $26.0 \pm 0.8$ \\
118   & WN6h        &             & $-60.0 \pm 2.9$ & $18.7 \pm 0.9$ \\
119   & WN6h        &             & $-50.7 \pm 4.1$ & $17.3 \pm 1.1$ \\
120   & WN9h        &             & $ -5.0 \pm 1.5$ & $ 6.1 \pm 0.3$ \\
130   & WN11h       &             & $ -1.3 \pm 0.2$ & $ 4.9 \pm 0.6$ \\
133   & WN11h       &             & $ -1.1 \pm 0.2$ & $ 4.9 \pm 0.6$ \\
\hline
\end{tabular}
\end{table}


\begin{figure*}
\begin{minipage}{165mm}
\includegraphics[width=78mm,angle=-90,trim= 120 20 35 20,clip]{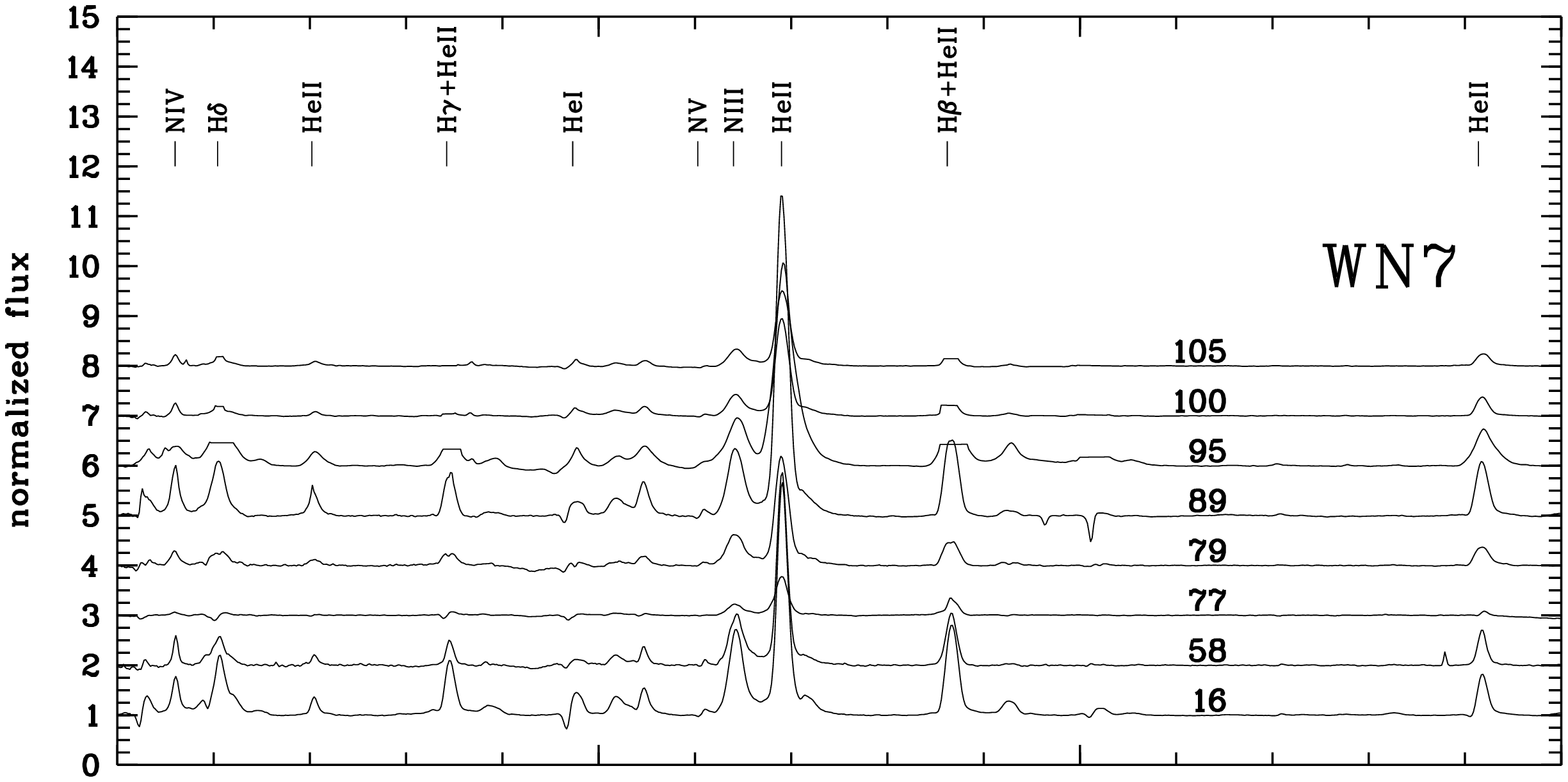}\vfill
\includegraphics[width=78mm,angle=-90,trim= 120 20 35 20,clip]{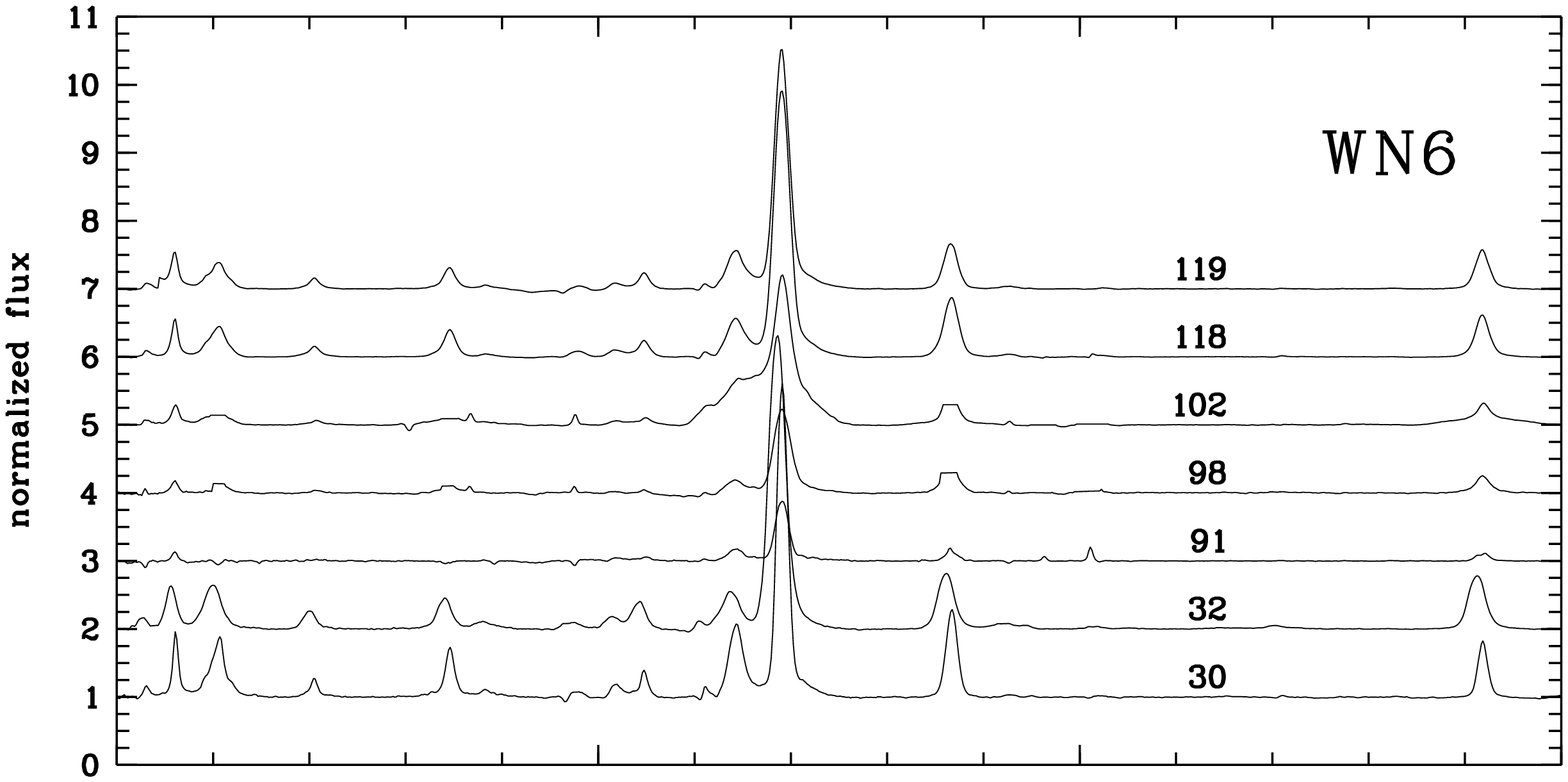}\vfill
\includegraphics[width=78mm,angle=-90,trim= 120 20 35 20,clip]{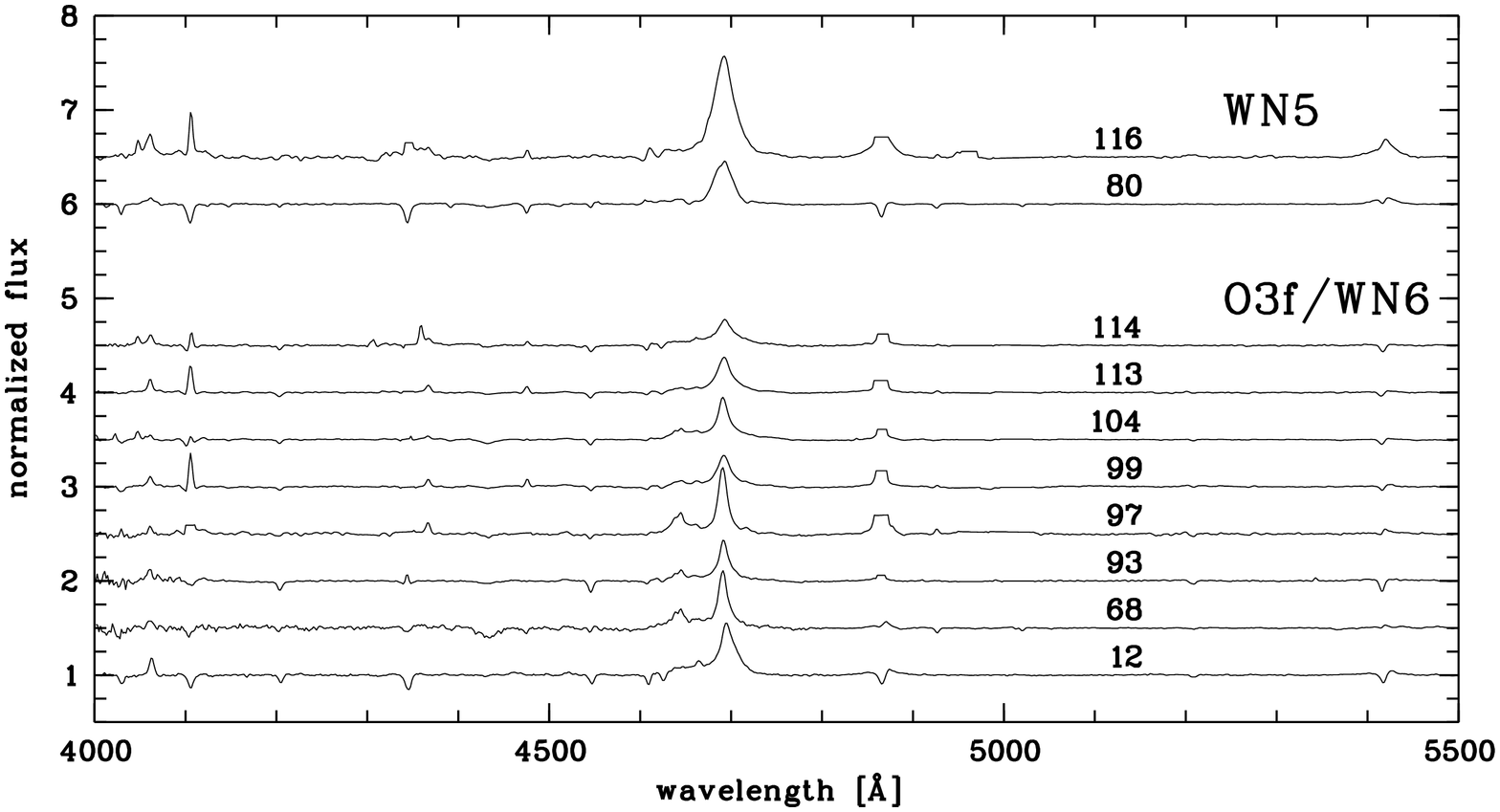}\vfill
\caption{Montages of mean spectra of our program stars. For clarity,
spectra have been shifted in flux. Note the different flux scales in
each plot. BAT99 numbers are indicated and important emission lines
identified.}
\label{spectypes}
\end{minipage}
\end{figure*}

\begin{figure*}
\begin{minipage}{165mm}
\includegraphics[width=78mm,angle=-90,trim= 120 20 35 20,clip]{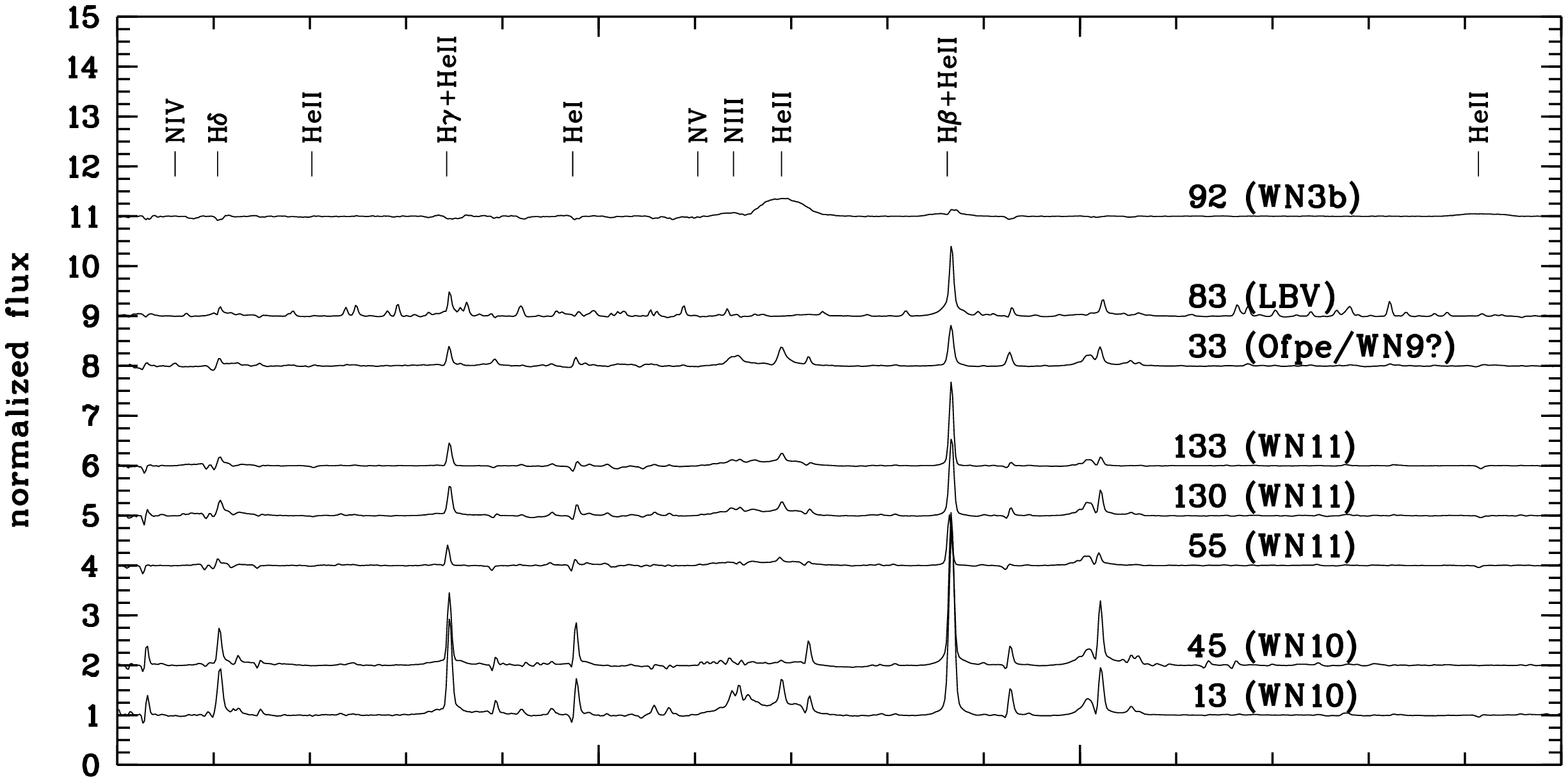}
\includegraphics[width=78mm,angle=-90,trim= 120 20 35 20,clip]{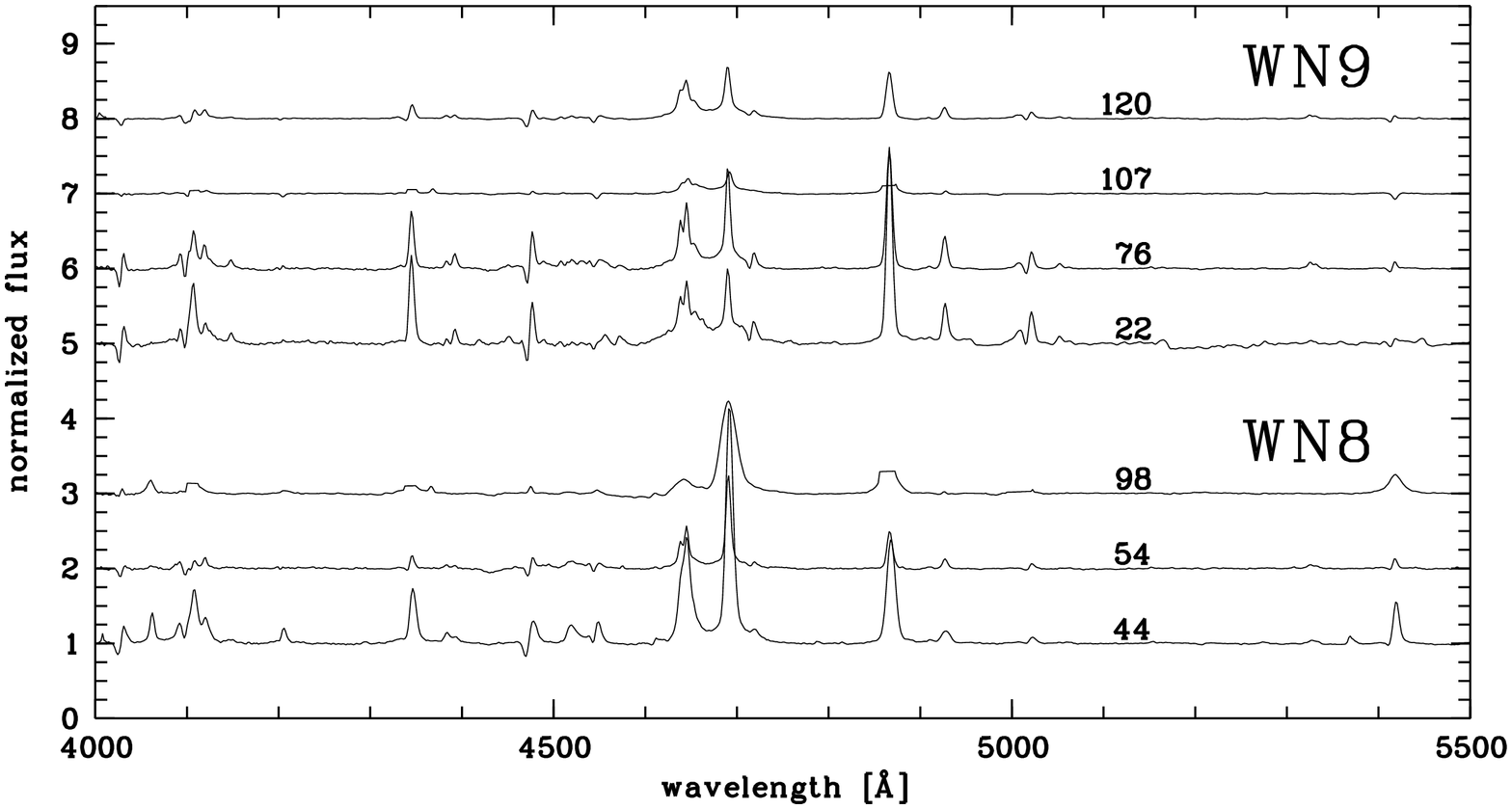}\vfill
\contcaption{}
\end{minipage}
\end{figure*}

\subsection{Analysis of the X-Ray Data}
\label{4.11}

\emph{Chandra} and \emph{ROSAT} archives were searched for X-ray
emission from the WNL stars in the LMC. We first extracted
\emph{Chandra} ACIS X-ray images in the 0.3--7.0 keV energy band, and
\emph{ROSAT} PSPC and HRI images in the full energy range of these
instruments, i.e., 0.1--2.4 keV for the PSPC and 0.1--2.0 keV for the
HRI.  We then compared these X-ray images with optical images
extracted from the Digitized Sky Survey\footnote{ The Digitized Sky
Survey (DSS) is based on photographic data obtained using the UK
Schmidt Telescope and the Oschin Schmidt Telescope on Palomar
Mountain. The UK Schmidt was operated by the Royal Observatory of
Edinburgh, with funding from the UK Science and Engineering Research
Council, until 1988 June, and thereafter by the Anglo-Australian
Observatory.  The Palomar Observatory Sky Survey was funded by the
National Geographic Society.  The Oschin Schmidt Telescope is operated
by the California Institute of Technology and Palomar Observatory.
The plates were processed into the present compressed digital form
with the permission of these institutes. The Digitized Sky Survey was
produced at the Space Telescope Science Institute under US government
grant NAGW-2166.} (DSS). We identified each WR star in the optical
images using the coordinates listed by BAT99, and then searched for
X-ray emission at the location of the WR star.

X-ray emission is detected in 15 WNL stars in the LMC, as indicated in
Table \ref{xrayall}.  To confirm these detections, we defined source
regions encompassing the X-ray sources at the location of the WR stars
and appropriate background regions. The background-subtracted count
number and count rates are also listed in Table \ref{xrayall}. For two
stars, BAT99-101/102 and BAT99-116, the background-subtracted count
number is large enough to render possible the analysis of their X-ray
spectra.  The spectral analysis has been performed adopting a
single-temperature MEKAL optically thin plasma emission model
(\citealt{Kaastra93}; \citealt{Liedahl95}), and the photoelectric
absorption model of \citet{Balucinska92} for the absorption along the
line of sight.  The chemical abundances of the X-ray-emitting gas and
for the absorbing material have been set to 0.33 $Z_{\sun}$.  The best
spectral fits indicate plasma temperatures of 0.9$\pm$0.2 and
5.0$^{+1.5}_{-1.0}$ keV and hydrogen column densities of
(9$\pm$2)$\times$10$^{21}$ and (9$\pm$2)$\times$10$^{21}$ cm$^{-2}$
for BAT99-101/102 and BAT99-116, respectively.  Further details of
these spectral fits are provided by \citet{Guerrero06}. The X-ray
luminosities in the 0.5--7.0 keV energy band of BAT99-101/102 and
BAT99-116 derived from these spectral fits are listed in Table
\ref{xrayall}. For the other 13 WNL stars detected in X-rays, their
X-ray luminosities listed in Table \ref{xrayall} have been derived
from their count rate, assuming that their X-ray emission follows a
single-temperature MEKAL optically thin plasma emission model with a
temperature of 1.6 keV and a hydrogen column density of
3$\times$10$^{21}$ cm$^{-2}$ (\citealt{Guerrero06}).

For the WNL stars undetected by \emph{Chandra} and \emph{ROSAT}
observations, we have derived count rate $3\sigma$ upper limits (see
Table \ref{xrayall}) using source regions with radii matching the PSF
of each observation. These limits assume the same single-temperature
MEKAL optically thin plasma emission model as above.


\begin{table*}
\centering
\footnotesize
\caption{List of our program stars for which X-ray data are available
from public archives.}
\label{xrayall}
\begin{tabular}{llccrrl}
\hline
BAT99 &  Instrument   &   Detection? & Count   &   Count rate or     &       $L_{x}$ or          & Comments\\
      &               &              & Number  & $3\sigma$ upper limit & $3\sigma$ upper limit   &\\
      &               &              & [cnts]  &    [cnts/s]         &   [ergs/s]                &\\
\hline
\hline
 12  &   ROSAT PSPC   &  No         &          &        $<1.1 \times 10^{-3}$  & $< 6.0 \times 10^{33}$ & binary(?)\\
 13  &   ROSAT PSPC   &  No         &          &        $<4.0 \times 10^{-4}$  & $< 2.2 \times 10^{33}$\\
 16  &   ROSAT PSPC   &  No         &          &        $<1.2 \times 10^{-3}$  & $< 6.6 \times 10^{33}$\\
 22  &   ROSAT PSPC   &  No         &          &        $<6.8 \times 10^{-4}$  & $< 3.7 \times 10^{33}$\\
 30  &   ROSAT PSPC   &  No         &          &        $<3.5 \times 10^{-4}$  & $< 1.9 \times 10^{33}$\\
 32  &   ROSAT PSPC   &  No         &          &        $<5.5 \times 10^{-4}$  & $< 3.0 \times 10^{33}$ & binary\\
 33  &   ROSAT HRI    &  No         &          &        $<3.2 \times 10^{-4}$  & $< 4.5 \times 10^{33}$\\
 44  &   ROSAT PSPC   &  No         &          &        $<4.3 \times 10^{-4}$  & $< 1.6 \times 10^{33}$\\
 45  &   ROSAT PSPC   &  No         &          &        $<3.3 \times 10^{-4}$  & $< 1.8 \times 10^{33}$\\
 54  &   ROSAT PSPC   &  No         &          &        $<2.7 \times 10^{-4}$  & $< 1.5 \times 10^{33}$\\
 55  &   ROSAT PSPC   &  No         &          &        $<2.8 \times 10^{-4}$  & $< 1.5 \times 10^{33}$\\
 58  &   ROSAT PSPC   &  No         &          &        $<1.0 \times 10^{-3}$  & $< 5.5 \times 10^{33}$\\
 68  &   ROSAT HRI    &  No         &          &        $<1.9 \times 10^{-4}$  & $< 2.7 \times 10^{33}$\\
 76  &   ROSAT PSPC   &  No         &          &        $<2.9 \times 10^{-4}$  & $< 1.6 \times 10^{33}$\\
 77  &   Chandra ACIS &  Yes        & $18\pm7$ &         $3.1 \times 10^{-4}$  & $(5\pm2) \times 10^{32}$  & binary\\
 79  &   Chandra ACIS &  Yes        & $19\pm7$ &         $3.1 \times 10^{-4}$  & $(5\pm2) \times 10^{32}$  & constant RV\\
 80  &   Chandra ACIS &  Yes        & $45\pm9$ &         $4.9 \times 10^{-4}$  & $(1.0\pm0.2) \times 10^{33}$ & constant RV\\
 83  &   Chandra ACIS &  No         &          &        $<1.3 \times 10^{-4}$  & $< 2.6 \times 10^{32}$\\
 89  &   Chandra ACIS &  No         &          &        $<3.7 \times 10^{-4}$  & $< 7.4 \times 10^{32}$\\
 91  &   Chandra ACIS &  No         &          &        $<5.0 \times 10^{-4}$  & $< 7.9 \times 10^{32}$\\
 92  &   Chandra ACIS &  Yes        & $12\pm4$ &         $6.7 \times 10^{-4}$  & $(1.3\pm0.4) \times 10^{33}$ & binary\\
 93  &   Chandra ACIS &  Yes        & $24\pm9$ &         $5.0 \times 10^{-4}$  & $(8\pm3) \times 10^{32}$ & constant RV\\
 95  &   Chandra ACIS &  No         &          &        $<3.2 \times 10^{-4}$  & $< 6.4 \times 10^{32}$  & binary\\
 96  &   Chandra ACIS &  No         &          &        $<3.5 \times 10^{-4}$  & $< 7.0 \times 10^{32}$\\
 97  &   Chandra ACIS &  No         &          &        $<3.1 \times 10^{-4}$  & $< 6.2 \times 10^{32}$\\
 98  &   Chandra ACIS &  No         &          &        $<3.6 \times 10^{-4}$  & $< 7.2 \times 10^{32}$\\
 99  &   Chandra ACIS &  Yes        & $77\pm9$ &         $4.4 \times 10^{-3}$  & $(1.3\pm0.2) \times 10^{34}$ & binary\\
 100  &   Chandra ACIS &  Yes        &  $8\pm3$ &         $4.1 \times 10^{-4}$  & $(8\pm3) \times 10^{32}$ & constant RV\\
101,102& Chandra ACIS &  Yes        &$330\pm20$&         $1.9 \times 10^{-2}$  & $(1.5\pm0.1) \times 10^{35}$ & slightly variable RV\\
103  &   Chandra ACIS &  Yes        &  $7\pm3$ &         $3.8 \times 10^{-4}$  & $(8\pm3) \times 10^{32}$ & binary\\
104  &   Chandra ACIS &  No         &          &        $<5.1 \times 10^{-4}$  & $< 1.0 \times 10^{33}$\\
105  &   Chandra ACIS &  Yes        & $21\pm5$ &         $1.2 \times 10^{-3}$  & $(2.4\pm0.6) \times 10^{33}$ & variable RV\\
107  &   Chandra ACIS &  Yes        & $16\pm4$ &         $9.2 \times 10^{-4}$  & $(1.8\pm0.5) \times 10^{33}$ & slightly variable RV\\
113  &   Chandra ACIS &  No         &          &        $<6.4 \times 10^{-4}$  & $< 1.3 \times 10^{33}$ & binary\\
114  &   Chandra ACIS &  Yes        &  $9\pm3$ &         $5.1 \times 10^{-4}$  & $(1.0\pm0.3) \times 10^{33}$ & slightly variable RV\\
116  &   Chandra ACIS &  Yes        &$860\pm30$&         $4.9 \times 10^{-2}$  & $(1.8\pm0.1) \times 10^{35}$ & variable RV\\
118  &   Chandra ACIS &  Yes        & $13\pm4$ &         $7.5 \times 10^{-4}$  & $(1.5\pm0.5) \times 10^{33}$ & variable RV\\
119  &   Chandra ACIS &  Yes        & $11\pm4$ &         $6.1 \times 10^{-4}$  & $(1.2\pm0.4) \times 10^{33}$ & binary\\
120  &   Chandra ACIS &  No         &          &        $<3.6 \times 10^{-3}$  & $< 5.7 \times 10^{33}$\\
130  &   ROSAT HRI    &  No         &          &        $<7.0 \times 10^{-4}$  & $< 1.0 \times 10^{34}$\\
133  &   ROSAT PSPC   &  No         &          &        $<7.1 \times 10^{-3}$  & $< 3.9 \times 10^{34}$\\

\hline
\end{tabular}
\end{table*}
\normalsize

\subsubsection{X-Rays and Binarity}

As can be seen from Table \ref{xrayall}, \emph{Chandra} ACIS is more
sensitive than \emph{ROSAT}'s instruments; the latter does not yield a
single detection of our program stars. We will therefore concentrate
on the \emph{Chandra} data for most of our discussion.

\emph{Chandra} observed 25 of our 41 program stars, detecting 15
sources. WR+O binaries are expected to be bright X-ray sources because
of the high-energy wind-wind collisions (WWCs) occurring in such
systems (e.g. \citealt{Pril76}). Indeed, phase-dependent profile
variations of the He\textsc{ii}$\lambda4686$ emission line are readily
observed in our binaries (not shown here, but see \citealt{Foell03b}
and \citealt{Bartz01}). However, of the seven binaries observed, only
five are detected, two of which, BAT99-99 and 119, are long-period
binaries. Two binary systems, BAT99-95 and 113, were missed, with
BAT99-95 apparently being a particularly faint X-ray source. One might
suspect that short-period, close binary systems produce fewer
observable X-rays, because the two winds have not yet reached their
respective terminal velocities, and/or the self-absorption in the wind
of generated X-ray photons might decrease the observable flux (see
e.g. \citealt{OwoCoh01}; \citealt{Ignace02}). Longer-period binaries,
on the other hand, have larger orbital separations, so their winds are
potentially faster and geometrically thinned, possibly decreasing
self-absorption of the X-rays and resulting in larger observed X-ray
fluxes. Along the same lines, one might also expect that binaries
containing hydrogen-depleted WNb stars generate more X-rays because
due to their higher mass-loss rates and faster winds, there is more
wind momentum available.

\begin{figure}
\includegraphics[width=60mm,angle=-90,trim= 0 0 0 0,clip]{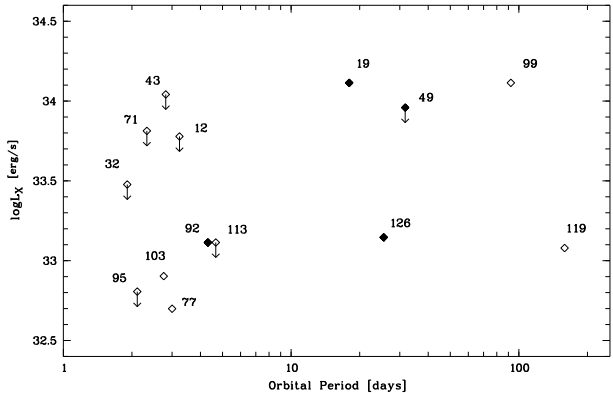}
\caption{X-ray luminosities of identified WN binaries in the LMC as
a function of their orbital period. Filled symbols denote broad-lined,
hydrogen-depleted WNb stars; empty symbols denote H-containing WN
stars. Upper limits are indicated by downward-pointing arrows, and
stars are named by their BAT99 numbers. As can be seen, WNb binaries
do not show higher X-ray luminosities as might have been expected from
the stronger winds of WNb stars.}
\label{xrayperiod}
\end{figure}

In order to test these speculations, we combined our data with those
of \citet{Foell03b} and plotted the observed X-ray flux \emph{versus}
the orbital period (Figure \ref{xrayperiod}). Non-detections of both
\emph{Chandra} and \emph{ROSAT} are given with their upper limit, by
downward-pointing arrows. As can be seen, there is no readily apparent
trend in the data which could support that longer-period or WNb
binaries indeed do display larger X-ray fluxes than shorter-period
systems or such which contain WR stars with lower mass-loss
rates. However, more data might help to clarify the situation.

Interestingly, BAT99-99 is X-ray brighter than BAT99-119, but this
could be due in part to the large eccentricity ($e \sim 0.7$; Schnurr
et al., in preparation) of BAT99-119. It is thus not unlikely that
\emph{Chandra} missed the moment of maximum X-ray luminosity for
BAT99-119 during periastron passage. Note however that the circular
orbit which was assumed for BAT99-99 might be incorrect, and that this
binary is eccentric, too. In fact, both systems merit a closer look
for orbit-related X-ray variability.

\subsubsection{X-Rays in Apparently Single Stars}
\label{xraysingle}

Of the remaining 10 stars detected by \emph{Chandra}, four have X-ray
luminosities which are comparable to those of the confirmed binaries
although they do not show RV variability: BAT99-79, 80, 93, 100. Six
do display RV variability: BAT99-101/102, 105, 107, 114, 116, and
118. Crowding might account for the recorded high X-ray fluxes, and
there still is the possibility that the X-rays originate in single WR
stars due to radiatively-induced instabilities in their winds
(\citealt{Lucy80}; \citealt{Willis96}). It can however not be ruled
out that these stars are long-period ($P>200$ days) binaries or
systems with very low inclination angles.

Two detected sources, BAT99-101/102 and 116, are extremely
luminous. BAT99-101/102 are two visually very close stars, and thus
\emph{Chandra}'s aperture integrated the combined flux. BAT99-101 is a
WC4 star (\citealt{Bartz01}) and potentially a binary
(\citealt{Bartz01}), while for BAT99-102, there is some confusion: As
detailed above, we find that BAT99-103 is the 2.76-day binary, not
102, as was reported by \citet{M89}. However, even if 102 is single,
there might be sufficient wind momentum confined in a very small
volume, and if both 101 and 102 are binary, then there most certainly
is enough WWC occurring to account for the observed X-ray flux.

BAT99-116, on the other hand, is visually isolated enough not to be
subject to such ambiguities. BAT99-116 is even more X-ray luminous
than the combined BAT99-101/102 system, and comparable to the X-ray
brightest WR stars known, the Galactic WN6ha stars NGC3603-C
(\citealt{Moff02}) and WR25 (\citealt{Seward82}). Both NGC3603-C
(Schnurr et al., in prep.) and WR25 (\citealt{Gamen06}) have been
identified as binaries, and they both contain two of the most luminous
WN stars known, so that there is a plausible explanation as to why
these objects are so X-ray bright. However, although 116 displays
significant RV variability (Section \ref{randomvar}), no indication of
binarity could be found. Unless the photometry published by BAT99 is
wrong, 116 is not particularly optically bright, either, $\sim$1.5 mag
fainter than BAT99-119, which is a confirmed binary. Since both stars
are similar enough (WN5ha for 116, and WN6ha for 119; see above) to
have the $\sim$same bolometric correction (no reddening considered,
but it is low and probably $\sim$similar for the two stars anyway),
116 is considerably less luminous and hence less massive than
119. Thus, if 116 is indeed single, its ratio $L_{x}/L_{bol}$ is
abnormally high, but if X-rays originate from WWCs in a binary, the
question remains how such faint objects (the total mass has now to be
split between two stars) can provide the required wind momenta. This
very intriguing system merits a closer look.

As to BAT99-118, the sheer bolometric luminosity of the object
(\citealt{CroDess98}) renders a two-star scenario more
likely. BAT99-118 closely resembles 119 in terms of both spectral type
and X-ray luminosity (the latter is, in fact, equal within the
errors). It is thus possible that BAT99-118 will turn out to be a
long-period binary as well, especially given the fact that it also
shows a significantly large RV scatter.

%
%

\section{Discussion}
\label{wnlsection5}

\subsection{Binary Detectability and Completeness}
\label{phasefill}

Before we can discuss the binary frequency among WNL stars in the LMC,
we have to address the question of how many binaries we missed. To do
so, let us consider some general aspects of binary detection. Most
approaches dealing with how to detect a binary through RV variations
are based on statistical tests which compare the observed RV scatter
$\sigma_{\rm RV}$ of a binary candidate to the (Gaussian) scatter of
an observed, constant comparison star (via a $\chi^{2}$ test), just as
we have done in Sec. \ref{randomvar}. How large does the RV amplitude
of a given binary have to be so that it can be identified as a
RV-variable star? The RV amplitude $K_{\rm RV}$ of the WR is given by

\noindent
\begin{eqnarray*}
K_{\rm WR} = 212.7 M_{\rm O}/(M_{\rm WR}+M_{\rm O})^{2/3} P^{-1/3} (1-e^{2})^{-1/2} \sin{i}, 
\end{eqnarray*}

\noindent
where $K_{1}$ is in kms$^{-1}$, all masses in $M_{\odot}$, and $P$ in
days. For a circular orbit ($e=0$) and \emph{continuous} sampling, the
RV scatter is $\sigma_{\rm RV} = K/\sqrt{2}$, and it has to be larger
than some detection threshold $\sigma_{\rm cut}$ for the star to be
considered as significantly variable. In Section \ref{randomvar}, we
defined the 99.9\%ile to be $\sigma_{\rm cut} = 22.6$ kms$^{-1}$.

\begin{figure}
\includegraphics[width=60mm,angle=-90,trim= 0 0 0
0,clip]{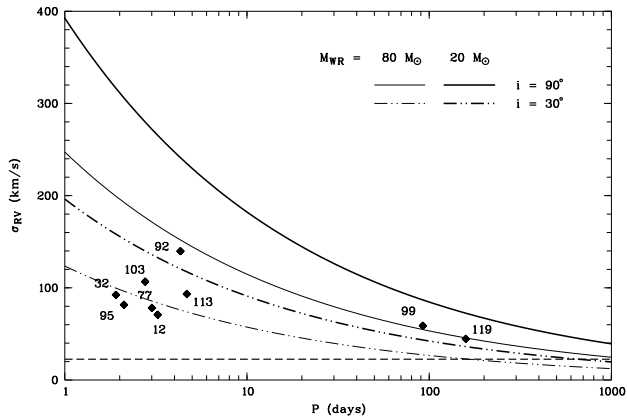}
\caption{Curves show RV scatter $\sigma_{\rm RV} = K / \sqrt{2}$ for
the WN component in two fictitious, continuously sampled, circular
binaries, computed for periods ranging from 1 to 1000 days. The mass
of the O-type secondary is fixed at 40 $M_{\odot}$, while the mass of
the WN component is $80 M_{\odot}$ (thin lines) or $20 M_{\odot}$
(thick lines), to represent the unevolved and the evolved cases. Solid
lines indicate $i = 90^{\circ}$, whereas dash-dotted lines are for $i
= 30^{\circ}$. The dashed, horizontal line indicates the 99.9\%
detection threshold of our $\chi^{2}$ test. Identified WNL binaries
with both their observed RV scatter and their BAT99 numbers are
given. For BAT99-119, we have used the period reported by Schnurr et
al. (in preparation).}
\label{detection}
\end{figure}

As already mentioned, some very luminous stars in our sample, such as
the WN5-7ha and O3f/WN6 stars, are not evolved, classical, He-burning
WN stars, but unevolved, very massive objects; when situated in binary
systems, they usually are (much) more massive than their companion,
while classical, evolved WN stars usually are the less massive
component. To cover the range of possible mass ratios, we have thus
fixed the mass of the O-type component to 40 $M_{\odot}$ and computed
the $\sigma_{\rm RV}$'s for two different masses of the WN component,
80 $M_{\odot}$ (for the unevolved case) and 20 $M_{\odot}$ (for the
evolved case), and for two different inclination angles, $i =
90^{\circ}$ and $i = 30^{\circ}$, assuming circular orbits. Values of
$\sigma_{\rm RV}$ are plotted versus the binary period $P$ in Figure
\ref{detection}.

It can clearly be seen that all binaries are well above our detection
threshold, and that we should in principle be able to see evolved WN
stars with periods somewhat longer than 100 days. In particular, there
is no reason why we should not be able to see systems with periods
between 5 and 90 days, of which there is an apparent lack in our
study. To miss these systems, their inclination angles would have to
be extremely low. However, as \citet{Foell03a} have shown, in a sample
with randomly distributed inclination angles there are 6.5 times more
binaries between $90^{\circ}$ and $30^{\circ}$ than there are between
$30^{\circ}$ and $0^{\circ}$; thus, it is very unlikely that there is
a significant number of low-inclination systems in our sample, in
particular if these harbour evolved (i.e., lower-mass) WN stars, which
are the primary focus of our study.

Let us further note that, while from our RV data alone, we were not
able to detect the binary nature of BAT99-119, because its period of
$\sim$160 days is too long and its orbit is too eccentric, we did
detect its large RV scatter, which firmly puts it above our variability
threshold. We were also able to identify BAT99-99's period of $\sim$92
days and even obtain an orbital solution (albeit a preliminary one),
despite the fact that this O3f*/WN6 star is an unevolved object, too,
and thus more massive than its companion.

Thus, we can be quite confident that systems with periods between 5
and 90 days are indeed absent in our program stars, and that we are
quite complete for evolved WN binaries with periods up to 200 days. Of
course, this estimate of our detection limits is very rough, because
elliptical orbits and discrete sampling will modify the individual
detection limit of a binary system. Since we have also searched for
periodicities in our RV data, the problem of binary detection is
shifted into the realm of Fourier statistics. (Note, however, that
cyclical variability in the RV data is a necessary, but not a
sufficient condition for binarity, because rotation or pulsation could
induce a periodically variable RV pattern.) In this case, the whole
situation becomes more intricate, since not only the total time span
$T$ covered by observations compared to the systemic periods $P$
becomes important, but also the distribution of data points in
phase. Detection limits do not directly depend on the ratio of the RV
ampitude $K$ of the primary star (the signal) to the measurement error
$\sigma$ (the noise) in data space, but on the S/N ratio achieved in
the frequency domain.

At a given number of data points covering a certain time span, short
periods generally have a large $T/P$ and better ``phase-filling'', and
therefore have a larger detection probability than longer periods
which suffer from holes in the phase coverage and a lower S/N per
frequency unit. Vice versa, the same detection probability as for
longer periods can be obtained for shorter periods but lower RV
amplitudes, which means that low-$K$ (low-inclination) binaries with a
given $K$ tend to be detected more easily if they are
short-period.

In the case of non-circular orbits, the eccentricity, the orientation
of the orbital ellipse, and the time of periastron passage seriously
affect the detection probability of periods for two reasons. First,
the orbital RV curve becomes distorted in a way that generates
harmonics in the Fourier spectrum. Power is transferred from the
fundamental to the harmonics, thereby lowering the peak of the
fundamental frequency. If the orbit is highly eccentric, the
fundamental peak might be pushed into the noise floor, and the period
will not be detected. Second, the distribution of the data points over
the orbital phase is even more critical, because the star spends most
of the time near apastron; if the passage of periastron, where the RVs
change rapidly, is missed, the RV curve will look flatter than it
really is. This is a problem for any $\sigma_{\rm RV}$-based
statistics in velocity space, because it is possible that observed RVs
display cyclical variability in the sense that a period has been
found, although the observed RV scatter is \emph{not} significantly
large.

The standard approach to deal with this problem is a Monte-Carlo
simulation of a population of artificial binaries by randomly drawing
orbital parameters following a predetermined distribution function,
and applying the desired test statistics in RV and/or frequency space,
or by actually full Keplerian fitting; for reasons of convenience and
computational expense, usually only the former is done. From the
detected fraction of artificial binaries, one can reconstruct the true
binary frequency among the artificial sample by statistical
means. (For an excellent description of this approach and all related
problems, see \citealt{Kouw06}).

The main problem, however, is to determine reasonable distribution
functions for the respective orbital parameters, because the results
of the simulation will obviously depend strongly on what initial
assumptions were made for the underlying, true binary population. For
their WNE stars in the LMC, \citet{Foell03b} used distribution
functions for Galactic O stars (based on statistics published by
\citealt{Mason98}; also see references therein); they found that 35\%
of the binaries in their sample were missed. For our study, the most
relevant assumption Foellmi et al. used was that the orbital periods
$P$ are distributed flat in log$P$, i.e. there are as many binaries
between 1 and 10 days as there are between 10 and 100 days, etc. In
our WNL sample, we have found eight binaries with periods shorter than
100 days. Of these eight binaries, only one, BAT99-99, has a period
between 10 and 100 days; the remaining seven binaries have periods
ranging between 1 and 10 days. Thus, without any sophisticated Monte
Carlo simulation, we can immediately determine that we would have
missed six binaries in the period range from 10 to 100 days, but of
course only if we believed that we are complete in the period range
from 1 to 10 days and that our chosen period distribution is correct.

But how sound is this assumption? Indeed, the distribution function
for pre(!)-RLOF, O+O binary periods is not suited for statistics of
post-RLOF populations, when one of the model predictions is that
orbital periods change due to binary interaction. Clearly, however,
any other choice of distribution function would produce a different
result. Moreover, our analysis above indicates that it is unlikely
that we missed binaries between 5 and 90 days, and of course one must
not forget that we operate with small numbers. Thus, even if we had a
way to obtain the true number of expected binaries from the number of
observed binaries, we would have to show first that the observed
number is statistically not consistent with the expected value;
otherwise, any correction would not be justified. Because of those
severe imponderabilities, we not carry out any statistical correction,
but consider the number of binaries that we have found as a lower
limit only. We will also consider that, whatever the detection bias
is, \citet{Bartz01}, \citet{Foell03b}, and this study suffer from it
in roughly the same way, given that we have carried out more or less
the same kind of observations. In regard to the achieved RV precision,
one can argue that the somewhat better precision obtained in this
study is compensated by the fact that classical WNL stars are expected
to be, on average, more massive than WNE and WC/WO stars, because they
had less time to shed mass by stellar winds; thus, RV amplitudes of
WNL binaries will, on average, be somewhat smaller than those of WNE
and WC/WO binaries. Since neither of the three studies used a more
sophisticated detection threshold than the RV scatter of (presumably)
constant, reference stars, we feel that all three studies have
$\sim$the same detection probability for periods up to $\sim$200
days. Thus, we consider the three studies to be directly comparable.

\subsection{The Binary Frequency among the WNL Stars in the LMC}

The present survey encompassed all 41 WNL stars listed in the BAT99
catalogue outside R136. The remaining 6 WNL stars in the core of R136
required higher spatial resolution, e.g. adaptive-optics assisted
spectroscopy with VLT/SINFONI; the work is in progress and results
will be published elsewhere (Schnurr et al., in preparation), but
preliminary results indicate that there is no short-period ($P \la 20$
days) binary among those stars. Foellmi et al. (2003b) reclassified
one of their program stars, BAT99-78, to WN6 (i.e. to WNL); in turn,
one of our stars, BAT99-92, was recognized to be an early-type WN3b
star (cf. Section \ref{reclass}). Thus, the respective numbers of WNE
and WNL stars which have been studied remain unchanged: There are 47
WNL stars in the LMC.

\begin{table}
\centering
\caption{Known binaries among the WNL population of the LMC, where the
O3f*/WN6A stars have been counted in. The improved period for
BAT99-119 was taken from Schnurr et al. (in prep).}
\label{binaries}
\begin{tabular}{lrll}
\hline
BAT99  &  Period & Spectral Type & Comments\\
       &  [days] &  (this study) & \\
\hline
12           &   3.24        & O3f*/WN6A   & new  \\
32           &   1.91        & WN6(h)      &       \\
77           &   3.00        & WN7ha       &       \\
95           &   2.11        & WN7         & new   \\
99           &  92.60        & O3f*/WN6-A  & new   \\
103          &   2.76        & WN6         &       \\
113          &   4.70        & O3f*/WN6-A  & new   \\
119          & 158.80        & WN6(h)      &       \\
\hline
\end{tabular}
\end{table}

In this study, we are only interested in binaries with periods up to
$\sim200$ days. A binary counts as detected only if an orbital
solution could be established. Using this criterion, eight stars have
been recognized as binaries (cf. Table \ref{binaries}): BAT99-12, 32,
77, 92, 95, 99, 103, 113. By combining our RV data with those of
\citet{M89} and with unpublished polarimetry, a ninth binary,
BAT99-119, was identified (Schnurr et al., in
preparation). \citet{M89} reported 6 binaries among the 14 WNL stars
he had studied. Our study has added four previously unknown WNL binary
systems to that list, and confirmed four known binaries and their
orbital periods. One of Moffat's (1989) binaries, BAT99-107, was
however found to be a single star. From significantly large RV
scatter, visual brightness or high X-ray luminosity, two additional
stars, BAT99-116 and 118, qualify as binary candidates in the
period-range of interest (i.e. up to 200 days), while more stars might
be binaries with (considerably) longer periods than that. With
BAT99-32 being a WNE binary, the confirmed WNL binary frequency is
$8/41 = 20\%$.

Interestingly, no binaries or RV variables can be found among the
WN8,9 stars in our sample -- as a matter of fact, they yield the
smallest RV scatter of our sample stars --, whereas WN6,7 stars do
show binaries among them. This seems to confirm the findings of Moffat
(1989), who suspected that this dichotomy reflected deeper differences
between WN6,7 and WN8,9 stars. He argued that Galactic WN8,9 stars
tended to be single runaways (an observation that we cannot confirm
from the systemic velocities measured for our WN8,9 stars in our
sample), and that in the LMC, WN8,9 stars tend to avoid clusters, very
much unlike WN6,7 star. While since \citet{M89}, the spectral type of
the Galactic binary WR12 (\citealt{Rauw96a}) was revised from WN7 to
WN8 by \citet{SSM96}, Moffat's (1989) conclusion still holds.

\subsection{Comparison with WNE and WC/WO stars}

Among the 61 WNE stars in the LMC, \citet{Foell03b} reported only five
certain binaries, two systems with unreliable orbital solutions (which
we count as identified), and two potential binaries; we have added to
this BAT99-92, so that the confirmed binary frequency among the WNE
stars in the LMC is now $8/61 = 13$\%. If one combines the results for
WNE and WNL stars, one obtains 102 WN stars studied among which
$8+8=16$ are confirmed binaries; thus, the confirmed binary frequency
is $16/102 = 15\%$. Among the 24 WC/WO stars, \citet{Bartz01} studied
23 and reported three confirmed binaries. The 24$^{th}$ star is very
faint and thus likely single. While they also reported 5 potential
binaries, even the one with the largest RV scatter, MG6, failed to be
significantly (99\% confidence level) variable; therefore, we consider
their remaining binary candidates to be constant as well. This brings
the binary frequency among 24 WC/WO stars in the LMC to 3; thus, the
binary frequency is $3/24 = 13\%$. Values are quoted in Table
\ref{popratios}. Note that the binary frequencies among the different
subgroups (WNL, WNE, and WC/WO) are statistically fully consistent
with each other.

\begin{table}
\centering
\small
\caption{Spectroscopic binary frequencies for three different
subgroups in the LMC.}
\label{popratios}
\begin{tabular}{lrl}
\hline
Subgroup   &      Frequency       & References\\
\hline
WNL     &   8/ 41 = 0.20 & this study\\
WNE     &   8/ 61 = 0.13 & Foellmi et al. (2003b)\\
WNL+WNE &  16/102 = 0.15 & \\
WC+WO   &   3/ 24 = 0.13 & Bartzakos et al. (2001)\\
\hline
WR all  &  19/126 = 0.15 & \\
\hline
\end{tabular}
\end{table}

\section{Summary and Conclusion}
\label{wnlsection6}

We have carried out spectroscopic monitoring of all 41 WNL stars in
the LMC that could be observed by conventional, ground-based
observations. Measured RV curves were used to identify binaries with
orbital periods from 1 to $\sim200$ days, because these systems were
expected to be post-RLOF candidates (cf
\citealt{Vanbev98}). Additionally, publicly available archive data
from X-ray satellite missions were searched for our program stars to
obtain X-ray luminosities. The results of our study can be summarized
as follows (see also Table \ref{binaries}):

\begin{itemize}

\item{We have identified four previously unknown binary systems:
  BAT99-12, 95, 99, and 113.}

\item{We confirmed the previously known binaries BAT99-32, 77,
  92. However, while we could reproduce the 2.76-day period that
  \citet{M89} had reported for star BAT99-102, we did so for star
  BAT99-103. It presently remains unknown whether \citet{M89} or we
  wrongly identified the binary.}

\item{We also confirmed that BAT99-119 is a binary; however, we had to
combine our RV data with those of \citet{M89} to do so, and it
required further combination with previously unpublished polarimetry
to identify the 159-day period of the system. For reasons of
completeness, we list the results here, but the complete study will be
published elsewhere (Schnurr et al., in prep).}

\item{One star, BAT99-107, had been suspected binary by \citet{M89},
but we were unable to reproduced that results. Therefore, we consider
105 to be single (i.e., not a binary with an orbital period in the
quoted range).}

\item{Two binary candidates were identified from their RV variability
and their X-ray luminosities, BAT99-116 and 118. Both systems merit a
closer look, because 116 is one of the brightest X-ray sources among
all WR stars, while 118 is the most luminous WR star and thus, the
most luminous unevolved star known in the Local Group
(cf. \citealt{CroDess98}).}

\item{One of our program stars, BAT99-92, a binary, was recognized to
be a WNE and not a WNL star.}

\item{Thus, our study brings the total number of known WNL binaries to
8, and the binary frequency among WNL stars in the LMC to 20\%, which
is fully consistent with the results for WC/WO stars
(\citealt{Bartz01}) and WNE stars (\citealt{Foell03b}); thus, there is
no statisticially significant differences between different WR
populations in the LMC. However, the overall binary frequency is only
half of what was predicted from model results by
\citet{MaedMey94}. The implications of this low binary frequency for
massive-star evolution will be discussed in a forthcoming paper.}

\end{itemize}

Remarkably, none of the WNL binaries contains a classical,
hydrogen-deficient, helium-burning WR star; instead, the WN components
are young, unevolved, objects (hot O3If/WN6 stars or more extreme
WN5-7ha stars), which most likely are very luminous and hence very
massive Of stars, and possibly even the most massive stars
known. These binaries offer the tremendous opportunity to directly
weigh these extreme stars using model-independent, Keplerian
orbits. Follow-up observations have partly been obtained and are
currently reduced, or are under way. The results of these observations
will be published elsewhere.

\section*{Acknowledgments}

OS is grateful to the various TACs for the generous allocation of
telescope time and for the very warm hospitality at all involved
observatories, as well as the quick, uncomplicated and competent help
of the night assistants, telescope operators, and observatory staff in
all situations. OS particularly thanks St\'ephane Vennes for
introduction at the Mount Stromlo 74-in telescope and his help during
the observing runs, and Marilena Salvo for introduction at the Siding
Spring 2.3-m telescope. OS also thanks the Mount Stromlo
administration for uncomplicated re-scheduling at Siding Spring after
the 74-in telescope at Stromlo was lost in the January 18, 2003
bushfires. OS is grateful to C\'edric Foellmi for having provided data
of the WNE stars, to Andr\'e-Nicolas Chen\'e for help and very
fruitful discussions, and to Paul Crowther for comments that led to
improvement of this manuscript. AFJM and NSL are grateful for
financial aid from the Natural Sciences and Engineering Council
(NSERC) of Canada.


\label{lastpage}

\end{document}